\documentclass[pra,aps,twocolumn,superscriptaddress]{revtex4}
\usepackage{hyperref}
\usepackage{bm}
\usepackage{epsfig}
\usepackage{amssymb}
\usepackage{graphicx}
\usepackage{mathrsfs}
\usepackage{amsmath}
\usepackage{epsfig}
\DeclareMathOperator{\arccosh}{arccosh}

\usepackage{color}
\usepackage[normalem]{ulem}
\usepackage{soul}
\begin{document}
\title{Gaussian intrinsic entanglement: An entanglement quantifier based on secret correlations}
%\title{Supplementary Information: Gaussian intrinsic entanglement}
\author{Ladislav Mi\v{s}ta, Jr.}
\affiliation{Department of Optics, Palack\' y University, 17.
listopadu 12,  771~46 Olomouc, Czech Republic}
\author{Richard Tatham}
\affiliation{Department of Optics, Palack\' y University, 17.
listopadu 12,  771~46 Olomouc, Czech Republic}
%\affiliation{???}
%\affiliation{School of Physics and Astronomy, University of St.
%Andrews, North Haugh, St. Andrews, Fife, KY16 9SS, Scotland}

\begin{abstract}
Intrinsic entanglement (IE) is a quantity which aims at
quantifying bipartite entanglement carried by a quantum state as
an optimal amount of the intrinsic information that can be
extracted from the state by measurement. We investigate in detail
the properties of a Gaussian version of IE, the so-called Gaussian
intrinsic entanglement (GIE). We show explicitly how GIE
simplifies to the mutual information of a distribution of outcomes
of measurements on a conditional state obtained by a measurement
on a purifying subsystem of the analyzed state, which is first
minimized over all measurements on the purifying subsystem and
then maximized over all measurements on the conditional state. By
constructing for any separable Gaussian state a purification and a
measurement on the purifying subsystem which projects the
purification onto a product state, we prove that GIE vanishes on
all Gaussian separable states. Via realization of quantum
operations by teleportation, we further show that GIE is
non-increasing under Gaussian local trace-preserving operations
and classical communication. For pure Gaussian states and a
reduction of the continuous-variable GHZ state, we calculate GIE
analytically and we show that it is always equal to the Gaussian
R\'{e}nyi-2 entanglement. We also extend the analysis of IE to a
non-Gaussian case by deriving an analytical lower bound on IE for
a particular form of the non-Gaussian continuous-variable Werner
state. Our results indicate that mapping of entanglement onto
intrinsic information is capable of transmitting also quantitative
properties of entanglement and that this property can be used for
introduction of a quantifier of Gaussian entanglement which is a
compromise between computable and physically meaningful
entanglement quantifiers.

\end{abstract}

\maketitle

\section{Introduction}

%%%%%%%%%%%%%%%%%%%%%%%%%%%%%%%%%%%%%%%%%%%%%%%%%%%%%%%%%%%%%%%%%%%%%%%%%%%%%%%%%%%%%%%%%%%%%%
Since the dawn of quantum information theory its development has
been guided by the findings of classical information theory.
Indeed, some key quantum information concepts including early
entanglement distillation protocols \cite{Bennett_96a}, quantum
error correction \cite{Shor_95} and some fundamental quantum
information inequalities \cite{Nielsen_00}, appeared initially as
nontrivial translations of their classical counterparts into the
language of quantum states. Naturally, the further independent
development of quantum information theory has led to the emergence
of concepts with no analogy in classical theory. This category
includes, for instance, bound entanglement \cite{MHorodecki_98},
entanglement distribution by separable states \cite{Cubitt_03} and
superactivation of entanglement \cite{Shor_03}. It is not
surprising then, that the opposite effect occurred when quantum
information started to enrich classical information theory with
new concepts such as bound information \cite{Gisin_00,Acin_04},
secrecy distribution by non-secret correlations \cite{Bae_09} and
a classical analogy to superactivation \cite{Pretticio_11}.

Classical analogies of quantum phenomena are almost exclusively
cryptographic analogies of some properties of quantum
entanglement. Entanglement is the key resource in quantum
information and it is synonymous with correlations among two or
more quantum systems which cannot be prepared by local operations
and classical communication (LOCC). The cryptographic parallels of
entanglement properties are carried by classical probability
distributions containing so called secret correlations
\cite{Maurer_93,Renner_03}. The correlations are a fundamental
resource in cryptography and appear in the scenario when two
honest parties, Alice and Bob, and an adversary Eve, share three
correlated random variables $A,B$ and $E$ obeying a probability
distribution $P(A,B,E)$. The distribution carries secret
correlations if it is impossible for Alice and Bob to create the
distribution by local operations and public communication
\cite{PC}. Owing to the apparent similarity with entanglement,
secret correlations can therefore be viewed as a classical analogy
to entanglement \cite{Collins_02}. In fact, secret correlations
and quantum entanglement are not just analogs but are directly
linked as the latter can be mapped onto the former as follows
\cite{Acin_05}. A third adversary party Eve, seemingly missing in
a quantum state $\rho_{AB}$, is associated with all information
which could potentially be carried by a third system $E$, i.e.,
the global state $|\Psi\rangle_{ABE}$ of the tripartite system is
a purification of the state $\rho_{AB}$
($\mbox{Tr}_{E}|\Psi\rangle_{ABE}\langle\Psi|=\rho_{AB}$). A given
quantum state $\rho_{AB}$ can then be mapped onto a probability
distribution $P(A,B,E)$ by performing measurements
$\Pi_{A},\Pi_{B}$ and $\Pi_{E}$ on subsystems $A,B$ and $E$ of the
purification as \cite{Acin_05}
%%%%%%%%%%%%%%%%%%%%%%%%%%%%%%%%%%%%%%%%%%%%%%%%%%%%%%%%%%%%%%%%%%
\begin{equation}\label{mapping}
P(A,B,E)=\mbox{Tr}(\Pi_{A}\otimes \Pi_{B}\otimes
\Pi_{E}|\Psi\rangle_{ABE}\langle\Psi|).
\end{equation}
%%%%%%%%%%%%%%%%%%%%%%%%%%%%%%%%%%%%%%%%%%%%%%%%%%%%%%%%%%%%%%%%%%

The presence of secret correlations in the obtained distribution
can be certified with the help of the so-called intrinsic
conditional information defined as \cite{Maurer_99}
%%%%%%%%%%%%%%%%%%%%%%%%%%%%%%%%%%%%%%%%%%%%%%%%%%%%%%%%%%%%%%%%%%%%%%%%%%%%%%%%%%%%%%%%%%%%%%%%%%%%
\begin{eqnarray}\label{intrinsicI}
I\left(A; B\downarrow E\right)=\mathop{\mbox{inf}}_{E\rightarrow
\tilde{E}}[I(A; B| \tilde{E})].
\end{eqnarray}
%%%%%%%%%%%%%%%%%%%%%%%%%%%%%%%%%%%%%%%%%%%%%%%%%%%%%%%%%%%%%%%%%%%%%%%%%%%%%%%%%%%%%%%%%%%%%%%%%%%%%
Here
%%%%%%%%%%%%%%%%%%%%%%%%%%%%%%%%%%%%%%%%%%%%%%%%%%%%%%%%%%%%%%%%%%%%%%%%%%%%%%%%%%%%%%%%%%%%%%%%%
\begin{eqnarray}\label{conditionalI}
I(A; B|E)&=&H(A,E)+H(B,E)-H(A,B,E)\nonumber\\
&&-H(E)
\end{eqnarray}
%%%%%%%%%%%%%%%%%%%%%%%%%%%%%%%%%%%%%%%%%%%%%%%%%%%%%%%%%%%%%%%%%%%%%%%%%%%%%%%%%%%%%%%%%%%%%%%%%
is the mutual information between $A$ and $B$ conditioned on $E$,
where $H(X)$ is the Shannon entropy \cite{Shannon_48}, and the
minimization is performed over all channels
$E\rightarrow\tilde{E}$ characterized by a conditional probability
distribution $P(\tilde{E}|E)$. The intrinsic information gives a
lower bound to the information of formation \cite{Renner_03}
quantifying the amount of secret bits \cite{secret bit} needed for
preparation of the distribution, and an upper bound to the rate at
which a secret key can be distilled from the distribution
\cite{Maurer_99} in the secret-key agreement protocol
\cite{Maurer_93}. More importantly, the distribution $P(A,B,E)$
contains secret correlations if and only if $I\left(A; B\downarrow
E\right)>0$ \cite{Renner_03,Bae_09}. Moving back to the mapping
(\ref{mapping}) one can then show using intrinsic information
(\ref{intrinsicI}) that provided that the state $\rho_{AB}$ is
entangled one can always find measurements $\Pi_{j}$ such that the
obtained distribution contains secret correlations
\cite{Gisin_00}. Moreover, the multipartite form of the mapping
(\ref{mapping}) is even capable of mapping more subtle properties
of entanglement such as its boundedness \cite{Acin_04}.

So far, the mapping (\ref{mapping}) has been investigated only
from the point of view of the ability to transmit qualitative
properties of quantum states onto classical probability
distributions. A natural step forward would therefore be to
elucidate whether the mapping can also preserve the quantitative
properties of input states. Specifically, it would be of interest
to know whether there is a function of a probability distribution
$P(A,B,E)$ associated with a quantum state $\rho_{AB}$ via mapping
(\ref{mapping}) which does not increase under any LOCC operation
on the state. This would mean that the composition of the mapping
and the function preserve the fundamental property that
entanglement does not increase under LOCC operations. This is,
however, important from a practical point of view because such a
function then can be used to quantify entanglement
\cite{Vidal_00}.

An interesting attempt to quantify entanglement with the mapping
(\ref{mapping}) has been put forward by Gisin and Wolf
\cite{Gisin_00}. They introduced the following optimized intrinsic
information
%%%%%%%%%%%%%%%%%%%%%%%%%%%%%%%%%%%%%%%%%%%%%%%%%%%%%%%%%%%%%%%%%%%%%%%%%%%%%%%%%%%%%%%%%%%%%%%%%%%%
\begin{eqnarray}\label{muadd}
\mu(\rho_{AB})=\mathop{\mbox{inf}}_{\left\{\Pi_{E},|\Psi\rangle\right\}}
\left\{\mathop{\mbox{sup}}_{\left\{\Pi_{A},\Pi_{B}\right\}}
\left[I\left(A;B\downarrow E\right)\right]\right\},
\end{eqnarray}
%%%%%%%%%%%%%%%%%%%%%%%%%%%%%%%%%%%%%%%%%%%%%%%%%%%%%%%%%%%%%%%%%%%%%%%%%%%%%%%%%%%%%%%%%%%%%%%%%%%%%
%%%%%%%%%%%%%%%%%%%%%%%%%%%%%%%%%%%%%%%%%%%%%%%%%%%%%%%%%%%%%%%%%%%%%%%%%%%%%%%%%%%%%%%%%%%%%%%%%%%%
%\begin{eqnarray}\label{mu}
%\mu(\rho_{AB})=\mathop{\mbox{inf}}_{\left\{|e\rangle,|\Psi\rangle\right\}}
%\left\{\mathop{\mbox{sup}}_{\left\{|a\rangle,|b\rangle\right\}}
%\left[I\left(A;B\downarrow E\right)\right]\right\},
%\end{eqnarray}
%%%%%%%%%%%%%%%%%%%%%%%%%%%%%%%%%%%%%%%%%%%%%%%%%%%%%%%%%%%%%%%%%%%%%%%%%%%%%%%%%%%%%%%%%%%%%%%%%%%%%
where the supremum is taken over all projective measurements
$\{\Pi_{A}=|A\rangle\langle A|\}$ and $\{\Pi_{B}=|B\rangle\langle
B|\}$ on subsystems $A$ and $B$, respectively, and the infimum is
taken over all purifications $|\Psi\rangle$ of the state
$\rho_{AB}$ and all positive operator-valued measures (POVM)
$\{\Pi_{E}\}$ on subsystem $E$. Further, in Ref.~\cite{Gisin_00}
it was shown that the quantity (\ref{muadd}) possesses some
properties of an entanglement measure such as equality to the von
Neumann entropy on pure states and convexity, and it was also
calculated analytically for two-qubit Werner states. The quantity
(\ref{muadd}) is particularly interesting because unlike most of
the other entanglement measures it is intimately related with a
meaningful protocol -- it is an upper bound in the secret-key
agreement protocol \cite{Maurer_93}. What is more, it may even
characterize secret correlations distillable to a secret key
provided that the conjectured bipartite nondistillable secret
correlations with a strictly positive intrinsic information (the
so-called bipartite bound information \cite{Gisin_00}) do not
exist. Despite this fact, the other properties of entanglement
measures have not been investigated for the quantity (\ref{muadd})
but it inspired the introduction of a different measure called
squashed entanglement \cite{Christandl_04}. In particular, the key
questions of whether the quantity (\ref{muadd}) is non-increasing
under LOCC operations and whether it can be calculated also for
other quantum states remain open.

To find answers to the latter questions can be a hard or even
intractable task owing to the apparent complexity of the quantity
(\ref{muadd}). Nevertheless, the quantity (\ref{muadd}) can still
inspire the introduction of a closely related quantity for which
the proof of monotonicity under LOCC operations as well as its
computation can be considerably easier. The quantity in question
is the so-called {\it intrinsic entanglement} (IE) defined as
\cite{Mista_14}
%%%%%%%%%%%%%%%%%%%%%%%%%%%%%%%%%%%%%%%%%%%%%%%%%%%%%%%%%%%%%%%%%%%%%%%%
\begin{equation}\label{Edownarrow}
E_{\downarrow}(\rho_{AB})=\mathop{\mbox{sup}}_{\left\{\Pi_{A},\Pi_{B}\right\}}
\left\{\mathop{\mbox{inf}}_{\left\{\Pi_{E},|\Psi\rangle\right\}}\left[I\left(A;
B\downarrow E\right)\right]\right\}.
\end{equation}
%%%%%%%%%%%%%%%%%%%%%%%%%%%%%%%%%%%%%%%%%%%%%%%%%%%%%%%%%%%%%%%%%%%%%%%%
In comparison with the quantity (\ref{muadd}) the order of
optimization in the definition of IE is reversed and hence
$E_{\downarrow}\leq\mu$ due to the max-min inequality
\cite{Boyd_04}. In fact, the two quantities may coincide if the
intrinsic information (\ref{intrinsicI}) together with the sets
$\{\Pi_{A},\Pi_{B}\}$ and $\{\Pi_{E},|\Psi\rangle\}$ possess the
strong max-min property \cite{Boyd_04} which guarantees that the
order of optimization in Eq.~(\ref{Edownarrow}) can be commuted.
The Ref.~\cite{Mista_14} further deals with a Gaussian version of
IE, the so-called Gaussian intrinsic entanglement (GIE). The GIE
is defined as in Eq.~(\ref{Edownarrow}), where all channels
$E\rightarrow\tilde{E}$ in Eq.~(\ref{intrinsicI}), and all quantum
states $\rho_{AB}$ and $|\Psi\rangle$, and measurements
$\{\Pi_{j}\}$, $j=A,B,E$, are assumed to be Gaussian. It is
further shown that GIE simplifies considerably to the optimized
mutual information of a distribution of outcomes of Gaussian
measurements on subsystems $A$ and $B$ of a conditional state
obtained by a Gaussian measurement on subsystem $E$ of a Gaussian
purification of the state $\rho_{AB}$. Next, it is proved that GIE
vanishes if and only if the state $\rho_{AB}$ is separable and
that it does not increase under Gaussian local trace-preserving
operations and classical communication (GLTPOCC). Finally, some
analytical formulae are obtained for GIE as well as IE. First, GIE
is calculated analytically for pure Gaussian states as well as for
a two-mode reduction of the three-mode CV GHZ sate \cite{Loock_00}
and it is shown that it always coincides with the Gaussian
R\'{e}nyi-2 (GR2) entanglement \cite{Adesso_12}. Second, an
analytical lower bound on IE is derived for a subset of the set of
the non-Gaussian continuous-variable Werner states
\cite{Mista_02}, which is given by convex mixtures of the two-mode
squeezed vacuum state and the vacuum state.
%Finally, an analytical formula for GIE of pure Gaussian states as
%well as two-mode reduction of the three-mode CV GHZ sate
%\cite{Loock_00} is derived and it is shown that it always
%coincides with the Gaussian R\'{e}nyi-2 (GR2) entanglement
%\cite{Adesso_12}.

The present paper accompanies the original paper on GIE
\cite{Mista_14}. It contains details of proofs of the properties
of GIE presented in Ref.~\cite{Mista_14}. Additionally, we also
provide two new results not mentioned in Ref.~\cite{Mista_14}.
First, we show that the monotonicity of GIE under GLTPOCC implies
the invariance of GIE with respect to Gaussian local unitaries.
Second, we prove that if we allow for non-Gaussian measurements
$\{\Pi_{A},\Pi_{B}\}$ in the definition of GIE we get a quantity
which is on pure Gaussian states equal to the entropy of
entanglement in analogy with the quantifier (\ref{muadd}) which is
also equal to the entropy of entanglement for pure states
\cite{Gisin_00}.

The paper is organized as follows. Section~\ref{sec_0} contains a
brief introduction into the formalism of Gaussian states. In
Section~\ref{sec_1} we show explicitly that for GIE the channel
$E\rightarrow\tilde{E}$ in Eq.~(\ref{intrinsicI}) can be
integrated into Eve's measurement. The next Section~\ref{sec_2}
contains a proof that in the definition of GIE (\ref{Edownarrow})
we can use a fixed purification and the minimization over all
Gaussian purifications can be omitted. Section~\ref{sec_3} then
presents the construction of a Gaussian measurement which projects
a Gaussian purification of a separable Gaussian state onto a
product state and Section~\ref{sec_4} is dedicated to a detailed
proof of the monotonicity of GIE under GLTPOCC operations.
Derivation of an analytical expression for GIE and proof of its
equality to GR2 entanglement is given for pure Gaussian states in
Section~\ref{sec_5} and for the two-mode reduction of the
three-mode CV GHZ state in Section~\ref{sec_6}. Finally, in
Section~\ref{sec_7} we derive an analytical lower bound on IE for
a subclass of the non-Gaussian continuous-variable Werner states.
Section~\ref{sec_8} contains conclusions.

%%%%%%%%%%%%%%%%%%%%%%%%%%%%%%%%%%%%%%%%%%%%%%%%%%%%%%%%%%%%%%%%%%%%%%%%%%%%%%%%%%%%%%%%%%%%%%%%%%%%%%%%%%%%%%%%%%
\section{Gaussian states}\label{sec_0}

In this paper we consider quantum systems with
infinite-dimensional Hilbert state spaces which can be physically
implemented by modes of the electromagnetic field. A system of $n$
modes can be conveniently described by a vector
of quadratures $\xi=(x_1,p_1,\ldots,x_n,p_n)^{T}$ whose components obey the
canonical commutation rules $[\xi_j,\xi_k]=i(\Omega_n)_{jk}$ with
%%%%%%%%%%%%%%%%%%%%%%%%%%%%%%%%%%%%%%%%%%%%%%%%%%%%%%%%%%%%%%%%%
\begin{eqnarray}\label{Omega}
\Omega_{n}=\bigoplus_{i=1}^{n}\left(\begin{array}{cc}
0 & 1 \\
-1 & 0\\
\end{array}\right)
\end{eqnarray}
%%%%%%%%%%%%%%%%%%%%%%%%%%%%%%%%%%%%%%%%%%%%%%%%%%%%%%%%%%%%%%%%%%%,
being the so-called symplectic matrix. According to definition, Gaussian states are
quantum states of modes, which possess a Gaussian Wigner function.
An $n$-mode Gaussian state $\rho$ is therefore fully characterized by a vector of
first moments $\langle\xi\rangle=\text{Tr}(\xi\rho)$,
and by a covariance matrix (CM) $\gamma$ with entries
$\gamma_{jk}=\langle\{\Delta\xi_j,\Delta\xi_k\}\rangle,$
where $\Delta\xi_j=\xi_j-\langle\xi_j\rangle$ and $\{A,B\}\equiv AB+BA$
is the anticommutator. The quantity GIE analyzed in this paper depends
only on the elements of the CM and thus the vector of the first moments
$\langle\xi\rangle$ is from now assumed to be zero for simplicity.
We use Gaussian unitary operations which are for $n$
modes represented at the level of CMs by a real $2n\times 2n$
symplectic matrix $S$ fulfilling $S\Omega_{n}S^{T}=\Omega_{n}$.
Recall also, that any CM $\gamma$ can be symplectically diagonalized, i.e.,
there exists a symplectic matrix $S$ that brings $\gamma$ to the Williamson normal
form \cite{Williamson_36}
%%%%%%%%%%%%%%%%%%%%%%%%%%%%%%%%%%%%%%%%%%%%%%%%%%%%%%%%%%%%%%%%%%%%%%%%%%%%%%%%%%
\begin{equation}\label{Williamson}
S\gamma S^{T}=\mbox{diag}\left(\nu_{1},\nu_{1},\ldots,\nu_{n},\nu_{n}\right),
\end{equation}
%%%%%%%%%%%%%%%%%%%%%%%%%%%%%%%%%%%%%%%%%%%%%%%%%%%%%%%%%%%%%%%%%%%%%%%%%%%%%%%%%%%
where $\nu_{1}\geq\ldots\geq\nu_{n}\geq 1$ are the symplectic eigenvalues of $\gamma$.

As for measurements we restrict ourselves to Gaussian measurements which can be implemented by
appending auxiliary vacuum modes, using passive and active linear optics (phase shifters, squeezers and beam splitters) and
homodyne detections. Any such measurement on $n$ modes is described by the following POVM \cite{Fiurasek_07}
%%%%%%%%%%%%%%%%%%%%%%%%%%%%%%%%%%%%%%%%%%%%%%%%%%%%%%%%%%%%%%%%%%%%%%%%%%%%%%%%%%%%%%%%%%%%%%%
\begin{equation}\label{POVMn}
\Pi(d)=\frac{1}{(2\pi)^{n}}D(d)\Pi_0 D^\dagger(d),
\end{equation}
%%%%%%%%%%%%%%%%%%%%%%%%%%%%%%%%%%%%%%%%%%%%%%%%%%%%%%%%%%%%%%%%%%%%%%%%%%%%%%%%%%%%%%%%%%%%%%%
where the seed element $\Pi_0$ is a normalized density matrix of a
generally mixed $n$-mode Gaussian state with zero first moments
and CM $\Gamma$, $D(d)=\exp(-id^{T}\Omega_{n}\xi)$ is the
displacement operator, and $d=(d_{1}^{(x)},d_{1}^{(p)},\ldots,d_{n}^{(x)},d_{n}^{(p)})^{T}\in\mathbb{R}_{2n}$
is a vector of measurement outcomes. From the normalization condition $\mbox{Tr}[\Pi_{0}]=1$ it follows
that the POVM (\ref{POVMn}) satisfies the completeness condition
%%%%%%%%%%%%%%%%%%%%%%%%%%%%%%%%%%%%%%%%%%%%%%%%%%%%%%%%%%%%%%%%%%%%%%%%%%%%%%%%%%%%%%%%
\begin{equation}\label{completeness}
\int_{\mathbb{R}_{2n}}\Pi(d){\rm d}^{2n}d=\openone,
\end{equation}
%%%%%%%%%%%%%%%%%%%%%%%%%%%%%%%%%%%%%%%%%%%%%%%%%%%%%%%%%%%%%%%%%%%%%%%%%%%%%%%%%%%%%%%%
where ${\rm d}^{2n}d=\Pi_{l=1}^{n}{\rm d}d_{l}^{(x)}{\rm d}d_{l}^{(p)}$.

In the present analysis of IE, Eq.~(\ref{Edownarrow}), we assume that the state
$\rho_{AB}\equiv\rho_{A_1\ldots A_{N}B_{1}\ldots B_{M}}$ is an $(N+M)$-mode Gaussian state
of $N$ modes $A_1,A_2,\ldots,A_{N}$ and $M$ modes
$B_{1},B_{2},\ldots,B_{M}$, which is described by the CM $\gamma_{AB}$. Further, we also
assume that $|\bar{\Psi}\rangle_{ABE}$ is an $(N+M+K)$-mode Gaussian purification
of the state $\rho_{AB}$, which contains $K$ purifying modes
$E_1,E_2,\ldots,E_K$, and which is described by the CM $\bar{\gamma}_{\pi}$.
By performing Gaussian measurements (\ref{POVMn}) with covariance
matrices (CMs) $\Gamma_{A},\Gamma_{B}$ and $\Gamma_{E}$ on subsystems $A,B$ and
$E$ of the purification $|\bar{\Psi}\rangle_{ABE}$, the mapping (\ref{mapping}) yields a
zero-mean Gaussian distribution $P(d_A,d_B,d_E)$ of measurement outcomes $d_A,d_B$ and $d_E$, which
is given by the formula
%%%%%%%%%%%%%%%%%%%%%%%%%%%%%%%%%%%%%%%%%%%%%%%%%%%%%%%%%%%%%%%%%%%%%%%%%%%%%%%%%%%%%%%%%%%%%%%
\begin{eqnarray}\label{Pinput}
P(d_A,d_B,d_E)=\frac{e^{-d^{T}\Sigma^{-1}d}}{\pi^{N+M+K}\sqrt{\mbox{det}\Sigma}},
\end{eqnarray}
%%%%%%%%%%%%%%%%%%%%%%%%%%%%%%%%%%%%%%%%%%%%%%%%%%%%%%%%%%%%%%%%%%%%%%%%%%%%%%%%%%%%%%%%%%%%%%%%%%
where $d=(d_{A}^{T},d_{B}^{T},d_{E}^{T})^{T}$ and
%%%%%%%%%%%%%%%%%%%%%%%%%%%%%%%%%%%%%%%%%%%%%%%%%%%%%%%%%%%%%%%%%%%%%%%%%%%%%%%%%%%%%%%%%%%%%%%%%%%%
%%%%%%%%%%%%%%%%%%%%%%%%%%%%%%%%%%%%%%%%%%%%%%%%%%%%%%%%%%%%%%%%%%%%%%%%%%%%%%%%%%%%%%%%%%%%%%%%%%%%
\begin{eqnarray}\label{Sigma}
\Sigma=\left(\begin{array}{cc}
\gamma_{AB}+\Gamma_{A}\oplus\Gamma_{B} & \bar{\gamma}_{ABE}\\
\bar{\gamma}_{ABE}^{T} & \bar{\gamma}_{E}+\Gamma_{E}\\
\end{array}\right)\equiv\left(\begin{array}{cc}
\alpha & \beta\\
\beta^{T} & \delta\\
\end{array}\right),
\end{eqnarray}
%%%%%%%%%%%%%%%%%%%%%%%%%%%%%%%%%%%%%%%%%%%%%%%%%%%%%%%%%%%%%%%%%%%%%%%%%%%%%%%%%%%%%%%%%%%%%%%%%%%%
is the CCM \cite{CCM} of the distribution expressed with respect to $AB|E$
splitting. Here $\gamma_{AB}, \bar{\gamma}_{ABE}$ and $\bar{\gamma}_{E}$ are blocks
of the CM $\bar{\gamma}_{\pi}$ of the purification $|\bar{\Psi}\rangle_{ABE}$, when
expressed with respect to the same splitting, i.e.,
%%%%%%%%%%%%%%%%%%%%%%%%%%%%%%%%%%%%%%%%%%%%%%%%%%%%%%%%%%%%%%%%%%%%
\begin{eqnarray}\label{bargammaABEsplittingadd}
\bar{\gamma}_{\pi}=\left(\begin{array}{cc}
\gamma_{AB} & \bar{\gamma}_{ABE} \\
\bar{\gamma}_{ABE}^{T} & \bar{\gamma}_{E} \\
\end{array}\right).
\end{eqnarray}
%%%%%%%%%%%%%%%%%%%%%%%%%%%%%%%%%%%%%%%%%%%%%%%%%%%%%%%%%%%%%%%%%

In what follows, we analyze a Gaussian version of the quantifier
(\ref{Edownarrow}), where the role of the distribution $P(A,B,E)$
is played by the Gaussian distribution (\ref{Pinput}).

%%%%%%%%%%%%%%%%%%%%%%%%%%%%%%%%%%%%%%%%%%%%%%%%%%%%%%%%%%%%%%%%%%%%%%%%%%%%%%%%%%%%%%%%%%%%%%%%%%%%%%%%%%%%%%%%%%
\section{Proof that any Gaussian channel can be integrated into Eve's measurement}\label{sec_1}

At the beginning we show that the quantity IE, Eq.~(\ref{Edownarrow}), greatly simplifies in the Gaussian
scenario. First, we prove that any Gaussian channel $E\rightarrow\tilde{E}$ appearing in Eq.~(\ref{intrinsicI})
can be always incorporated into Eve's measurement.

The proof goes as follows. We assume that the channel $E\rightarrow\tilde{E}$ in Eq.~(\ref{intrinsicI})
is a Gaussian channel $d_{E}\rightarrow\tilde{d}_{E}$ mapping a $2K\times1$ column vector $d_{E}$ onto an
$L\times1$ column vector $\tilde{d}_{E}$, where $d_{E}$ contains measurement outcomes
of a measurement on Eve's $K$ modes of an $(N+M+K)$-mode purification $|\bar{\Psi}\rangle_{ABE}$
of the state $\rho_{AB}$. Such a channel is described by a linear transformation
%%%%%%%%%%%%%%%%%%%%%%%%%%%%%%%%%%%%%%%%%%%%%%%%%%%%%%%%%%%%%%%%%%%%%%%%%%%%%%%%%%%%%%%%%%%%%%%%%%%%%%%%%%
\begin{equation}\label{dtilde}
\tilde{d}_{E}=Xd_{E}+y,
\end{equation}
%%%%%%%%%%%%%%%%%%%%%%%%%%%%%%%%%%%%%%%%%%%%%%%%%%%%%%%%%%%%%%%%%%%%%%%%%%%%%%%%%%%%%%%%%%%%%%%%%%%%%%%%%%%
where $X$ is a fixed real $L\times2K$ matrix and
$y=(y_1,y_2,\ldots,y_L)^{T}$ is an $L\times1$ random column vector
distributed with a zero mean Gaussian distribution characterized
by an $L\times L$ CCM $Y$ with elements $Y_{ij}=\langle
\{y_{i},y_{j}\}\rangle$, $i,j=1,\ldots,L$. The input to the channel is a vector
$d_{E}$ of Eve's measurement outcomes, which is distributed
according to a zero mean Gaussian distribution with a fixed CCM
$\delta=\bar{\gamma}_{E}+\Gamma_{E}$ given in Eq.~(\ref{Sigma}). The channel is therefore
fully characterized by a joint Gaussian distribution
$P(d_{E},\tilde{d}_{E})$ with zero mean and a CCM of the form
%%%%%%%%%%%%%%%%%%%%%%%%%%%%%%%%%%%%%%%%%%%%%%%%%%%%%%%%%%%%%%%%%%%%%%%%%%%%%%%%%%%%%%%%%%%%%%%%%%%%
\begin{eqnarray}\label{chi}
\chi=\left(\begin{array}{cc}
\delta & \delta X^T\\
X\delta & X\delta X^T+Y \\
\end{array}\right).
\end{eqnarray}
%%%%%%%%%%%%%%%%%%%%%%%%%%%%%%%%%%%%%%%%%%%%%%%%%%%%%%%%%%%%%%%%%%%%%%%%%%%%%%%%%%%%%%%%%%%%%%%%%%%%
The input Gaussian distribution $P(d_A,d_B,d_{E})$, Eq.~(\ref{Pinput}), is then
transformed by the channel as
%%%%%%%%%%%%%%%%%%%%%%%%%%%%%%%%%%%%%%%%%%%%%%%%%%%%%%%%%%%%%%%%%%%%%%%%%%%%%%%%%%%%%%%%%%%%%%%%%%%%
\begin{equation}\label{Ptilde}
\tilde{P}(d_{A},d_{B},\tilde{d}_{E})=\int
P(\tilde{d}_{E}|d_{E})P(d_{A},d_{B},d_{E})d^{2K}d_{E},
\end{equation}
%%%%%%%%%%%%%%%%%%%%%%%%%%%%%%%%%%%%%%%%%%%%%%%%%%%%%%%%%%%%%%%%%%%%%%%%%%%%%%%%%%%%%%%%%%%%%%%%%%%%
where
%%%%%%%%%%%%%%%%%%%%%%%%%%%%%%%%%%%%%%%%%%%%%%%%%%%%%%%%%%%%%%%%%%%%%%%%%%%%%%%%%%%%%%%%%%%%%
\begin{equation}\label{Pconditional}
P(\tilde{d}_{E}|d_{E})=\frac{P(d_{E},\tilde{d}_{E})}{P(d_{E})}=\frac{e^{-\left(\tilde{d}_{E}-Xd_{E}\right)^{T}Y^{-1}
\left(\tilde{d}_{E}-Xd_{E}\right)}}{\sqrt{\pi^L\mbox{det}Y}}
\end{equation}
%%%%%%%%%%%%%%%%%%%%%%%%%%%%%%%%%%%%%%%%%%%%%%%%%%%%%%%%%%%%%%%%%%%%%%%%%%%%%%%%%%%%%%%%%%%%%
is a conditional Gaussian probability distribution of the channel.
We now substitute into the right-hand side (RHS) of
Eq.~(\ref{Ptilde}) for the distribution $P(d_{A},d_{B},d_{E})$
from Eq.~(\ref{Pinput}), which gives the output distribution
(\ref{Ptilde}) in the form
%%%%%%%%%%%%%%%%%%%%%%%%%%%%%%%%%%%%%%%%%%%%%%%%%%%%%%%%%%%%%%%%%%%%%%%%%%%%%%%%%%%%%%%%%%%%%%%
\begin{eqnarray}\label{Poutput}
\tilde{P}(d_{A},d_{B},\tilde{d}_{E})=\frac{e^{-\tilde{d}^{T}{\tilde{\Sigma}}^{-1}\tilde{d}}}
{\pi^{N+M+\frac{L}{2}}\sqrt{\mbox{det}\tilde{\Sigma}}},
\end{eqnarray}
%%%%%%%%%%%%%%%%%%%%%%%%%%%%%%%%%%%%%%%%%%%%%%%%%%%%%%%%%%%%%%%%%%%%%%%%%%%%%%%%%%%%%%%%%%%%%%%%%%
where $\tilde{d}=(d_{A}^{T},d_{B}^{T},\tilde{d}_{E}^{T})^{T}$ and
%%%%%%%%%%%%%%%%%%%%%%%%%%%%%%%%%%%%%%%%%%%%%%%%%%%%%%%%%%%%%%%%%%%%%%%%%%%%%%%%%%%%%%%%%%%%%%%%%%%%
\begin{eqnarray}\label{tildeSigma}
\tilde{\Sigma}=\left(\begin{array}{cc}
\alpha & \beta X^{T}\\
X\beta^{T} & X\delta X^{T}+Y\\
\end{array}\right),
\end{eqnarray}
%%%%%%%%%%%%%%%%%%%%%%%%%%%%%%%%%%%%%%%%%%%%%%%%%%%%%%%%%%%%%%%%%%%%%%%%%%%%%%%%%%%%%%%%%%%%%%%%%%%%
where the matrices $\alpha,\beta$ and $\delta$ are defined in Eq.~(\ref{Sigma}).

The figure of merit considered in this paper is the conditional
mutual information $I(A;B|E)$ of the output distribution
(\ref{Poutput}) which coincides with the standard mutual
information $I(A;B)$ of the corresponding conditional distribution
$\tilde{P}(d_A,d_B|\tilde{d}_{E})$ \cite{Mista_14}. The latter
distribution is Gaussian and the mutual information depends on its
CCM which is given by the Schur complement \cite{Horn_85} of the
CCM (\ref{tildeSigma}),
%%%%%%%%%%%%%%%%%%%%%%%%%%%%%%%%%%%%%%%%%%%%%%%%%%%%%%%%%%%%%%%%%%%%%%%%%%%%%%%%%%%%%%%%%%%%%%
\begin{eqnarray}\label{Schur1}
\sigma_{AB}=\alpha-\beta X^{T}(X\delta X^{T}+Y)^{-1}X\beta^{T},
\end{eqnarray}
%%%%%%%%%%%%%%%%%%%%%%%%%%%%%%%%%%%%%%%%%%%%%%%%%%%%%%%%%%%%%%%%%%%%%%%%%%%%%%%%%%%%%%%%%%%%%%%
where the inverse is to be understood generally as the
pseudoinverse.

Now we prove that for any channel (\ref{dtilde}) there is a
measurement on Eve's modes characterized by a CM
$\tilde{\Gamma}_{E}$ such that
%%%%%%%%%%%%%%%%%%%%%%%%%%%%%%%%%%%%%%%%%%%%%%%%%%%%%%%%%%%%%%%%%%%%%%%%%%%%%%%%%%%%%%%%%%%%%%
\begin{eqnarray}\label{sigmaintegrated}
\sigma_{AB}=\alpha-\beta(\bar{\gamma}_{E}+\tilde{\Gamma}_{E})^{-1}\beta^{T}.
\end{eqnarray}
%%%%%%%%%%%%%%%%%%%%%%%%%%%%%%%%%%%%%%%%%%%%%%%%%%%%%%%%%%%%%%%%%%%%%%%%%%%%%%%%%%%%%%%%%%%%%%%
As a result, without loss of generality, we can omit the
minimization appearing in Eq.~(\ref{intrinsicI}) and the intrinsic
conditional information $I\left(A; B\downarrow E\right)$ in the
definition (\ref{Edownarrow}) can thus be replaced with the standard conditional
mutual information $I(A;B|E)$, Eq.~(\ref{conditionalI}).

Our proof utilizes the singular value decomposition \cite{Horn_85}
of the matrix $X$,
%%%%%%%%%%%%%%%%%%%%%%%%%%%%%%%%%%%%%%%%%%%%%%%%%%%%%%%%%%%%%%%%%%%%%%%%%%%%%%%%%%%%%%%%%%%%%%
\begin{equation}\label{SVD}
X=\mathcal{U}\mathcal{S}\mathcal{V}^{T},
\end{equation}
%%%%%%%%%%%%%%%%%%%%%%%%%%%%%%%%%%%%%%%%%%%%%%%%%%%%%%%%%%%%%%%%%%%%%%%%%%%%%%%%%%%%%%%%%%%%%%%
where $\mathcal{U}$ is an $L\times L$ real orthogonal matrix,
$\mathcal{V}$ is a $2K\times 2K$ real orthogonal matrix and
$\mathcal{S}$ is an $L\times 2K$ rectangular diagonal matrix of
the form
%%%%%%%%%%%%%%%%%%%%%%%%%%%%%%%%%%%%%%%%%%%%%%%%%%%%%%%%%%%%%%%%%%%%%%%%%%%%%%%%%%%%%%%%%%%%%%%%%%%%
\begin{eqnarray}\label{calS}
\mathcal{S}=\left(\begin{array}{cc}
s_{Q} & \mathbb{O}_{Q\times(2K-Q)}\\
\mathbb{O}_{(L-Q)\times Q} & \mathbb{O}_{(L-Q)\times(2K-Q)}\\
\end{array}\right),
\end{eqnarray}
%%%%%%%%%%%%%%%%%%%%%%%%%%%%%%%%%%%%%%%%%%%%%%%%%%%%%%%%%%%%%%%%%%%%%%%%%%%%%%%%%%%%%%%%%%%%%%%%%%%
where $\mathbb{O}_{I\times J}$ is an $I\times J$ zero matrix,
$s_{Q}=\mbox{diag}(\varsigma_{1},\varsigma_{2},\ldots,\varsigma_{Q})$
is a $Q\times Q$ diagonal matrix with the strictly positive singular values
$\varsigma_{1}\geq\varsigma_{2}\geq\ldots\geq\varsigma_{Q}>0$ on the
diagonal and $Q=\mbox{rank} X$. Inserting Eq.~(\ref{SVD}) into
Eq.~(\ref{Schur1}) one obtains
%%%%%%%%%%%%%%%%%%%%%%%%%%%%%%%%%%%%%%%%%%%%%%%%%%%%%%%%%%%%%%%%%%%%%%%%%%%%%%%%%%%%%%%%%%%%%%
\begin{eqnarray}\label{Schur2}
\sigma_{AB}=\alpha-\beta
\mathcal{V}\mathcal{S}^{T}(\mathcal{S}\mathcal{V}^{T}\delta
\mathcal{V}\mathcal{S}^{T}+\mathcal{U}^{T}Y\mathcal{U})^{-1}\mathcal{S}\mathcal{V}^{T}\beta^{T}.
\end{eqnarray}
%%%%%%%%%%%%%%%%%%%%%%%%%%%%%%%%%%%%%%%%%%%%%%%%%%%%%%%%%%%%%%%%%%%%%%%%%%%%%%%%%%%%%%%%%%%%%%%
Making use of Eq.~(\ref{calS}) one can further express the
$L\times L$ matrix $\mathcal{S}\mathcal{V}^{T}\delta
\mathcal{V}\mathcal{S}^{T}$, appearing in the round brackets in
Eq.~(\ref{Schur2}), as
%%%%%%%%%%%%%%%%%%%%%%%%%%%%%%%%%%%%%%%%%%%%%%%%%%%%%%%%%%%%%%%%%%%%%%%%%%%%%%%%%%%%%%%%%%%%%%%%%%%%
\begin{equation}\label{Exp1}
\mathcal{S}\mathcal{V}^{T}\delta
\mathcal{V}\mathcal{S}^{T}=\tau_{Q}\omega_{Q}\tau_{Q},
\end{equation}
%%%%%%%%%%%%%%%%%%%%%%%%%%%%%%%%%%%%%%%%%%%%%%%%%%%%%%%%%%%%%%%%%%%%%%%%%%%%%%%%%%%%%%%%%%%%%%%%%%%%
where $\tau_{Q}$ is an $L\times L$ matrix of the form
%%%%%%%%%%%%%%%%%%%%%%%%%%%%%%%%%%%%%%%%%%%%%%%%%%%%%%%%%%%%%%%%%%%%%%%%%%%%%%%%%%%%%%%%%%%%%%%%%%%
\begin{equation}\label{tauQ}
\tau_{Q}=s_{Q}\oplus\openone_{L-Q}
\end{equation}
%%%%%%%%%%%%%%%%%%%%%%%%%%%%%%%%%%%%%%%%%%%%%%%%%%%%%%%%%%%%%%%%%%%%%%%%%%%%%%%%%%%%%%%%%%%%%%%%%%%%
and
%%%%%%%%%%%%%%%%%%%%%%%%%%%%%%%%%%%%%%%%%%%%%%%%%%%%%%%%%%%%%%%%%%%%%%%%%%%%%%%%%%%%%%%%%%%%%%%%%%%%
\begin{equation}\label{omegaQ}
\omega_{Q}=(\mathcal{V}^{T}\delta\mathcal{V})_{Q}\oplus\mathbb{O}_{(L-Q)\times(L-Q)},
\end{equation}
%%%%%%%%%%%%%%%%%%%%%%%%%%%%%%%%%%%%%%%%%%%%%%%%%%%%%%%%%%%%%%%%%%%%%%%%%%%%%%%%%%%%%%%%%%%%%%%%%%%%
where $(\mathcal{V}^{T}\delta\mathcal{V})_{Q}$ is the first
$Q\times Q$ block of the matrix
$\mathcal{V}^{T}\delta\mathcal{V}$ and $\openone_{I}$ is an $I\times I$ identity matrix.
Substitution for
$\mathcal{S}\mathcal{V}^{T}\delta \mathcal{V}\mathcal{S}^{T}$ in
Eq.~(\ref{Schur2}) from Eq.~(\ref{Exp1}) and utilizing the formula
%%%%%%%%%%%%%%%%%%%%%%%%%%%%%%%%%%%%%%%%%%%%%%%%%%%%%%%%%%%%%%%%%%%%%%%%%%%%%%%%%%%%%%%%%%%%%%%%%%%%
\begin{eqnarray}\label{calI}
\tau_{Q}^{-1}\mathcal{S}=\left(\begin{array}{cc}
\openone_{Q} & \mathbb{O}_{Q\times(2K-Q)}\\
\mathbb{O}_{(L-Q)\times Q} & \mathbb{O}_{(L-Q)\times(2K-Q)}\\
\end{array}\right)\equiv\mathcal{I},
\end{eqnarray}
%%%%%%%%%%%%%%%%%%%%%%%%%%%%%%%%%%%%%%%%%%%%%%%%%%%%%%%%%%%%%%%%%%%%%%%%%%%%%%%%%%%%%%%%%%%%%%%%%%%
further yields the matrix (\ref{Schur2}) in the form
%%%%%%%%%%%%%%%%%%%%%%%%%%%%%%%%%%%%%%%%%%%%%%%%%%%%%%%%%%%%%%%%%%%%%%%%%%%%%%%%%%%%%%%%%%%%%%
\begin{eqnarray}\label{Schur3}
\sigma_{AB}=\alpha-\beta
\mathcal{V}\mathcal{I}^{T}(\omega_{Q}+Y_{\mathcal{U},s_{Q}})^{-1}\mathcal{I}\mathcal{V}^{T}\beta^{T},
\end{eqnarray}
%%%%%%%%%%%%%%%%%%%%%%%%%%%%%%%%%%%%%%%%%%%%%%%%%%%%%%%%%%%%%%%%%%%%%%%%%%%%%%%%%%%%%%%%%%%%%%%
where
%%%%%%%%%%%%%%%%%%%%%%%%%%%%%%%%%%%%%%%%%%%%%%%%%%%%%%%%%%%%%%%%%%%%%%%%%%%%%%%%%%%%%%%%%%%%%%%
\begin{equation}\label{YcalUsigmaQ}
Y_{\mathcal{U},s_{Q}}=\tau_{Q}^{-1}\mathcal{U}^{T}Y\mathcal{U}\tau_{Q}^{-1}
\end{equation}
%%%%%%%%%%%%%%%%%%%%%%%%%%%%%%%%%%%%%%%%%%%%%%%%%%%%%%%%%%%%%%%%%%%%%%%%%%%%%%%%%%%%%%%%%%%%%%%
is an $L\times L$ positive-semidefinite matrix. Substitution for
the matrix $\mathcal{I}$ from Eq.~(\ref{calI}) into
Eq.~(\ref{Schur3}) further yields
%%%%%%%%%%%%%%%%%%%%%%%%%%%%%%%%%%%%%%%%%%%%%%%%%%%%%%%%%%%%%%%%%%%%%%%%%%%%%%%%%%%%%%%%%%%%%%
\begin{eqnarray}\label{Schur4}
\sigma_{AB}=\alpha-\beta
\mathcal{V}\mathcal{W}\mathcal{V}^{T}\beta^{T},
\end{eqnarray}
%%%%%%%%%%%%%%%%%%%%%%%%%%%%%%%%%%%%%%%%%%%%%%%%%%%%%%%%%%%%%%%%%%%%%%%%%%%%%%%%%%%%%%%%%%%%%%%
with
%%%%%%%%%%%%%%%%%%%%%%%%%%%%%%%%%%%%%%%%%%%%%%%%%%%%%%%%%%%%%%%%%%%%%%%%%%%%%%%%%%%%%%%%%%%%%%%
\begin{eqnarray}\label{calW}
\mathcal{W}=\left(\begin{array}{cc}
w_{Q} & \mathbb{O}\\
\mathbb{O}^{T} & \tilde{\mathbb{O}}\\
\end{array}\right)
\end{eqnarray}
%%%%%%%%%%%%%%%%%%%%%%%%%%%%%%%%%%%%%%%%%%%%%%%%%%%%%%%%%%%%%%%%%%%%%%%%%%%%%%%%%%%%%%%%%%%%%%%%%%%%
being a $2K\times 2K$ matrix, where we have defined
$\mathbb{O}\equiv\mathbb{O}_{Q\times (2K-Q)}$,
$\tilde{\mathbb{O}}=\mathbb{O}_{(2K-Q)\times(2K-Q)}$, and
%%%%%%%%%%%%%%%%%%%%%%%%%%%%%%%%%%%%%%%%%%%%%%%%%%%%%%%%%%%%%%%%%%%%%%%%%%%%%%%%%%%%%%%%%%%%%%%%%%%%
\begin{eqnarray}\label{wQ}
w_{Q}\equiv\left[\left(\omega_{Q}+Y_{\mathcal{U},s_{Q}}\right)^{-1}\right]_{Q}
\end{eqnarray}
%%%%%%%%%%%%%%%%%%%%%%%%%%%%%%%%%%%%%%%%%%%%%%%%%%%%%%%%%%%%%%%%%%%%%%%%%%%%%%%%%%%%%%%%%%%%%%%%%%%%
is the first $Q\times Q$ block of the matrix
$\left(\omega_{Q}+Y_{\mathcal{U},s_{Q}}\right)^{-1}$. If we
express the matrix (\ref{YcalUsigmaQ}) in the block form
%%%%%%%%%%%%%%%%%%%%%%%%%%%%%%%%%%%%%%%%%%%%%%%%%%%%%%%%%%%%%%%%%%%%%%%%%%%%%%%%%%%%%%%%%%%%%%%%%%%%
\begin{eqnarray}\label{Yblock}
Y_{\mathcal{U},s_{Q}}=\left(\begin{array}{cc}
\mathcal{A} & \mathcal{C}\\
\mathcal{C}^{T} & \mathcal{B}\\
\end{array}\right),
\end{eqnarray}
%%%%%%%%%%%%%%%%%%%%%%%%%%%%%%%%%%%%%%%%%%%%%%%%%%%%%%%%%%%%%%%%%%%%%%%%%%%%%%%%%%%%%%%%%%%%%%%%%%%%
where $\mathcal{A}$ is a $Q\times Q$ block, $\mathcal{C}$ is a
$Q\times(L-Q)$ block and $\mathcal{B}$ is an $(L-Q)\times(L-Q)$
block, we can write
%%%%%%%%%%%%%%%%%%%%%%%%%%%%%%%%%%%%%%%%%%%%%%%%%%%%%%%%%%%%%%%%%%%%%%%%%%%%%%%%%%%%%%%%%%%%%%%%%%%%
\begin{eqnarray}\label{finalblock}
w_{Q}=\left[\left(\mathcal{V}^{T}\delta\mathcal{V}\right)_{Q}
+\mathcal{A}-\mathcal{C}\mathcal{B}^{-1}\mathcal{C}^{T}\right]^{-1},
\end{eqnarray}
%%%%%%%%%%%%%%%%%%%%%%%%%%%%%%%%%%%%%%%%%%%%%%%%%%%%%%%%%%%%%%%%%%%%%%%%%%%%%%%%%%%%%%%%%%%%%%%%%%%%
where we have used the blockwise inversion \cite{Horn_85}
%%%%%%%%%%%%%%%%%%%%%%%%%%%%%%%%%%%%%%%%%%%%%%%%%%%%%%%%%%%%%%%%%%%%%%%%%%%%%%%%%%%%%%%%%%%%%%%%%%%%
\begin{widetext}
\begin{eqnarray}\label{blockwise}
\left(\begin{array}{cc}
A & C\\
C^{T} & B\\
\end{array}\right)^{-1}=\left(\begin{array}{cc}
\left(A-CB^{-1}C^{T}\right)^{-1} & A^{-1}C\left(C^{T}A^{-1}C-B\right)^{-1}\\
\left(C^{T}A^{-1}C-B\right)^{-1}C^{T}A^{-1} & \left(B-C^{T}A^{-1}C\right)^{-1}\\
\end{array}\right).
\end{eqnarray}
\end{widetext}
%%%%%%%%%%%%%%%%%%%%%%%%%%%%%%%%%%%%%%%%%%%%%%%%%%%%%%%%%%%%%%%%%%%%%%%%%%%%%%%%%%%%%%%%%%%%%%%%%%%%

Repeated use of the formula \cite{Horn_85}
%%%%%%%%%%%%%%%%%%%%%%%%%%%%%%%%%%%%%%%%%%%%%%%%%%%%%%%%%%%%%%%%%%%%%%%%%%%%%%%%%%%%%%%%%%%%%%%%%%%%
\begin{widetext}
\begin{equation}\label{blockinversion}
\left(A-CB^{-1}C^{T}\right)^{-1}=A^{-1}+A^{-1}C\left(B-C^{T}A^{-1}C\right)^{-1}C^{T}A^{-1}
\end{equation}
\end{widetext}
%%%%%%%%%%%%%%%%%%%%%%%%%%%%%%%%%%%%%%%%%%%%%%%%%%%%%%%%%%%%%%%%%%%%%%%%%%%%%%%%%%%%%%%%%%%%%%%%%%%%
further reveals, that the $2K\times2K$ matrix $\mathcal{W}$ given in Eq.~(\ref{calW})
can be obtained as a limit of the $2K\times 2K$ matrix
%%%%%%%%%%%%%%%%%%%%%%%%%%%%%%%%%%%%%%%%%%%%%%%%%%%%%%%%%%%%%%%%%%%%%%%%%%%%%%%%%%%%%%%%%%%%%%%%%%%%
\begin{eqnarray}\label{calWx}
\mathcal{W}_{x}=\left[\mathcal{V}^{T}\delta\mathcal{V}+\left(\begin{array}{cc}
\mathcal{A}-\mathcal{C}\mathcal{B}^{-1}\mathcal{C}^{T} & \mathbb{O}\\
\mathbb{O}^{T} & x\openone\\
\end{array}\right)\right]^{-1},
\end{eqnarray}
%%%%%%%%%%%%%%%%%%%%%%%%%%%%%%%%%%%%%%%%%%%%%%%%%%%%%%%%%%%%%%%%%%%%%%%%%%%%%%%%%%%%%%%%%%%%%%%%%%%%
when $x\rightarrow+\infty$, $x\geq0$, and
$\openone\equiv\openone_{2K-Q}$. The Schur complement
(\ref{Schur4}) is then obtained from the matrix
%%%%%%%%%%%%%%%%%%%%%%%%%%%%%%%%%%%%%%%%%%%%%%%%%%%%%%%%%%%%%%%%%%%%%%%%%%%%%%%%%%%%%%%%%%%%%%
\begin{eqnarray}\label{Schur4x}
\sigma_{AB,x}\equiv\alpha-\beta
\mathcal{V}\mathcal{W}_{x}\mathcal{V}^{T}\beta^{T}
\end{eqnarray}
%%%%%%%%%%%%%%%%%%%%%%%%%%%%%%%%%%%%%%%%%%%%%%%%%%%%%%%%%%%%%%%%%%%%%%%%%%%%%%%%%%%%%%%%%%%%%%%
in the limit for $x\rightarrow+\infty$. By substitution we get
immediately
%%%%%%%%%%%%%%%%%%%%%%%%%%%%%%%%%%%%%%%%%%%%%%%%%%%%%%%%%%%%%%%%%%%%%%%%%%%%%%%%%%%%%%%%%%%%%%
\begin{eqnarray}\label{Schurfinal}
\sigma_{AB,x}\equiv\alpha-\beta\mathcal{Z}_{x}^{-1}\beta^{T},
\end{eqnarray}
%%%%%%%%%%%%%%%%%%%%%%%%%%%%%%%%%%%%%%%%%%%%%%%%%%%%%%%%%%%%%%%%%%%%%%%%%%%%%%%%%%%%%%%%%%%%%%%
with
%%%%%%%%%%%%%%%%%%%%%%%%%%%%%%%%%%%%%%%%%%%%%%%%%%%%%%%%%%%%%%%%%%%%%%%%%%%%%%%%%%%%%%%%%%%%%%%
\begin{eqnarray}\label{calZx}
\mathcal{Z}_{x}=\bar{\gamma}_{E}+\Gamma_{E}+\mathcal{V}\left(\begin{array}{cc}
\mathcal{A}-\mathcal{C}\mathcal{B}^{-1}\mathcal{C}^{T} & \mathbb{O}\\
\mathbb{O}^{T} & x\openone\\
\end{array}\right)\mathcal{V}^{T},
\end{eqnarray}
%%%%%%%%%%%%%%%%%%%%%%%%%%%%%%%%%%%%%%%%%%%%%%%%%%%%%%%%%%%%%%%%%%%%%%%%%%%%%%%%%%%%%%%%%%%%%%%
where we have used the equality $\delta=\bar{\gamma}_{E}+\Gamma_{E}$. The last matrix in the
latter equation is positive-semidefinite and therefore the matrix
%%%%%%%%%%%%%%%%%%%%%%%%%%%%%%%%%%%%%%%%%%%%%%%%%%%%%%%%%%%%%%%%%%%%%%%%%%%%%%%%%%%%%%%%%%%%%%%
\begin{equation}\label{tildeGammaE}
\tilde{\Gamma}_{E}^{x}=\Gamma_{E}+\mathcal{V}\left(\begin{array}{cc}
\mathcal{A}-\mathcal{C}\mathcal{B}^{-1}\mathcal{C}^{T} & \mathbb{O}\\
\mathbb{O}^{T} & x\openone\\
\end{array}\right)\mathcal{V}^{T}
\end{equation}
%%%%%%%%%%%%%%%%%%%%%%%%%%%%%%%%%%%%%%%%%%%%%%%%%%%%%%%%%%%%%%%%%%%%%%%%%%%%%%%%%%%%%%%%%%%%%%%
represents a legitimate CM of a Gaussian quantum state.
Consequently, a Gaussian measurement (\ref{POVMn}) on Eve's system
described by a CM $\Gamma_{E}$ followed by a Gaussian channel (\ref{dtilde})
characterized by the matrices $X$ and $Y$ on the outcomes of the
measurement can be replaced with another Gaussian
measurement with the CM
$\tilde{\Gamma}_{E}=\tilde{\Gamma}_{E}^{x\rightarrow+\infty}$
which concludes the proof.

%%%%%%%%%%%%%%%%%%%%%%%%%%%%%%%%%%%%%%%%%%%%%%%%%%%%%%%%%%%%%%%%%%%%%%%%%%%%%%%%%%%%%%%%%%%%%%%%%%%%%
\section{Proof that minimization over purifications can be omitted}\label{sec_2}

The next step of simplification of IE, Eq.~(\ref{Edownarrow}), is the proof that in the Gaussian scenario,
without loss of generality, we can use in Eq.~(\ref{Edownarrow}) a fixed minimal purification \cite{Caruso_11} of the state $\rho_{AB}$,
i.e., a purification containing minimum possible number of purifying modes.
Moreover, we also show that the minimization over all Gaussian purifications can be integrated into a minimization over
Eve's measurement.

According to the assumption, the state $\rho_{AB}$ is an $(N+M)$-mode Gaussian state, where subsystem
$A$ consists of $N$ modes and subsystem $B$ consists of $M$ modes.
The minimal purification of such a state is an $(N+M+R)$-mode pure Gaussian state
$|\Psi\rangle_{ABE}$ satisfying
$\mbox{Tr}_{E}|\Psi\rangle_{ABE}\langle\Psi|=\rho_{AB}$, for which the
purifying subsystem $E$ consists of $R\leq N+M$ modes, where $R$
is the number of symplectic eigenvalues of the CM $\gamma_{AB}$ of the state $\rho_{AB}$, that are strictly
greater than one \cite{Caruso_11}. When expressed with respect to
the $AB|E$ splitting the CM $(\equiv\gamma_{\pi})$ of the minimal
purification reads as
%%%%%%%%%%%%%%%%%%%%%%%%%%%%%%%%%%%%%%%%%%%%%%%%%%%%%%%%%%%%%%%%%%%%
\begin{eqnarray}\label{gammaABEsplitting}
\gamma_{\pi}=\left(\begin{array}{cc}
\gamma_{AB} & \gamma_{ABE} \\
\gamma_{ABE}^{T} & \gamma_{E} \\
\end{array}\right),
\end{eqnarray}
%%%%%%%%%%%%%%%%%%%%%%%%%%%%%%%%%%%%%%%%%%%%%%%%%%%%%%%%%%%%%%%%%
where
%%%%%%%%%%%%%%%%%%%%%%%%%%%%%%%%%%%%%%%%%%%%%%%%%%%%%%%%%%%%%%%%%%%%
\begin{eqnarray}\label{gammaEgammaABE}
\gamma_{E}=\bigoplus_{i=1}^{R}\nu_{i}\openone_{2},\quad
\gamma_{ABE}=S^{-1}\left(\begin{array}{c}
\bigoplus_{i=1}^{R}\sqrt{\nu_{i}^{2}-1}\sigma_{z} \\
\mathbb{O}_{2(N+M-R)\times2R} \\
\end{array}\right).\nonumber\\
\end{eqnarray}
%%%%%%%%%%%%%%%%%%%%%%%%%%%%%%%%%%%%%%%%%%%%%%%%%%%%%%%%%%%%%%%%%
Here, $\sigma_{z}=\mbox{diag}(1,-1)$ is the Pauli diagonal matrix and $S$ is
the symplectic matrix that brings the CM $\gamma_{AB}$ to the Williamson normal form (\ref{Williamson}), where $n=N+M$
and $\nu_{1}\geq\nu_{2}\geq\ldots\geq\nu_{R}>\nu_{R+1}=\ldots=\nu_{N+M}=1$.

In Eq.~(\ref{Edownarrow}) we consider the minimization over all Gaussian
purifications of the investigated Gaussian state
$\rho_{AB}$. For any such purification $|\bar{\Psi}\rangle_{ABE}$ with $K$-mode purifying subsystem $E$,
there is a Gaussian unitary transformation $U_{E}(\bar{S}_{E})$ on Eve's modes which connects the purification
$|\bar{\Psi}\rangle_{ABE}$ with the minimal purification $|\Psi\rangle_{ABE}$ by the formula \cite{Giedke_03,Magnin_10}
%%%%%%%%%%%%%%%%%%%%%%%%%%%%%%%%%%%%%%%%%%%%%%%%%%%%%%%%%%%%%%%%%%
\begin{eqnarray}\label{barPsi}
|\bar{\Psi}\rangle_{ABE}=U_{E}^{\dag}(\bar{S}_{E})|\Psi\rangle_{ABE}|\{0\}\rangle_{E_{R+1}\ldots E_{K}}.
\end{eqnarray}
%%%%%%%%%%%%%%%%%%%%%%%%%%%%%%%%%%%%%%%%%%%%%%%%%%%%%%%%%%%%%%%%%%
Here, $|\{0\}\rangle_{E_{R+1}\ldots
E_{K}}\equiv\otimes_{i=1}^{K-R}|0\rangle_{E_{R+i}}$ is the product
of $K-R$ ancillary vacuum modes that Eve can use, and the operator
$U_{E}(\bar{S}_{E})$ symplectically diagonalizes reduced state
$\bar{\rho}_{E}=\mbox{Tr}_{AB}(|\bar{\Psi}\rangle_{ABE}\langle\bar{\Psi}|)$
of Eve's subsystem $E$. Denoting the CM of the purification
$|\bar{\Psi}\rangle_{ABE}$ as $\bar{\gamma}_{\pi}$ one can express
the transformation (\ref{barPsi}) on the level of CMs in the form
%%%%%%%%%%%%%%%%%%%%%%%%%%%%%%%%%%%%%%%%%%%%%%%%%%%%%%%%%%%%%%%%%%%%
\begin{eqnarray}\label{bargammapi}
\bar{\gamma}_{\pi}=\left[\openone_{AB}\oplus
\bar{S}_{E}^{-1}\right]\gamma_{\pi}\oplus\openone_{E_{R+1}\ldots
E_{K}}\left[\openone_{AB}\oplus\left(\bar{S}_{E}^{T}\right)^{-1}\right],\nonumber\\
\end{eqnarray}
%%%%%%%%%%%%%%%%%%%%%%%%%%%%%%%%%%%%%%%%%%%%%%%%%%%%%%%%%%%%%%%%%%%%
where $\openone_{AB}$ is a $2(N+M)\times2(N+M)$ identity matrix,
$\openone_{E_{R+1}\ldots E_{K}}$ is a $2(K-R)\times2(K-R)$
identity matrix, and $\bar{S}_{E}$ is the $2K\times 2K$ symplectic
matrix symplectically diagonalizing the local CM $\bar{\gamma}_{E}$ of
Eve's subsystem, i.e.,
%%%%%%%%%%%%%%%%%%%%%%%%%%%%%%%%%%%%%%%%%%%%%%%%%%%%%%%%%%%%%%%%%%%%
\begin{equation}\label{SE}
\bar{S}_{E}\bar{\gamma}_{E}\bar{S}_{E}^{T}=\gamma_{E}\oplus\openone_{E_{R+1}\ldots E_{K}},
\end{equation}
%%%%%%%%%%%%%%%%%%%%%%%%%%%%%%%%%%%%%%%%%%%%%%%%%%%%%%%%%%%%%%%%%%%%
where $\gamma_{E}$ is the diagonal $2R\times2R$ CM of the reduced state of
subsystem $E$ of the minimal purification $|\Psi\rangle_{ABE}$ given in Eq.~(\ref{gammaEgammaABE}).
Expressing now the CMs $\gamma_{\pi}$ and $\bar{\gamma}_{\pi}$ with respect
to the $A|B|E$ splitting,
%%%%%%%%%%%%%%%%%%%%%%%%%%%%%%%%%%%%%%%%%%%%%%%%%%%%%%%%%%%%%%%%%%%%
\begin{eqnarray}\label{bargammapi1}
\bar{\gamma}_{\pi}=\left(\begin{array}{ccc}
\gamma_{A} & \omega_{AB} & \bar{\gamma}_{AE}\\
\omega_{AB}^{T} & \gamma_{B} & \bar{\gamma}_{BE}\\
\bar{\gamma}_{AE}^{T} & \bar{\gamma}_{BE}^{T} & \bar{\gamma}_{E}\\
\end{array}\right),
\end{eqnarray}
%%%%%%%%%%%%%%%%%%%%%%%%%%%%%%%%%%%%%%%%%%%%%%%%%%%%%%%%%%%%%%%%%
and
%%%%%%%%%%%%%%%%%%%%%%%%%%%%%%%%%%%%%%%%%%%%%%%%%%%%%%%%%%%%%%%%%%%%
\begin{eqnarray}\label{gammapi}
\gamma_{\pi}=\left(\begin{array}{ccc}
\gamma_{A} & \omega_{AB} & \gamma_{AE}\\
\omega_{AB}^{T} & \gamma_{B} & \gamma_{BE}\\
\gamma_{AE}^{T} & \gamma_{BE}^{T} & \gamma_{E}\\
\end{array}\right),
\end{eqnarray}
%%%%%%%%%%%%%%%%%%%%%%%%%%%%%%%%%%%%%%%%%%%%%%%%%%%%%%%%%%%%%%%%%
one gets from Eq.~(\ref{bargammapi}) for the $2(N+M)\times 2K$ block $\left(\bar{\gamma}_{AE}^{T},
\bar{\gamma}_{BE}^{T}\right)^{T}$ the expression
%%%%%%%%%%%%%%%%%%%%%%%%%%%%%%%%%%%%%%%%%%%%%%%%%%%%%%%%%%%%%%%%%
\begin{eqnarray}\label{bargammaABE}
\left(\begin{array}{c}
\bar{\gamma}_{AE} \\
\bar{\gamma}_{BE}\\
\end{array}\right)=\left(\begin{array}{cc}
\gamma_{AE} & \mathbb{O}_{2N\times2(K-R)}\\
\gamma_{BE} & \mathbb{O}_{2M\times2(K-R)}\\
\end{array}\right)\left(\bar{S}_{E}^{T}\right)^{-1}.
\end{eqnarray}
%%%%%%%%%%%%%%%%%%%%%%%%%%%%%%%%%%%%%%%%%%%%%%%%%%%%%%%%%%%%%%%%%
Further, by inverting Eq.~(\ref{SE}) we can also
express the CM $\bar{\gamma}_{E}$ as
%%%%%%%%%%%%%%%%%%%%%%%%%%%%%%%%%%%%%%%%%%%%%%%%%%%%%%%%%%%%%%%%%%%%
\begin{equation}\label{bargammaE}
\bar{\gamma}_{E}=\bar{S}_{E}^{-1}\left(\gamma_{E}\oplus\openone_{E_{R+1}...E_{K}}\right)\left(\bar{S}_{E}^{T}\right)^{-1}.
\end{equation}
%%%%%%%%%%%%%%%%%%%%%%%%%%%%%%%%%%%%%%%%%%%%%%%%%%%%%%%%%%%%%%%%%%%%

As any Gaussian channel on Eve's measurement outcomes can be integrated into Eve's measurement the CCM relevant to the optimization of the conditional mutual information is given by the Schur
complement
%%%%%%%%%%%%%%%%%%%%%%%%%%%%%%%%%%%%%%%%%%%%%%%%%%%%%%%%%%%%%%%%%%%%%%%%%%%%%%%%%%%%%%%%%%%%%%
\begin{eqnarray}\label{barsigmaintegrated}
\bar{\sigma}_{AB}=\alpha-\left(\begin{array}{c}
\bar{\gamma}_{AE} \\
\bar{\gamma}_{BE}\\
\end{array}\right)
(\bar{\gamma}_{E}+\bar{\Gamma}_{E})^{-1}\left(\begin{array}{c}
\bar{\gamma}_{AE} \\
\bar{\gamma}_{BE}\\
\end{array}\right)^{T}\nonumber\\
\end{eqnarray}
%%%%%%%%%%%%%%%%%%%%%%%%%%%%%%%%%%%%%%%%%%%%%%%%%%%%%%%%%%%%%%%%%%%%%%%%%%%%%%%%%%%%%%%%%%%%%%%
of the CCM
%%%%%%%%%%%%%%%%%%%%%%%%%%%%%%%%%%%%%%%%%%%%%%%%%%%%%%%%%%%%%%%%%%%%%%%%%%%%%%%%%%%%%%%%%%%%%%%%%%
\begin{eqnarray}\label{barSigma}
\bar{\Sigma}=\bar{\gamma}_{\pi}+\Gamma_{A}\oplus\Gamma_{B}\oplus\bar{\Gamma}_{E}\equiv\left(\begin{array}{cc}
\alpha & \beta\\
\beta^{T} & \bar{\delta} \\
\end{array}\right),
\end{eqnarray}
%%%%%%%%%%%%%%%%%%%%%%%%%%%%%%%%%%%%%%%%%%%%%%%%%%%%%%%%%%%%%%%%%%%%%%%%%%%%%%%%%%%%%%%%%%%%%%%%%%%%
where $\Gamma_{A},\Gamma_{B}$ and $\bar{\Gamma}_{E}$ are CMs of
the measurements on Alice's, Bob's and Eve's subsystems and the
last $2\times2$ block matrix is the expression of the matrix
$\bar{\Sigma}$ with respect to $AB|E$ splitting. Inserting from
Eq.~(\ref{bargammaABE}) into Eq.~(\ref{barsigmaintegrated}) one
gets after some algebra
%%%%%%%%%%%%%%%%%%%%%%%%%%%%%%%%%%%%%%%%%%%%%%%%%%%%%%%%%%%%%%%%%%%%%%%%%%%%%%%%%%%%%%%%%%%%%%%%%%%%
%%%%%%%%%%%%%%%%%%%%%%%%%%%%%%%%%%%%%%%%%%%%%%%%%%%%%%%%%%%%%%%%%%%%%%%%%%%%%%%%%%%%%%%%%%%%%%
\begin{eqnarray}\label{RHS}
\bar{\sigma}_{AB}=\alpha-\left(\begin{array}{c}
\gamma_{AE} \\
\gamma_{BE}\\
\end{array}\right)\mathcal{T}_{R}
\left(\begin{array}{c}
\gamma_{AE} \\
\gamma_{BE}\\
\end{array}\right)^{T},\nonumber\\
\end{eqnarray}
%%%%%%%%%%%%%%%%%%%%%%%%%%%%%%%%%%%%%%%%%%%%%%%%%%%%%%%%%%%%%%%%%%%%%%%%%%%%%%%%%%%%%%%%%%%%%%%
where $\mathcal{T}_{R}$ is the first $2R\times2R$ diagonal block of the
$2K\times2K$ matrix
%%%%%%%%%%%%%%%%%%%%%%%%%%%%%%%%%%%%%%%%%%%%%%%%%%%%%%%%%%%%%%%%%%%%%%%%%%%%%%%%%%%%%%%%%%%%%%%
\begin{equation}\label{calT}
\mathcal{T}=\left[\gamma_{E}\oplus\openone_{E_{R+1}...E_{K}}+\bar{\Gamma}_{E}(\bar{S}_{E})\right]^{-1},
\end{equation}
%%%%%%%%%%%%%%%%%%%%%%%%%%%%%%%%%%%%%%%%%%%%%%%%%%%%%%%%%%%%%%%%%%%%%%%%%%%%%%%%%%%%%%%%%%%%%%%
and $\bar{\Gamma}_{E}(\bar{S}_{E})=\bar{S}_{E}\bar{\Gamma}_{E}\bar{S}_{E}^{T}$
is a $2K\times2K$ CM of another of Eve's measurements. If we express
finally the latter matrix in the block form
%%%%%%%%%%%%%%%%%%%%%%%%%%%%%%%%%%%%%%%%%%%%%%%%%%%%%%%%%%%%%%%%%%%%%%%%%%%%%%%%%%%%%%%%%%%%%%%%%%%%
\begin{eqnarray}\label{tildeGammablock}
\bar{\Gamma}_{E}(\bar{S}_{E})=\left(\begin{array}{cc}
\bar{A} & \bar{C}\\
\bar{C}^{T} & \bar{B}\\
\end{array}\right),
\end{eqnarray}
%%%%%%%%%%%%%%%%%%%%%%%%%%%%%%%%%%%%%%%%%%%%%%%%%%%%%%%%%%%%%%%%%%%%%%%%%%%%%%%%%%%%%%%%%%%%%%%%%%%%
with $\bar{A}$ being a $2R\times2R$ matrix and $\bar{B}$ being a
$2(K-R)\times2(K-R)$ matrix we can express the block $\mathcal{T}_{R}$
using formula (\ref{blockwise}) as
%%%%%%%%%%%%%%%%%%%%%%%%%%%%%%%%%%%%%%%%%%%%%%%%%%%%%%%%%%%%%%%%%%%%%%%%%%%%%%%%%%%%%%%%%%%%%%%
\begin{equation}\label{calTL}
\mathcal{T}_{R}=\left[\gamma_{E}+\bar{A}-\bar{C}(\bar{B}+\openone)^{-1}\bar{C}^{T}\right]^{-1}.
\end{equation}
%%%%%%%%%%%%%%%%%%%%%%%%%%%%%%%%%%%%%%%%%%%%%%%%%%%%%%%%%%%%%%%%%%%%%%%%%%%%%%%%%%%%%%%%%%%%%%%
The matrix
%%%%%%%%%%%%%%%%%%%%%%%%%%%%%%%%%%%%%%%%%%%%%%%%%%%%%%%%%%%%%%%%%%%%%%%%%%%%%%%%%%%%%%%%%%%%%%%
\begin{equation}\label{GammaE}
\Gamma_{E}\equiv\bar{A}-\bar{C}(\bar{B}+\openone)^{-1}\bar{C}^{T}
\end{equation}
%%%%%%%%%%%%%%%%%%%%%%%%%%%%%%%%%%%%%%%%%%%%%%%%%%%%%%%%%%%%%%%%%%%%%%%%%%%%%%%%%%%%%%%%%%%%%%%
can be viewed as a CM of an $R$-mode conditional Gaussian state
obtained by projecting the last $K-R$ modes of a
$K$-mode Gaussian state with CM (\ref{tildeGammablock}) onto a coherent state and
therefore $\Gamma_E$ is a legitimate CM of a physical quantum state.
Consequently, one finally gets for the matrix (\ref{barsigmaintegrated}) the following equation:
%%%%%%%%%%%%%%%%%%%%%%%%%%%%%%%%%%%%%%%%%%%%%%%%%%%%%%%%%%%%%%%%%%%%%%%%%%%%%%%%%%%%%%%%%%%%%%
\begin{eqnarray}\label{barsigmafinal}
\bar{\sigma}_{AB}=\alpha-\left(\begin{array}{c}
\gamma_{AE} \\
\gamma_{BE}\\
\end{array}\right)
(\gamma_{E}+\Gamma_{E})^{-1}\left(\begin{array}{c}
\gamma_{AE} \\
\gamma_{BE}\\
\end{array}\right)^{T}=\sigma_{AB}.\nonumber\\
\end{eqnarray}
%%%%%%%%%%%%%%%%%%%%%%%%%%%%%%%%%%%%%%%%%%%%%%%%%%%%%%%%%%%%%%%%%%%%%%%%%%%%%%%%%%%%%%%%%%%%%%%
Thus, for any Gaussian purification and any Gaussian measurement on subsystem
$E$, the matrix (\ref{barsigmaintegrated}) can be obtained from the minimal
purification with CM (\ref{gammaABEsplitting}) and a Gaussian
measurement with CM (\ref{GammaE}) on Eve's part of the
purification. Hence, when calculating the quantity defined in Eq.~(\ref{Edownarrow})
in the Gaussian scenario we can consider only a fixed minimal purification and
we can omit the minimization with respect to all Gaussian purifications, which accomplishes the proof.

Having found simplifications of IE in the Gaussian scenario we are
now in the position to incorporate them into the definition
(\ref{Edownarrow}). Let us consider a Gaussian state $\rho_{AB}$
and its minimal purification with CM (\ref{gammaABEsplitting})
which has been mapped by Gaussian measurements with CMs
$\Gamma_{A},\Gamma_{B}$ and $\Gamma_{E}$ onto the Gaussian
distribution of the form (\ref{Pinput}). As we have already said,
the intrinsic information in Eq.~(\ref{Edownarrow}) can be
replaced with the standard conditional mutual information
(\ref{conditionalI}), which coincides with the standard mutual
information ($\equiv I_{c}(A;B)$) of the corresponding conditional
distribution. The latter distribution possesses the CCM in the
form of the Schur complement \cite{Horn_85}
%%%%%%%%%%%%%%%%%%%%%%%%%%%%%%%%%%%%%%%%%%%%%%%%%%%%%%%%%%%%%%%%
%\begin{widetext}
\begin{eqnarray}\label{sigma}
\sigma_{AB}&=&\gamma_{AB}+\Gamma_{A}\oplus\Gamma_{B}
-\gamma_{ABE}\left(\gamma_{E}+\Gamma_{E}\right)^{-1}\gamma_{ABE}^{T},\nonumber\\
\end{eqnarray}
%\end{widetext}
%%%%%%%%%%%%%%%%%%%%%%%%%%%%%%%%%%%%%%%%%%%%%%%%%%%%%%%%%%%%%%%%%%%%%
where $\gamma_{AB},\gamma_{ABE}$ and $\gamma_{E}$ are submatrices
of the CM $\gamma_{\pi}$ of the minimal purification of the state
$\rho_{AB}$, which are defined in Eq.~(\ref{gammaEgammaABE}). From
the formula for mutual information of a bivariate Gaussian
distribution \cite{Gelfand_57} it follows further that
$I_{c}(A;B)=f(\gamma_{\pi},\Gamma_{A},\Gamma_{B},\Gamma_{E})$,
where
%%%%%%%%%%%%%%%%%%%%%%%%%%%%%%%%%%%%%%%%%%%%%%%%%%%%%%%%%%%%%%%%%%%%%
\begin{equation}\label{f}
f\left(\gamma_{\pi},\Gamma_{A},\Gamma_{B},\Gamma_{E}\right)=\frac{1}{2}\ln\left(\frac{\mbox{det}\sigma_{A}\mbox{det}\sigma_{B}}{\mbox{det}\sigma_{AB}}\right)
\end{equation}
%%%%%%%%%%%%%%%%%%%%%%%%%%%%%%%%%%%%%%%%%%%%%%%%%%%%%%%%%%%%%%%%%%%%%
with $\sigma_{A,B}$ being local submatrices of CCM (\ref{sigma}).
If we use now the definition of IE, Eq.~(\ref{Edownarrow}), and we
take into account that we can omit minimization over all
purifications, we arrive finally at the following formula for GIE
\cite{Mista_14}
%%%%%%%%%%%%%%%%%%%%%%%%%%%%%%%%%%%%%%%%%%%%%%%%%%%%%%%%%%%%%%%%%%%%%
\begin{equation}\label{Gaussianmu}
E_{\downarrow}^{G}\left(\rho_{AB}\right)=\mathop{\mbox{sup}}_{\Gamma_{A},\Gamma_{B}}
\mathop{\mbox{inf}}_{\Gamma_{E}}f\left(\gamma_{\pi},\Gamma_{A},\Gamma_{B},\Gamma_{E}\right).
\end{equation}
%%%%%%%%%%%%%%%%%%%%%%%%%%%%%%%%%%%%%%%%%%%%%%%%%%%%%%%%%%%%%%%%%%%%%

Before going further let us note one consequence stemming from the fact
that for any purification with $K$ purifying modes described by the CM
(\ref{bargammapi1}) and any measurement on the modes with CM
$\bar{\Gamma}_{E}$ we can find a measurement with CM (\ref{GammaE}) on the minimal
purification giving the same matrix (\ref{barsigmaintegrated}).
This implies that for any two purifications containing a generally different
and not necessarily minimal number of purifying modes, we can find
measurements on the purifying subsystems which yield the same
matrix (\ref{barsigmaintegrated}). To show this, consider two purifications
with CMs $\bar{\gamma}_{\pi}$ and $\gamma_{\pi}'$ which contain $K$ and
$K'$ purifying modes, respectively, where $K\leq K'$. By using
Eq.~(\ref{bargammapi}) for both the CMs $\bar{\gamma}_{\pi}$ and $\gamma_{\pi}'$
one finds that they are connected by a similar equation,
%%%%%%%%%%%%%%%%%%%%%%%%%%%%%%%%%%%%%%%%%%%%%%%%%%%%%%%%%%%%%%%%%%%%
\begin{eqnarray}\label{tildegammapi}
\gamma_{\pi}'=\left[\openone_{AB}\oplus
\mathscr{S}_{E}^{-1}\right]\bar{\gamma}_{\pi}\oplus\openone_{2(K'-K)}\left[\openone_{AB}\oplus\left(\mathscr{S}_{E}^{T}\right)^{-1}\right].\nonumber\\
\end{eqnarray}
%%%%%%%%%%%%%%%%%%%%%%%%%%%%%%%%%%%%%%%%%%%%%%%%%%%%%%%%%%%%%%%%%%%%
Here, $\openone_{2(K'-K)}=\openone_{E_{K+1}...E_{K'}}$ and the
symplectic matrix
$\mathscr{S}_{E}=[\bar{S}_{E}^{-1}\oplus\openone_{2(K'-K)}]S_{E}'$
satisfies
$\mathscr{S}_{E}\gamma_{E}'\mathscr{S}_{E}^{T}=\bar{\gamma}_{E}\oplus\openone_{2(K'-K)}$
and it consists of symplectic matrices $\bar{S}_{E}$ and $S'_{E}$
which symplectically diagonalize the local CMs $\bar{\gamma}_{E}$
and $\gamma_{E}'$ of CMs $\bar{\gamma}_{\pi}$ and $\gamma_{\pi}'$,
respectively, corresponding to subsystem $E$. Making use of the
formula (\ref{tildegammapi}) we can now repeat the procedure
leading from Eq.~(\ref{bargammapi}) to Eq.~(\ref{barsigmafinal})
to show that for the purification with CM $\gamma_{\pi}'$ and an
arbitrary measurement with CM $\Gamma_{E}'$ on subsystem $E$ there
always exists a measurement with CM $\bar{\Gamma}_{E}$ on the
purification with CM $\bar{\gamma}_{\pi}$ for which it holds that
$\sigma_{AB}'=\bar{\sigma}_{AB}$. If we perform, on the other
hand, on the subsystem $E$ of the purification with CM
$\gamma_{\pi}'$ the measurement with CM
$\Gamma_{E}'\equiv\mathscr{S}_{E}^{-1}[\bar{\Gamma}_{E}\oplus\openone_{2(K'-K)}](\mathscr{S}_{E}^{T})^{-1}$,
one finds easily that the matrix $\sigma_{AB}'$ is equal to the
matrix $\bar{\sigma}_{AB}$ corresponding to the purification with
CM $\bar{\gamma}_{\pi}$ and the measurement with CM
$\bar{\Gamma}_{E}$. Therefore, without loss of generality we can
consider in the formula (\ref{Gaussianmu})
an arbitrary fixed purification, i.e., a fixed purification containing an arbitrary
number of purifying modes, and we can restrict ourselves to
minimizing only over all Gaussian measurements on the purifying
modes. This property proves to be useful in the proof of the
monotonicity of the GIE under GLTPOCC, which is given later.

%%%%%%%%%%%%%%%%%%%%%%%%%%%%%%%%%%%%%%%%%%%%%%%%%%%%%%%%%%%%%%%%%%%%%%%%%%%%%%%%%%%%%%%%%%%%%%%%%%%%%%%%%%%%%%%%
\section{Gaussian measurement projecting purification of a separable Gaussian state onto a product state}\label{sec_3}

A basic property of any entanglement measure is that it
vanishes on all separable states \cite{Horodecki_09}. In
Ref.~\cite{Gisin_00} it was shown that for any separable state
whatever measurements are performed by Alice and Bob there is
always Eve's measurement such that the conditional mutual
information (\ref{conditionalI}) of the probability distribution
(\ref{mapping}) vanishes. Inspired by the proof of the latter
statement we show here, that also the GIE is zero for all
separable Gaussian states.

The vanishing of the GIE on separable Gaussian states is a direct
consequence of the fact that for any separable Gaussian state
$\rho_{AB}^{\rm sep}$ there is a Gaussian measurement on the
purifying system $E$ of the minimal purification of the state,
that projects modes $A$ and $B$ onto a pure product state. Indeed,
by performing such a measurement on subsystem $E$ of the minimal
purification of a separable state $\rho_{AB}^{\rm sep}$ one finds
that the conditional distribution $P(d_{A},d_{B}|d_{E})$
factorizes as $P(d_{A},d_{B}|d_{E})=P(d_{A}|d_{E})P(d_{B}|d_{E})$
for any measurement on subsystems $A$ and $B$. Consequently, the
conditional mutual information (\ref{conditionalI}) and therefore
also GIE are equal to zero.

It remains to find the measurement mentioned above. The sought
measurement can be constructed after consideration of a
measurement on another purification created using the separability
criterion \cite{Werner_01}. According to the separability
criterion a Gaussian state with CM $\gamma_{AB}^{\rm sep}$ is
separable if and only if there exist pure-state CMs
$\gamma_{A,B}^{p}$ such that the matrix $Q\equiv\gamma_{AB}^{\rm
sep}-\gamma_{A}^{p}\oplus\gamma_{B}^{p}\geq0$. If $V$ denotes the
orthogonal matrix diagonalizing the matrix $Q$, i.e., $V^T Q
V=\mbox{diag}(\lambda_1,\lambda_{2},\ldots,\lambda_P,0,\ldots,0)$,
where $\lambda_{i}$, $i=1,\ldots,P$ are the strictly positive
eigenvalues of the matrix $Q$, the state $\rho_{AB}^{\rm sep}$ can
be expressed as
%%%%%%%%%%%%%%%%%%%%%%%%%%%%%%%%%%%%%%%%%%%%%%%%%%%%%%%%%%%%%%%%%%%%%
\begin{eqnarray}\label{rhoABsep}
\rho_{AB}^{\rm sep}&=&\int
p(\mathbf{r})\bigotimes_{j=A,B}^{}D_{j}[(\mathscr{V}\mathbf{r})_{j}]|\gamma_{j}^{p}\rangle_{j}\langle\gamma_{j}^{p}|
D^{\dag}_{j}[(\mathscr{V}\mathbf{r})_{j}]\nonumber\\
&&\times\Pi_{l=1}^{P}dr_{l}.
\end{eqnarray}
%%%%%%%%%%%%%%%%%%%%%%%%%%%%%%%%%%%%%%%%%%%%%%%%%%%%%%%%%%%%%%%%%%%%
%%%%%%%%%%%%%%%%%%%%%%%%%%%%%%%%%%%%%%%%%%%%%%%%%%%%%%%%%%%%%%%%%%%%%
%\begin{eqnarray}\label{rhoABsep}
%\rho_{AB}^{\rm sep}&=&\int
%\frac{e^{-\sum_{i=1}^{K}\frac{r_{i}^2}{\lambda_{i}}}}{\sqrt{\pi^K\Pi_{i=1}^{K}\lambda_{i}}}\bigotimes_{j=A,B}^{}D_{j}[(\mathcal{V}r)_{j}]|\omega_{j}\rangle_{j}\langle\omega_{j}|
%D^{\dag}_{j}[(\mathcal{V}r)_{j}]\nonumber\\
%&&\times\Pi_{l=1}^{K}dr_{l},
%\end{eqnarray}
%%%%%%%%%%%%%%%%%%%%%%%%%%%%%%%%%%%%%%%%%%%%%%%%%%%%%%%%%%%%%%%%%%%%
%$p(r)=\mbox{exp}\left(-\sum_{i=1}^{K}r_{i}^2/\lambda_{i}\right)/\sqrt{\pi^K\Pi_{i=1}^{K}\lambda_{i}}$
Here $D_j(d_j)$ stands for the $J_{j}$-mode displacement operator
performing phase-space displacement of the subsystem $j$ by
$d_j=(d_{j_{1}}^{(x)},d_{j_{1}}^{(p)},\ldots,d_{j_{J_{j}}}^{(x)},d_{j_{J_{j}}}^{(p)})^{T}$
with
$\xi_{j}=(x_{j_{1}},p_{j_{1}},\ldots,x_{j_{J_{j}}},p_{j_{J_{j}}})^{T}$
being the vector of the quadratures of the subsystem,
$p(\mathbf{r})\equiv\Pi_{i=1}^{P}\mbox{exp}\left(-r_{i}^2/\lambda_{i}\right)/\sqrt{\pi\lambda_{i}}$,
$|\gamma_{A,B}^{p}\rangle_{A,B}$ are pure states with CMs
$\gamma_{A,B}^{p}$ and zero displacements, $\mathscr{V}$ is the
$2(N+M)\times P$ matrix composed of the first $P$ columns of the
matrix $V$, $\mathbf{r}=(r_1,r_2,\ldots,r_P)^{T}$,
$(\mathscr{V}\mathbf{r})_{A}=[(\mathscr{V}\mathbf{r})_{1},(\mathscr{V}\mathbf{r})_{2},\ldots,(\mathscr{V}\mathbf{r})_{2N}]^{T}$,
and
$(\mathscr{V}\mathbf{r})_{B}=[(\mathscr{V}\mathbf{r})_{2N+1},(\mathscr{V}\mathbf{r})_{2N+2},\ldots,(\mathscr{V}\mathbf{r})_{2(N+M)}]^{T}$.
Now we construct a new $(N+M+P)$-mode purification by encoding the
displacements $r_{j}$, $j=1,\ldots,P$ into the eigenvectors
$|r_j\rangle_{E_j}$ of position quadratures of $P$ purifying modes
$E_{1},E_{2},\ldots,E_{P}$ as
%%%%%%%%%%%%%%%%%%%%%%%%%%%%%%%%%%%%%%%%%%%%%%%%%%%%%%%%%%%%%%%%%%%%%
\begin{eqnarray}\label{tildepsipi}
|\tilde\Psi\rangle_{ABE}&=&\int\sqrt{p(\mathbf{r})}\bigotimes_{j=A,B}^{}D_{j}[(\mathscr{V}\mathbf{r})_{j}]|\gamma_{j}^{p}\rangle_{j}|\mathbf{r}\rangle_{E}\nonumber\\
&&\times\Pi_{l=1}^{P}dr_{l},
\end{eqnarray}
%%%%%%%%%%%%%%%%%%%%%%%%%%%%%%%%%%%%%%%%%%%%%%%%%%%%%%%%%%%%%%%%%%%%
where
$|\mathbf{r}\rangle_{E}\equiv|r_1\rangle_{E_1}|r_2\rangle_{E_2}\ldots|r_P\rangle_{E_P}$.
By measuring position quadratures on all modes of the subsystem $E$
with the outcome $\mathbf{r}'$ one gets the following product
conditional state:
%%%%%%%%%%%%%%%%%%%%%%%%%%%%%%%%%%%%%%%%%%%%%%%%%%%%%%%%%%%%%%%%%%%%%
\begin{eqnarray}\label{product}
D_{A}[(\mathscr{V}\mathbf{r'})_{A}]|\gamma_{A}^{p}\rangle_{A}D_{B}[(\mathscr{V}\mathbf{r'})_{B}]|\gamma_{B}^{p}\rangle_{B}.
\end{eqnarray}
%%%%%%%%%%%%%%%%%%%%%%%%%%%%%%%%%%%%%%%%%%%%%%%%%%%%%%%%%%%%%%%%%%%%
At this point we have shown that there is a Gaussian measurement
that can be performed on the $P$ modes of the $(N+M+P)$-mode pure
state $|\tilde\Psi\rangle_{ABE}$ that leaves Alice's and Bob's
modes separable. As the $(N+M+R)$-mode minimal purification
$|\Psi\rangle_{ABE}$ and the $(N+M+P)$-mode purification
$|\tilde\Psi\rangle_{ABE}$ both possess the same reduced state
$\rho_{AB}^{\rm sep}$, there is a Gaussian unitary transformation
$U_{E}(\tilde{S}_{E})$  which transforms the purification
(\ref{tildepsipi}) into the minimal purification as
\cite{Giedke_03,Magnin_10}
%%%%%%%%%%%%%%%%%%%%%%%%%%%%%%%%%%%%%%%%%%%%%%%%%%%%%%%%%%%%%%%%%%
\begin{eqnarray}\label{US}
U_{E}(\tilde{S}_{E})|\tilde\Psi\rangle_{ABE}=|\Psi\rangle_{ABE}|\left\{0\right\}\rangle_{E_{R+1}\ldots
E_{P}}.
\end{eqnarray}
%%%%%%%%%%%%%%%%%%%%%%%%%%%%%%%%%%%%%%%%%%%%%%%%%%%%%%%%%%%%%%%%%%
Here, $|\left\{0\right\}\rangle_{E_{R+1}\ldots
E_{P}}\equiv|0\rangle_{E_{R+1}}|0\rangle_{E_{R+2}}\ldots|0\rangle_{E_{P}}$
is the product of $P-R$ vacuum states and $R\leq P$ is the number
of symplectic eigenvalues of the CM $\gamma_{AB}^{\rm sep}$ that
are strictly greater than one. The operator $U_{E}(\tilde{S}_{E})$ on the
$P$ modes $E_{1},E_{2},\ldots,E_{P}$ corresponds to the symplectic
transformation $\tilde{S}_{E}$ symplectically diagonalizing the
$2P\times2P$ CM $\tilde{\gamma}_{E}$ of the subsystem $E$ of the
purification (\ref{tildepsipi}), i.e.,
$\tilde{S}_{E}\tilde{\gamma}_{E}\tilde{S}_{E}^{T}=\mbox{diag}(\nu_{1},\nu_{1},\ldots,\nu_{R},\nu_{R},1,1,\ldots,1,1)$,
where $\nu_{1},\nu_{2},\ldots,\nu_{R}$ are symplectic eigenvalues
of the CM $\gamma_{AB}^{\rm sep}$ which are strictly greater than
one. Thus, by appending $P-R$ vacuum states $|0\rangle_{E_j}$,
$j=R+1,R+2,\ldots,P$ to the minimal purification $|\Psi\rangle_{ABE}$,
applying the Gaussian unitary $U_{E}^{\dag}(\tilde{S}_{E})$ to the
subsystem $E$ and projecting the subsystem onto the position
eigenstate $|\mathbf r'\rangle_{E}$ we get the product state
(\ref{product}). Simple algebra reveals that this measurement
can be rewritten as a projection of $R$ modes
$E_{1},E_{2},\ldots,E_{R}$ of the minimal purification
$|\Psi\rangle_{ABE}$ onto an unnormalized (and generally unnormalizable) Gaussian state
%%%%%%%%%%%%%%%%%%%%%%%%%%%%%%%%%%%%%%%%%%%%%%%%%%%%%%%%%%%%%%%%%%%%%%%%%%%%%%
\begin{eqnarray}\label{Pi0primed}
\Pi_{0}'=\langle\left\{0\right\}|U_{E}(\tilde{S}_{E})|\mathbf r=0\rangle_{E}\langle\mathbf r=0|U_{E}^{\dag}(\tilde{S}_{E})|\left\{0\right\}\rangle,
\end{eqnarray}
%%%%%%%%%%%%%%%%%%%%%%%%%%%%%%%%%%%%%%%%%%%%%%%%%%%%%%%%%%%%%%%%%%%%%%%%%%%%%%%
displaced by some factor dependent on the elements of the vector
$\mathbf{r}'$ and symplectic matrix $\tilde{S}_{E}$, where in
equation (\ref{Pi0primed}) we have omitted the subscripts of the
state $|\left\{0\right\}\rangle_{E_{R+1}\ldots E_{P}}$ for brevity. Now, let us define a
normalized $R$-mode zero mean Gaussian state
%%%%%%%%%%%%%%%%%%%%%%%%%%%%%%%%%%%%%%%%%%%%%%%%%%%%%%%%%%%%%%%%%%%%%%%%%%%%%%
\begin{eqnarray}\label{Pi0normalized}
\Pi_{0}=\frac{\langle\left\{0\right\}|U_{E}(\tilde{S}_{E})|\mathbf{s}\rangle_{E}^{(x)}\langle\mathbf{s}|U_{E}^{\dag}(\tilde{S}_{E})|\left\{0\right\}\rangle}
{\mbox{Tr}\left[\langle\left\{0\right\}|U_{E}(\tilde{S}_{E})|\mathbf{s}\rangle_{E}^{(x)}\langle\mathbf{s}|U_{E}^{\dag}(\tilde{S}_{E})|\left\{0\right\}\rangle\right]},
\end{eqnarray}
%%%%%%%%%%%%%%%%%%%%%%%%%%%%%%%%%%%%%%%%%%%%%%%%%%%%%%%%%%%%%%%%%%%%%%%%%%%%%%%
which is obtained by replacing the $P$-mode position eigenvector
$|\mathbf r=0\rangle_{E}$ in the state in Eq.~(\ref{Pi0primed})
with a $P$-mode zero mean position squeezed vacuum state
$|\mathbf{s}\rangle_{E}^{(x)}\equiv|s_{1}\rangle_{E_{1}}^{(x)}|s_{2}\rangle_{E_{2}}^{(x)}\ldots|s_{P}\rangle_{E_{P}}^{(x)}$,
where $|s_{j}\rangle_{E_{j}}^{(x)}$ is the zero mean position
squeezed vacuum state of mode $E_{j}$ with the squeezing parameter
$s_{j}$. It is now obvious that by performing a Gaussian
measurement
$\Pi_E(d_E)=D_E(d_E)\Pi_{0}D_E^\dagger(d_E)/(2\pi)^{R}$ on the
subsystem $E$ of the minimal purification $|\Psi\rangle_{ABE}$, we
project the purification onto a product state
%%%%%%%%%%%%%%%%%%%%%%%%%%%%%%%%%%%%%%%%%%%%%%%%%%%%%%%%%%%%%%%%%%%%%
\begin{eqnarray}\label{productfinal}
D_{A}(d_{A}')|\gamma_{A}^{p}\rangle_{A}D_{B}(d_{B}')|\gamma_{B}^{p}\rangle_{B},
\end{eqnarray}
%%%%%%%%%%%%%%%%%%%%%%%%%%%%%%%%%%%%%%%%%%%%%%%%%%%%%%%%%%%%%%%%%%%%
in the limit of infinite squeezing parameters $s_j$. The vectors
of displacements $d_{A}'$ and $d_{B}'$ are linear combinations of
the elements of the vector $d_{E}$ of the measurement outcomes
\cite{Giedke_02}. We have therefore found that for any separable
Gaussian state of two subsystems $A$ and $B$, there is a Gaussian
measurement (\ref{POVMn}) on the purifying part of the state that
projects the minimal purification onto a product of pure local
states of subsystems $A$ and $B$ as we set out to prove.
%%%%%%%%%%%%%%%%%%%%%%%%%%%%%%%%%%%%%%%%%%%%%%%%%%%%%%%%%%%%%%%%%%%%%%%%%%%%%%%%%5

\section{Monotonicity of GIE under Gaussian local trace-preserving operations and classical communication}\label{sec_4}

The most important property of any good entanglement measure
is its monotonicity \cite{Vidal_00}, which means that the measure
does not increase under LOCC operations. Specifically, a good
Gaussian entanglement measure should not increase under (generally
probabilistic) Gaussian local operations and classical
communication (GLOCC) \cite{Wolf_04}. In this section we prove
that the GIE defined in Eq.~(\ref{Gaussianmu}) is non-increasing
under the subset of GLOCC given by GLTPOCC. This means, that if
the operation $(\equiv \mathcal{E})$ transforms the input Gaussian
state $\rho_{AB}^{\cal I}$ onto a state $\rho_{AB}^{\mathcal{E}}$,
then
%%%%%%%%%%%%%%%%%%%%%%%%%%%%%%%%%%%%%%%%%%%%%%%%%%%%%%%%%%%%%%%%%%%%%
\begin{equation}\label{monotonicity2}
E_{\downarrow}^{G}\left(\rho_{AB}^{\cal I}\right)\geq
E_{\downarrow}^{G}\left(\rho_{AB}^{\mathcal{E}}\right).
\end{equation}
%%%%%%%%%%%%%%%%%%%%%%%%%%%%%%%%%%%%%%%%%%%%%%%%%%%%%%%%%%%%%%%%%%%%%

It was shown in the previous section that for two different
purifications having in general a differing number of modes, one
can always find measurements on Eve's modes of either purification
that yield the same matrix (\ref{sigma}). Therefore, for CMs
$\gamma_{\pi}$ and $\Gamma_{E}$ in Eq.~(\ref{sigma}) we can
consider a CM of an arbitrary (not necessarily minimal)
purification and a CM of a measurement on Eve's modes of this
purification. In the following paragraph we prove the monotonicity
of GIE under GLTPOCC by using a suitable non-minimal purification
of the output state $\rho_{AB}^\mathcal{E}$.

A trace-preserving operation $\mathcal{E}$ transforms the input
state $\rho_{AB}^{\cal I}$ to a state
%%%%%%%%%%%%%%%%%%%%%%%%%%%%%%%%%%%%%%%%%%%%%%%%%%%%%%%%%%%%%%%%%%%%%%
\begin{equation}\label{rhoM}
\rho_{AB}^\mathcal{E}=\mbox{Tr}_{\rm
in}\left[\chi\left(\rho_{AB}^{\cal
I}\right)^{T}\otimes\openone_{\rm out}\right],
\end{equation}
%%%%%%%%%%%%%%%%%%%%%%%%%%%%%%%%%%%%%%%%%%%%%%%%%%%%%%%%%%%%%%%%%%%%%%
where $\chi$ is a positive-semidefinite operator representing the
operation \cite{Jamiolkowski_72} on the tensor product
$\mathcal{H}_{AB}\otimes\mathcal{H}_{\rm out}$ of the input
Hilbert space $\mathcal{H}_{AB}$ and the output Hilbert space
$\mathcal{H}_{\rm out}$, $\openone_{\rm out}$ is the identity
operator on the output Hilbert space and $\mbox{Tr}_{\rm in}$ is
the trace over the input Hilbert space. The map preserves the
trace of the input state, i.e., $\mbox{Tr}_{\rm
in}[\rho_{AB}^{\cal I}]=\mbox{Tr}_{\rm
out}[\rho_{AB}^\mathcal{E}]$, which imposes the following
constraint on the state $\chi$
%%%%%%%%%%%%%%%%%%%%%%%%%%%%%%%%%%%%%%%%%%%%%%%%%%%%%%%%%%%%%%%%%%%%%%
\begin{equation}\label{TPconstraint}
\mbox{Tr}_{\rm out}[\chi]=\openone_{\rm in},
\end{equation}
%%%%%%%%%%%%%%%%%%%%%%%%%%%%%%%%%%%%%%%%%%%%%%%%%%%%%%%%%%%%%%%%%%%%%%%
where $\mbox{Tr}_{\rm out}$ is the trace over the output Hilbert space
and $\openone_{\rm in}$ is the identity operator on the input Hilbert space.

Let us denote for the state $\rho_{AB}^{\cal I}$ its minimal
purification $|\Psi\rangle_{ABE_{\rho}}$ with the CM
$\gamma_{\pi}^{\cal I}$. Let us further denote the measurements on
subsystems $A$, $B$ and $E$ that achieve the optimum in
Eq.~(\ref{Gaussianmu}) as $\Pi_{A}(d_{A})$, $\Pi_{B}(d_{B})$ and
$\Pi_{E_{\rho}}(d_{E_{\rho}})$ and the corresponding CMs as
$\Gamma_{A}^{\cal I}, \Gamma_{B}^{\cal I}$ and $\Gamma_{E}^{\cal
I}$, respectively. That is,
%%%%%%%%%%%%%%%%%%%%%%%%%%%%%%%%%%%%%%%%%%%%%%%%%%%%%%%%%%%%%%%%%%%%%
\begin{equation}\label{Eq10}
E_{\downarrow}^{G}\left(\rho_{AB}^{\cal
I}\right)=f\left(\gamma_{\pi}^{\cal I},\Gamma_{A}^{\cal
I},\Gamma_{B}^{\cal I},\Gamma_{E}^{\cal I}\right).
\end{equation}
%%%%%%%%%%%%%%%%%%%%%%%%%%%%%%%%%%%%%%%%%%%%%%%%%%%%%%%%%%%%%%%%%%%%%
Likewise, the purification of the state $\rho_{AB}^{\mathcal{E}}$
is denoted as $|\Psi^{\mathcal{E}}\rangle_{ABE}$ and it has the CM
$\gamma_{\pi}^{\mathcal{E}}$. The measurements on subsystems $A$,
$B$ and $E$, which achieve the optimum in Eq.~(\ref{Gaussianmu})
are denoted as $\Pi_{A}^{\mathcal{E}}(d_{A})$,
$\Pi_{B}^{\mathcal{E}}(d_{B})$ and $\Pi_{E}^{\mathcal{E}}(d_{E})$
and they have the CMs $\Gamma_{A}^{\mathcal E},
\Gamma_{B}^{\mathcal E}$ and $\Gamma_{E}^{\mathcal{E}}$,
respectively. That is,
%%%%%%%%%%%%%%%%%%%%%%%%%%%%%%%%%%%%%%%%%%%%%%%%%%%%%%%%%%%%%%%%%%%%%
\begin{equation}\label{Eq1M}
E_{\downarrow}^{G}\left(\rho_{AB}^{\mathcal{E}}\right)=f\left(\gamma_{\pi}^{\mathcal{E}},\Gamma_{A}^{\mathcal{E}},\Gamma_{B}^{\mathcal{E}},\Gamma_{E}^{\mathcal{E}}\right).
\end{equation}
%%%%%%%%%%%%%%%%%%%%%%%%%%%%%%%%%%%%%%%%%%%%%%%%%%%%%%%%%%%%%%%%%%%%%
To prove the inequality (\ref{monotonicity2}) we will now find a
suitable non-minimal purification of the state (\ref{rhoM}). The
purification can be constructed using the trick that any Gaussian
operation on a known state can be implemented via teleportation
\cite{Giedke_02,Fiurasek_02}. First, we prepare an $(N+M+N_{\rm
out}+M_{\rm out})$-mode state $\chi_{A_{\rm in}B_{\rm in}A_{\rm
out}B_{\rm out}}$ of $N$-mode subsystem $A_{\rm in}$, $M$-mode
subsystem $B_{\rm in}$, $N_{\rm out}$-mode subsystem $A_{\rm out}$
and $M_{\rm out}$-mode subsystem $B_{\rm out}$, which represents
the operation $\mathcal{E}$. Next, subsystems $A$ and $B$ of the
input state $\rho_{AB}^{\cal I}$ are teleported by a standard
continuous-variable teleportation \cite{Braunstein_98}, where the
state $\chi_{A_{\rm in}B_{\rm in}A_{\rm out}B_{\rm out}}$ serves
as a quantum channel. The sender performs Bell measurements on
pairs of subsystems $(A,A_{\rm in})$ and $(B,B_{\rm in})$ and
sends the outcomes of the measurements to the receiver who
appropriately displaces his subsystems $A_{\rm out}$ and $B_{\rm
out}$. As a result he obtains the output state $\rho_{A_{\rm
out}B_{\rm out}}^{\mathcal{E}}$ of the operation $\mathcal{E}$.
Let us now consider a pure state
%%%%%%%%%%%%%%%%%%%%%%%%%%%%%%%%%%%%%%%%%%%%%%%%%%%%%%%%%%%%%%%%%%%%%%%%%%%%%%%%
\begin{eqnarray}\label{Phi}
|\Phi\rangle=|\Psi\rangle_{ABE_{\rho}}|\chi\rangle_{A_{\rm
in}B_{\rm in}A_{\rm out}B_{\rm out}E_{\chi}}
\end{eqnarray}
%%%%%%%%%%%%%%%%%%%%%%%%%%%%%%%%%%%%%%%%%%%%%%%%%%%%%%%%%%%%%%%%%%%%%%%%%%%%%%%%
formed as a product of the minimal purification
$|\Psi\rangle_{ABE_{\rho}}$ (with CM $\gamma_{\pi}^{\cal I}$) of
the input state $\rho_{AB}^{\cal I}$ and a suitable purification
$|\chi\rangle_{A_{\rm in}B_{\rm in}A_{\rm out}B_{\rm
out}E_{\chi}}$ of the state $\chi_{A_{\rm in}B_{\rm in}A_{\rm
out}B_{\rm out}}$, which will be specified later. Now we perform
Bell measurements on the pairs of subsystems $(A,A_{\rm in})$ and
$(B,B_{\rm in})$. A Bell measurement on a pair of modes $(j,j_{\rm
in})$, $j=A_{1},\ldots, A_{N},B_{1},\ldots,B_{M}$, is formally
described by the set of rank-one operators $\{|\beta_{j}\rangle_{j
j_{\rm in}}\langle\beta_{j}|\}_{\beta_{j}\in\mathbb{C}}$, where
$\beta_{j}$ is the measurement outcome, $\mathbb{C}$ is the set of
complex numbers and \cite{Hofmann_00}
%%%%%%%%%%%%%%%%%%%%%%%%%%%%%%%%%%%%%%%%%%%%%%%%%%%%%%%%%%%%%%%%%%%%%%%%%%%%%%%%%%%
\begin{eqnarray}\label{beta}
|\beta_{j}\rangle_{j j_{\rm
in}}=\sum_{n=0}^{\infty}\mathcal{D}_{j}(\beta_{j})|n\rangle_{j}|n\rangle_{j_{\rm
in}},
\end{eqnarray}
%%%%%%%%%%%%%%%%%%%%%%%%%%%%%%%%%%%%%%%%%%%%%%%%%%%%%%%%%%%%%%%%%%%%%%%%%%%%%%%%%%
where $\mathcal{D}_{j}(\beta_{j})=\mbox{exp}(\beta_{j}
a_{j}^{\dag}-\beta_{j}^{\ast}a_{j})=D_{j}[\sqrt{2}(\mbox{Re}\beta_{j},\mbox{Im}\beta_{j})^{T}]$
is the displacement operator on mode $j$, $a_{j} (a_{j}^{\dag})$
is the annihilation (creation) operator of the mode, and
$|n\rangle$, $n=0,1,\ldots$, are the Fock states. If we now
perform the Bell measurements on pairs of modes $(A_{1},A_{1 \rm
in}),\ldots,(A_{N},A_{N \rm in})$ and $(B_{1},B_{1 \rm
in}),\ldots,(B_{M},B_{M \rm in})$ of the state (\ref{Phi})
followed by compensation of the displacements exactly as in the
implementation of a generic Gaussian operation by teleportation
\cite{Fiurasek_02} we obtain a pure state of the form
%%%%%%%%%%%%%%%%%%%%%%%%%%%%%%%%%%%%%%%%%%%%%%%%%%%%%%%%%%%%%%%%%%%%%%%%%%%%%%%%
\begin{widetext}
\begin{eqnarray}\label{PsiM}
|\Psi^{\mathcal{E}}\rangle_{A_{\rm out}B_{\rm
out}E_{\rho}E_{\chi}}=\frac{1}{\sqrt{p_{0}}}{}_{A A_{\rm
in}}\langle \tilde{\{0\}}| _{B B_{\rm in}}\langle \tilde{\{0\}}
|\Psi\rangle_{ABE_{\rho}}|\chi\rangle_{A_{\rm in}B_{\rm in}A_{\rm
out}B_{\rm out}E_{\chi}},
\end{eqnarray}
\end{widetext}
%%%%%%%%%%%%%%%%%%%%%%%%%%%%%%%%%%%%%%%%%%%%%%%%%%%%%%%%%%%%%%%%%%%%%%%%%%%%%%%%
where $\sqrt{p_{0}}$ is the normalization factor, and where we
have defined $|\tilde{\{0\}}\rangle_{jj_{\rm
in}}\equiv|\beta_{j1}=0\rangle_{j_{1}j_{1 \rm
in}}\ldots|\beta_{jJ_{j}}=0\rangle_{j_{J_{j}}j_{J_{j}\rm in}}$,
$j=A,B$, where $J_{A}=N$ and $J_{B}=M$ is the number of modes of
subsystem $A$ and $B$, respectively. The state (\ref{PsiM})
satisfies
$\mbox{Tr}_{E_{\rho}E_{\chi}}(|\Psi^{\mathcal{E}}\rangle_{A_{\rm
out}B_{\rm
out}E_{\rho}E_{\chi}}\langle\Psi^{\mathcal{E}}|)=\rho_{A_{\rm
out}B_{\rm out}}^{\mathcal{E}}$ and therefore it is the sought
suitable purification of the state $\rho_{AB}^{\mathcal{E}}$.
Consequently, the prescription
%%%%%%%%%%%%%%%%%%%%%%%%%%%%%%%%%%%%%%%%%%%%%%%%%%%%%%%%%%%%%%%%%%%%%%%%%%%%%%%%
\begin{widetext}
\begin{eqnarray}\label{PMoptimal}
P(d_{A},d_{B},d_{E_{\rho}},d_{E_{\chi}})=\mbox{Tr}\left[|\Psi^{\mathcal{E}}\rangle\langle\Psi^{\mathcal{E}}|\Pi_{A_{\rm
out}}^{\mathcal{E}}(d_{A})\otimes \Pi_{B_{\rm
out}}^{\mathcal{E}}(d_{B})\otimes\Pi_{E_{\rho}E_{\chi}}^{\mathcal{E}}(d_{E_{\rho}},d_{E_{\chi}})\right],
\end{eqnarray}
\end{widetext}
%%%%%%%%%%%%%%%%%%%%%%%%%%%%%%%%%%%%%%%%%%%%%%%%%%%%%%%%%%%%%%%%%%%%%%%%%%%%%%%%
defines the optimal distribution whose conditional mutual
information equals to
$E_{\downarrow}^{G}\left(\rho_{AB}^{\mathcal{E}}\right)$ where
$\Pi_{A_{\rm out}}^{\mathcal{E}}(d_{A})$, $\Pi_{B_{\rm
out}}^{\mathcal{E}}(d_{B})$ and
$\Pi_{E_{\rho}E_{\chi}}^{\mathcal{E}}(d_{E_{\rho}},d_{E_{\chi}})$
are optimal measurements with CMs $\Gamma_{A}^{\mathcal E},
\Gamma_{B}^{\mathcal E}$ and $\Gamma_{E}^{\mathcal{E}}$. Here, a
different symbol
$\Pi_{E_{\rho}E_{\chi}}^{\mathcal{E}}(d_{E_{\rho}},d_{E_{\chi}})$
for the optimal measurement $\Pi_{E}^{\mathcal{E}}(d_{E})$ has
been used to express the fact that it acts on two purifying
subsystems $E_{\rho}$ and $E_{\chi}$. Here and in what follows we
also omit the indices of the purification (\ref{PsiM}) for
brevity.

Now we will construct a suitable purification
$|\chi\rangle_{A_{\rm in}B_{\rm in}A_{\rm out}B_{\rm
out}E_{\chi}}$ of the state $\chi_{A_{\rm in}B_{\rm in}A_{\rm
out}B_{\rm out}}$ representing the Gaussian map $\mathcal{E}$. As
the map can be created by local operations and classical
communication, the corresponding Gaussian state $\chi_{A_{\rm
in}B_{\rm in}A_{\rm out}B_{\rm out}}$ is separable across $A_{\rm
in}A_{\rm out}|B_{\rm in}B_{\rm out}$ splitting \cite{Giedke_02}.
For the $2(N+N_{\rm out}+M+M_{\rm out})$-dimensional CM
$\gamma_{A_{\rm in}A_{\rm out}B_{\rm in}B_{\rm out}}^{\chi}$ of
the state there therefore exists local $2(N+N_{\rm
out})$-dimensional CM $\gamma_{A_{\rm in}A_{\rm out}}^{\chi}$ and
$2(M+M_{\rm out})$-dimensional CM $\gamma_{B_{\rm in}B_{\rm
out}}^{\chi}$ corresponding to generally mixed Gaussian states
$\chi_{A_{\rm in}A_{\rm out}}$ and $\chi_{B_{\rm in}B_{\rm out}}$
of the subsystems $A$ and $B$ such that \cite{Werner_01}
%%%%%%%%%%%%%%%%%%%%%%%%%%%%%%%%%%%%%%%%%%%%%%%%%%%%%%%%%%%%%%
\begin{eqnarray}\label{O}
O\equiv\gamma_{A_{\rm in}A_{\rm out}B_{\rm in}B_{\rm
out}}^{\chi}-\gamma_{A_{\rm in}A_{\rm
out}}^{\chi}\oplus\gamma_{B_{\rm in}B_{\rm out}}^{\chi}\geq0.
\end{eqnarray}
%%%%%%%%%%%%%%%%%%%%%%%%%%%%%%%%%%%%%%%%%%%%%%%%%%%%%%%%%%%%%%
Repeating the algorithm leading to Eq.~(\ref{rhoABsep}) for the
case of the state $\chi_{A_{\rm in}B_{\rm in}A_{\rm out}B_{\rm
out}}$ we then arrive at the following expression of the state
%%%%%%%%%%%%%%%%%%%%%%%%%%%%%%%%%%%%%%%%%%%%%%%%%%%%%%%%%%%%%%%%%%%%%
\begin{widetext}
\begin{eqnarray}\label{chiABsep}
\chi_{A_{\rm in}B_{\rm in}A_{\rm out}B_{\rm out}}&=&\int
q(\mathbf{r})D(\mathscr{W}\mathbf{r})\left(\chi_{A_{\rm in}A_{\rm
out}}\otimes\chi_{B_{\rm in}B_{\rm out}}\right)
D^{\dag}(\mathscr{W}\mathbf{r})d^{P'}\!\!{\bf r}.
\end{eqnarray}
\end{widetext}
%%%%%%%%%%%%%%%%%%%%%%%%%%%%%%%%%%%%%%%%%%%%%%%%%%%%%%%%%%%%%%%%%%%%
Here $d^{P'}\!\!{\bf r}\equiv\Pi_{l=1}^{P'}dr_{l}$,
$q(\mathbf{r})\equiv\Pi_{i=1}^{P'}\mbox{exp}\left(-r_{i}^2/o_{i}\right)/\sqrt{\pi
o_{i}}$ with $o_{i}$, $i=1,2,\ldots,P'$ being all strictly
positive eigenvalues of the matrix $O$, $\mathscr{W}$ is the
$2(N+N_{\rm out}+M+M_{\rm out})\times P'$ matrix composed of the
first $P'$ columns of the matrix $W$ which diagonalizes the matrix
$O$ as
$W^{T}OW=\mbox{diag}\left(o_1,o_2,\ldots,o_{P'},0,\ldots,0\right)$,
and $\mathbf{r}=(r_1,r_2,\ldots,r_{P'})^{T}$. Further, for CM
$\gamma_{j_{\rm in}j_{\rm out}}^{\chi}$, $j=A,B$ there always
exist pure-state CMs $\gamma_{j_{\rm in}j_{\rm out}}^{\chi,p}$
such that \cite{Giedke_01}
%%%%%%%%%%%%%%%%%%%%%%%%%%%%%%%%%%%%%%%%%%%%%%%%%%%%%%%%%%%%%%%%%%%%%
\begin{equation}\label{Tj}
T_{j}\equiv\gamma_{j_{\rm in}j_{\rm out}}^{\chi}-\gamma_{j_{\rm
in}j_{\rm out}}^{\chi,p}\geq0,\quad j=A,B.
\end{equation}
%%%%%%%%%%%%%%%%%%%%%%%%%%%%%%%%%%%%%%%%%%%%%%%%%%%%%%%%%%%%%%%%%%%%%
Denoting as $R_j$, $j=A,B$, the orthogonal matrix bringing the
matrix $T_{j}$ to the diagonal form, i.e.,
$R_{j}^{T}T_{j}R_{j}=\mbox{diag}\left(t_1^{j},t_2^{j},\ldots,t_{P_{j}}^{j},0,\ldots,0\right)$,
where $t_{l}^{j}$, $l=1,2,\ldots,P_{j}$ are the strictly positive
eigenvalues of the matrix $T_{j}$, we can further express the
local Gaussian states $\chi_{j_{\rm in}j_{\rm out}}$ as
%%%%%%%%%%%%%%%%%%%%%%%%%%%%%%%%%%%%%%%%%%%%%%%%%%%%%%%%%%%%%%%%%%%%%
\begin{widetext}
\begin{eqnarray}\label{chij}
\chi_{j_{\rm in}j_{\rm out}}&=&\int
q_{j}(\mathbf{r}_{j})D_{j}(\mathscr{R}_{j}\mathbf{r}_{j})|\gamma_{j_{\rm
in}j_{\rm out}}^{\chi,p}\rangle_{j_{\rm in}j_{\rm
out}}\langle\gamma_{j_{\rm in}j_{\rm out}}^{\chi,p}|
D^{\dag}_{j}(\mathscr{R}_{j}\mathbf{r}_{j})d^{P_j}{\bf r}_{j}.
\end{eqnarray}
\end{widetext}
%%%%%%%%%%%%%%%%%%%%%%%%%%%%%%%%%%%%%%%%%%%%%%%%%%%%%%%%%%%%%%%%%%%%
Here $d^{P_j}{\bf r}_{j}\equiv\Pi_{l=1}^{P_j}dr_{jl}$,
$q_{j}(\mathbf{r}_{j})\equiv\Pi_{i=1}^{P_j}\mbox{exp}\left(-r_{ji}^2/t_{i}^{j}\right)/\sqrt{\pi
t_{i}^{j}}$, $\mathscr{R}_{A}$ is the $2(N+N_{\rm out})\times
P_{A}$ matrix composed of the first $P_{A}$ columns of the matrix
$R_{A}$, $\mathscr{R}_{B}$ is the $2(M+M_{\rm out})\times P_{B}$
matrix composed of the first $P_{B}$ columns of the matrix
$R_{B}$, and
$\mathbf{r}_{j}=(r_{j1},r_{j2},\ldots,r_{jP_{j}})^{T}$. Inserting
now into Eq.~(\ref{chiABsep}) for the states $\chi_{A_{\rm
in}A_{\rm out}}$ and $\chi_{B_{\rm in}B_{\rm out}}$ from
Eq.~(\ref{chij}) we get
%%%%%%%%%%%%%%%%%%%%%%%%%%%%%%%%%%%%%%%%%%%%%%%%%%%%%%%%%%%%%%%%%%%%%
\begin{widetext}
\begin{eqnarray}\label{chiABsep2}
\chi_{A_{\rm in}B_{\rm in}A_{\rm out}B_{\rm out}}&=&\int
q(\mathbf{r})D(\mathscr{W}\mathbf{r})\left[\bigotimes_{j=A,B}
q_{j}(\mathbf{r}_{j})D_{j}(\mathscr{R}_{j}\mathbf{r}_{j})|\gamma_{j_{\rm
in}j_{\rm out}}^{\chi,p}\rangle_{j_{\rm in}j_{\rm
out}}\langle\gamma_{j_{\rm in}j_{\rm
out}}^{\chi,p}|D^{\dag}_{j}(\mathscr{R}_{j}\mathbf{r}_{j})\right]
D^{\dag}(\mathscr{W}\mathbf{r})d^{P_A}{\bf r}_{A}d^{P_B}{\bf
r}_{B}d^{P'}\!\!{\bf
r}.\nonumber\\
\end{eqnarray}
\end{widetext}
%%%%%%%%%%%%%%%%%%%%%%%%%%%%%%%%%%%%%%%%%%%%%%%%%%%%%%%%%%%%%%%%%%%%
By encoding the vectors of displacements ${\bf r},{\bf r}_{A}$ and
${\bf r}_{B}$ into eigenvectors $|{\bf r}\rangle_{E_{O}},|{\bf
r}_{A}\rangle_{E_{A}}$ and $|{\bf r}_{B}\rangle_{E_{B}}$ of
position quadratures of Eve's $(P'+P_A+P_B)$-mode subsystem
$E_{\chi}\equiv(E_{O}E_{A}E_{B})$ we obtain finally the sought
purification
%%%%%%%%%%%%%%%%%%%%%%%%%%%%%%%%%%%%%%%%%%%%%%%%%%%%%%%%%%%%%%%%%%%%%
\begin{widetext}
\begin{eqnarray}\label{ketchi}
|\chi\rangle&=&\int\sqrt{q(\mathbf{r})q_{A}(\mathbf{r}_{A})q_{B}(\mathbf{r}_{B})}D(\mathscr{W}\mathbf{r})\bigotimes_{j=A,B}
D_{j}(\mathscr{R}_{j}\mathbf{r}_{j})|\gamma_{j_{\rm in}j_{\rm
out}}^{\chi,p}\rangle_{j_{\rm in}j_{\rm out}}|{\bf
r}\rangle_{E_{O}}|{\bf r}_{A}\rangle_{E_{A}}|{\bf
r}_{B}\rangle_{E_{B}}d^{P_A}{\bf r}_{A}d^{P_B}{\bf
r}_{B}d^{P'}\!\!{\bf r},
\end{eqnarray}
\end{widetext}
%%%%%%%%%%%%%%%%%%%%%%%%%%%%%%%%%%%%%%%%%%%%%%%%%%%%%%%%%%%%%%%%%%%%
where we have omitted the indices $A_{\rm in}B_{\rm in}A_{\rm
out}B_{\rm out}E_{\chi}$ of the purification $|\chi\rangle$ for
brevity.

A specific feature of the state (\ref{ketchi}) is that by a simple
measurement on the purifying subsystem $E_{\chi}$ we can project
the state onto a displaced product state $\chi_{A_{\rm in}A_{\rm
out}}\otimes\chi_{B_{\rm in}B_{\rm out}}$ of the subsystems
$(A_{\rm in},A_{\rm out})$ and $(B_{\rm in},B_{\rm out})$. More
precisely, consider the following measurement on Eve's subsystem
$E_{\chi}$:
%%%%%%%%%%%%%%%%%%%%%%%%%%%%%%%%%%%%%%%%%%%%%%%%%%%%%%%%%%%%%%%%%%%%
\begin{equation}\label{PiEchi}
\tilde{\Pi}_{E_{\chi}}^{\cal{E}}({\bf r}')=|{\bf
r}'\rangle_{E_{O}}\langle{\bf
r}'|\otimes\openone_{E_{A}}\otimes\openone_{E_{B}},
\end{equation}
%%%%%%%%%%%%%%%%%%%%%%%%%%%%%%%%%%%%%%%%%%%%%%%%%%%%%%%%%%%%%%%%%%%%
%%%%%%%%%%%%%%%%%%%%%%%%%%%%%%%%%%%%%%%%%%%%%%%%%%%%%%%%%%%%%%%%%%%%
%\begin{equation}\label{PiEchi}
%\tilde{\Pi}_{E_{\chi}}^{\cal{E}}(d_{E_{\chi}})=|{\bf
%r}'\rangle_{E_{O}}\langle{\bf
%r}'|\otimes\openone_{E_{A}}\otimes\openone_{E_{B}},
%\end{equation}
%%%%%%%%%%%%%%%%%%%%%%%%%%%%%%%%%%%%%%%%%%%%%%%%%%%%%%%%%%%%%%%%%%%%
which describes the projection of subsystem $E_{O}$ onto a
$P'$-mode position eigenvector $|{\bf r}'\rangle_{E_{O}}$ and
projection of subsystems $E_{A}$ and $E_{B}$ onto maximally mixed
states, which gives Eve no information on the state of the two
subsystems. Recall, that the latter measurements on subsystems
$E_{A}$ and $E_{B}$ can be seen as Gaussian measurements
(\ref{POVMn}) with seed elements given by thermal states in the
limit of infinite temperature. By performing the measurement
(\ref{PiEchi}) on the subsystem $E_{\chi}$ of the purification
(\ref{ketchi}) we then arrive using Eq.~(\ref{chij}) at the
conditional state of the form:
%%%%%%%%%%%%%%%%%%%%%%%%%%%%%%%%%%%%%%%%%%%%%%%%%%%%%%%%%%%%%%%%%%%%%%%%%%%%%%%%%%%%%%%%%%%%%%%%%%%%%%%%%%%%
%\begin{widetext}
\begin{equation}\label{chis}
\mbox{Tr}_{E_{\chi}}\left[|\chi\rangle\langle
\chi|\tilde{\Pi}_{E_{\chi}}^{\mathcal{E}}({\bf r}')\right]=q({\bf
r}')\!\!\!\bigotimes_{j=A,B}\chi_{j_{\rm in}j_{\rm
out}}[(\mathscr{W}\mathbf{r}')_{j_{\rm in}j_{\rm out}}],
\end{equation}
%\end{widetext}
%%%%%%%%%%%%%%%%%%%%%%%%%%%%%%%%%%%%%%%%%%%%%%%%%%%%%%%%%%%%%%%%%%%%%%%%%%%%%%%%%%%%%%%%%%%%%%%%%%%%%%%%%%%%
%%%%%%%%%%%%%%%%%%%%%%%%%%%%%%%%%%%%%%%%%%%%%%%%%%%%%%%%%%%%%%%%%%%%%%%%%%%%%%%%%%%%%%%%%%%%%%%%%%%%%%%%%%%%
%\begin{widetext}
%\begin{equation}\label{chis}
%\mbox{Tr}_{E_{\chi}}\left[|\chi\rangle\langle
%\chi|\tilde{\Pi}_{E_{\chi}}^{\mathcal{E}}(d_{E_{\chi}})\right]=q({\bf
%r}')\!\!\!\bigotimes_{j=A,B}\chi_{j_{\rm in}j_{\rm
%out}}[(\mathscr{W}\mathbf{r}')_{j_{\rm in}j_{\rm out}}],
%\end{equation}
%\end{widetext}
%%%%%%%%%%%%%%%%%%%%%%%%%%%%%%%%%%%%%%%%%%%%%%%%%%%%%%%%%%%%%%%%%%%%%%%%%%%%%%%%%%%%%%%%%%%%%%%%%%%%%%%%%%%%
which is the desired product state with respect to the $A_{\rm
in}A_{\rm out}|B_{\rm in}B_{\rm out}$ splitting. Here
%%%%%%%%%%%%%%%%%%%%%%%%%%%%%%%%%%%%%%%%%%%%%%%%%%%%%%%%%%%%%%%%%%%%%%%%%%%%%%%%%%%%%%%%%%%%%%%%%%%%%%%%%%%%
\begin{widetext}
\begin{equation}\label{chisj}
\chi_{j_{\rm in}j_{\rm out}}[(\mathscr{W}\mathbf{r}')_{j_{\rm
in}j_{\rm out}}]\equiv D_{{j_{\rm in}j_{\rm
out}}}[(\mathscr{W}\mathbf{r}')_{j_{\rm in}j_{\rm
out}}]\chi_{j_{\rm in}j_{\rm out}}D^{\dag}_{{j_{\rm in}j_{\rm
out}}}[(\mathscr{W}\mathbf{r}')_{j_{\rm in}j_{\rm out}}],
\end{equation}
\end{widetext}
%%%%%%%%%%%%%%%%%%%%%%%%%%%%%%%%%%%%%%%%%%%%%%%%%%%%%%%%%%%%%%%%%%%%%%%%%%%%%%%%%%%%%%%%%%%%%%%%%%%%%%%%%%%%
where
%%%%%%%%%%%%%%%%%%%%%%%%%%%%%%%%%%%%%%%%%%%%%%%%%%%%%%%%%%%%%%%%%%%%%%%%%%%%%%%%%%%%%%%%%%%%%%%%%%%%%%%%%%%%
\begin{widetext}
\begin{eqnarray}
(\mathscr{W}\mathbf{r}')_{A_{\rm in}A_{\rm
out}}&\equiv&((\mathscr{W}\mathbf{r}')_{A_{\rm
in}1},(\mathscr{W}\mathbf{r}')_{A_{\rm
in}2},\ldots,(\mathscr{W}\mathbf{r}')_{A_{\rm
in}2N},(\mathscr{W}\mathbf{r}')_{A_{\rm
out}1},(\mathscr{W}\mathbf{r}')_{A_{\rm
out}2},\ldots,(\mathscr{W}\mathbf{r}')_{A_{\rm out}2N_{\rm
out}})^{T},\nonumber\\
(\mathscr{W}\mathbf{r}')_{B_{\rm in}B_{\rm
out}}&\equiv&((\mathscr{W}\mathbf{r}')_{B_{\rm
in}1},(\mathscr{W}\mathbf{r}')_{B_{\rm
in}2},\ldots,(\mathscr{W}\mathbf{r}')_{B_{\rm
in}2M},(\mathscr{W}\mathbf{r}')_{B_{\rm
out}1},(\mathscr{W}\mathbf{r}')_{B_{\rm
out}2},\ldots,(\mathscr{W}\mathbf{r}')_{B_{\rm out}2M_{\rm
out}})^{T}.
\end{eqnarray}
\end{widetext}
%%%%%%%%%%%%%%%%%%%%%%%%%%%%%%%%%%%%%%%%%%%%%%%%%%%%%%%%%%%%%%%%%%%%%%%%%%%%%%%%%%%%%%%%%%%%%%%%%%%%%%%%%%%%
%\begin{widetext}
%\begin{equation}\label{chis}
%\mbox{Tr}_{E_{\chi}}\left[|\chi\rangle\langle
%\chi|\tilde{\Pi}_{E_{\chi}}^{\mathcal{E}}(d_{E_{\chi}})\right]=q({\bf
%r}')D(\mathscr{W}\mathbf{r}')\left(\chi_{A_{\rm in}A_{\rm
%out}}\otimes\chi_{B_{\rm in}B_{\rm
%out}}\right)D^{\dag}(\mathscr{W}\mathbf{r}'),
%\end{equation}
%\end{widetext}
%%%%%%%%%%%%%%%%%%%%%%%%%%%%%%%%%%%%%%%%%%%%%%%%%%%%%%%%%%%%%%%%%%%%%%%%%%%%%%%%%%%%%%%%%%%%%%%%%%%%%%%%%%%%

Before going further let us note, that any trace-preserving
Gaussian operation $\mathcal{E}$ is represented by an unphysical
(infinitely squeezed) density matrix $\chi$. This is because the
matrix is obtained by an action of the operation $\mathcal{E}$ on
one part of an unphysical maximally entangled state
$|\tilde{\Phi}\rangle$ \cite{Jamiolkowski_72}. The unphysical
states can nevertheless be dealt with rigorously in the context of
positive forms \cite{Holevo_11} or by using the limiting procedure
proposed in Ref.~\cite{Giedke_02}. The latter approach consists of
the replacement of the state $|\tilde{\Phi}\rangle$ by its
physical approximation $|\tilde{\Phi}(r)\rangle$ given by a tensor
product of identical two-mode squeezed vacuum states with
squeezing parameter $r$. The operation $\mathcal{E}$ is then
represented by a quantum state $\chi(r)$ obtained by action of the
operation on one part of the state $|\tilde{\Phi}(r)\rangle$,
which is a physical approximation of the exact state $\chi$. For a
quantum operation $\mathcal{E}$, which can be prepared by local
operations and classical communication, the density matrix
$\chi(r)$ is separable and hence the above formulas remain valid
also for quantum state $\chi(r)$. The sought exact result is
recovered and the limiting procedure is thus accomplished by
taking the limit $r\rightarrow\infty$ at the end of our
calculations.

Returning to the monotonicity proof consider now the
probability density
%%%%%%%%%%%%%%%%%%%%%%%%%%%%%%%%%%%%%%%%%%%%%%%%%%%%%%%%%%%%%%%%%%%%%%%%%%%%%%%%
\begin{widetext}
\begin{eqnarray}\label{tildeP}
\tilde{P}(d_{A},d_{B},d_{E_{\rho}},{\bf
r}')=\mbox{Tr}\left[|\Psi^{\mathcal{E}}\rangle\langle\Psi^{\mathcal{E}}|\Pi_{A_{\rm
out}}^{\mathcal{E}}(d_{A})\otimes \Pi_{B_{\rm
out}}^{\mathcal{E}}(d_{B})\otimes\Pi_{E_{\rho}}(d_{E_{\rho}})\otimes\tilde{\Pi}_{E_{\chi}}^{\mathcal{E}}({\bf
r}')\right],
\end{eqnarray}
\end{widetext}
%%%%%%%%%%%%%%%%%%%%%%%%%%%%%%%%%%%%%%%%%%%%%%%%%%%%%%%%%%%%%%%%%%%%%%%%%%%%%%%%
%%%%%%%%%%%%%%%%%%%%%%%%%%%%%%%%%%%%%%%%%%%%%%%%%%%%%%%%%%%%%%%%%%%%%%%%%%%%%%%%
%\begin{widetext}
%\begin{eqnarray}\label{tildeP}
%\tilde{P}(d_{A},d_{B},d_{E_{\rho}},d_{E_{\chi}})=\mbox{Tr}\left[|\Psi^{\mathcal{E}}\rangle\langle\Psi^{\mathcal{E}}|\Pi_{A_{\rm
%out}}^{\mathcal{E}}(d_{A})\otimes \Pi_{B_{\rm
%out}}^{\mathcal{E}}(d_{B})\otimes\Pi_{E_{\rho}}(d_{E_{\rho}})\otimes\tilde{\Pi}_{E_{\chi}}^{\mathcal{E}}(d_{E_{\chi}})\right],
%\end{eqnarray}
%\end{widetext}
%%%%%%%%%%%%%%%%%%%%%%%%%%%%%%%%%%%%%%%%%%%%%%%%%%%%%%%%%%%%%%%%%%%%%%%%%%%%%%%%
which is obtained from the probability density (\ref{PMoptimal})
by replacing the optimal measurement
$\Pi_{E_{\rho}E_{\chi}}^{\mathcal{E}}(d_{E_{\rho}},d_{E_{\chi}})$
with a product measurement
$\Pi_{E_{\rho}}(d_{E_{\rho}})\otimes\tilde{\Pi}_{E_{\chi}}^{\mathcal{E}}({\bf
r}')$. Here, $\Pi_{E_{\rho}}(d_{E_{\rho}})$ is the optimal
measurement with the CM $\Gamma_{E}^{\cal I}$ on the minimal
purification $|\Psi\rangle_{ABE_{\rho}}$ of the state
$\rho_{AB}^{\cal I}$ and
$\tilde{\Pi}_{E_{\chi}}^{\mathcal{E}}({\bf r}')$ is the
measurement (\ref{PiEchi}) with the CM
$\tilde{\Gamma}_{E_{\chi}}^{\mathcal{E}}$, which projects the
purification (\ref{ketchi}) onto the product state (\ref{chis}).
At given CMs $\gamma_{\pi}^{\mathcal{E}},\Gamma_{A}^{\mathcal{E}}$
and $\Gamma_{B}^{\mathcal{E}}$ the product measurement with the CM
$\Gamma_{E}^{\cal I}\oplus\tilde{\Gamma}_{E_{\chi}}^{\mathcal{E}}$
does not generally minimize the function
$f\left(\gamma_{\pi}^{\mathcal{E}},\Gamma_{A}^{\mathcal{E}},\Gamma_{B}^{\mathcal{E}},\Gamma_{E}\right)$
with respect to the CM $\Gamma_{E}$ and hence the function $f$
corresponding to the distribution
$\tilde{P}(d_{A},d_{B},d_{E_{\rho}},{\bf r}')$,
Eq.~(\ref{tildeP}), satisfies
%%%%%%%%%%%%%%%%%%%%%%%%%%%%%%%%%%%%%%%%%%%%%%%%%%%%%%%%%%%%%%%%%%%%%
\begin{equation}\label{Ineq1}
E_{\downarrow}^{G}\left(\rho_{AB}^{\mathcal{E}}\right)\leq
f\left(\gamma_{\pi}^{\mathcal{E}},\Gamma_{A}^{\mathcal{E}},\Gamma_{B}^{\mathcal{E}},\Gamma_{E}^{\cal
I}\oplus\tilde{\Gamma}_{E_{\chi}}^{\mathcal{E}}\right).
\end{equation}
%%%%%%%%%%%%%%%%%%%%%%%%%%%%%%%%%%%%%%%%%%%%%%%%%%%%%%%%%%%%%%%%%%%%%
What is more, one can show that there exist Gaussian measurements
$\tilde{\Pi}_{A}(\tilde{d}_{A})$ and
$\tilde{\Pi}_{B}(\tilde{d}_{B})$ on the subsystems $A$ and $B$ of
the normalized conditional state
%%%%%%%%%%%%%%%%%%%%%%%%%%%%%%%%%%%%%%%%%%%%%%%%%%%%%%%%%%%%%%%%%%%%%
\begin{equation}\label{rhoABErho}
\rho_{AB|E_{\rho}}(d_{E_{\rho}})=\frac{\mbox{Tr}_{E_{\rho}}\left[|\Psi\rangle_{ABE_{\rho}}\langle\Psi|\Pi_{E_{\rho}}(d_{E_{\rho}})\right]}{P(d_{E_{\rho}})}
\end{equation}
%%%%%%%%%%%%%%%%%%%%%%%%%%%%%%%%%%%%%%%%%%%%%%%%%%%%%%%%%%%%%%%%%%%%
obtained by the optimal measurement $\Pi_{E_{\rho}}(d_{E_{\rho}})$
on subsystem $E_{\rho}$ of the minimal purification
$|\Psi\rangle_{ABE_{\rho}}$, which are characterized by the CMs
$\tilde{\Gamma}_{A}$ and $\tilde{\Gamma}_{B}$, such that the
conditional distribution
%%%%%%%%%%%%%%%%%%%%%%%%%%%%%%%%%%%%%%%%%%%%%%%%%%%%%%%%%%%%%%%%%%%%%%%%%%%%%%%%
%\begin{widetext}
\begin{eqnarray}\label{Psuboptimal}
\tilde{p}(\tilde{d}_{A},\tilde{d}_{B}|d_{E_{\rho}})=\mbox{Tr}\left[\rho_{AB|E_{\rho}}(d_{E_{\rho}})\tilde{\Pi}_{A}(\tilde{d}_{A})\otimes
\tilde{\Pi}_{B}(\tilde{d}_{B})\right]\nonumber\\
\end{eqnarray}
%\end{widetext}
%%%%%%%%%%%%%%%%%%%%%%%%%%%%%%%%%%%%%%%%%%%%%%%%%%%%%%%%%%%%%%%%%%%%%%%%%%%%%%%%
yields the function $f$ defined in Eq.~(\ref{f}), which is greater
or equal than the function on the RHS of Ineq.~(\ref{Ineq1}),
i.e.,
%%%%%%%%%%%%%%%%%%%%%%%%%%%%%%%%%%%%%%%%%%%%%%%%%%%%%%%%%%%%%%%%%%%%%
%\begin{widetext}
\begin{equation}\label{Ineq2}
f\left(\gamma_{\pi}^{\mathcal{E}},\Gamma_{A}^{\mathcal{E}},\Gamma_{B}^{\mathcal{E}},\Gamma_{E}^{\cal
I}\oplus\tilde{\Gamma}_{E_{\chi}}^{\mathcal{E}}\right)\leq
f\left(\gamma_{\pi}^{\cal
I},\tilde{\Gamma}_{A},\tilde{\Gamma}_{B},\Gamma_{E}^{\cal
I}\right).
\end{equation}
%\end{widetext}
%%%%%%%%%%%%%%%%%%%%%%%%%%%%%%%%%%%%%%%%%%%%%%%%%%%%%%%%%%%%%%%%%%%%%
This can be shown as follows. The function on the RHS of
Ineq.~(\ref{Ineq1}) is the mutual information of the conditional
Gaussian distribution $\tilde{P}(d_{A},d_{B}|d_{E_{\rho}},{\bf
r}')=\tilde{P}(d_{A},d_{B},d_{E_{\rho}},{\bf
r}')/\tilde{P}(d_{E_{\rho}},{\bf r}')$, where the distribution
$\tilde{P}(d_{A},d_{B},d_{E_{\rho}},{\bf r}')$ is given in
Eq.~(\ref{tildeP}). The conditional distribution is the
distribution of outcomes of Gaussian measurements with CMs
$\Gamma_{A}^{\mathcal{E}}$ and $\Gamma_{B}^{\mathcal{E}}$ on
subsystems $A$ and $B$ of the conditional state
($\equiv\rho_{AB|E}^{\mathcal{E}}$) obtained by Gaussian
measurement
$\Pi_{E_{\rho}}(d_{E_{\rho}})\otimes\tilde{\Pi}_{E_{\chi}}^{\mathcal{E}}({\bf
r}')$ (with CM $\Gamma_{E}^{\cal
I}\oplus\tilde{\Gamma}_{E_{\chi}}^{\mathcal{E}}$) on the
purification (\ref{PsiM}), where the state $|\chi\rangle$ is given
in Eq.~(\ref{ketchi}). Substituting from Eqs.~(\ref{PsiM}),
(\ref{ketchi}) and (\ref{chis}) into the explicit expression for
the (unnormalized) conditional state
%%%%%%%%%%%%%%%%%%%%%%%%%%%%%%%%%%%%%%%%%%%%%%%%%%%%%%%%%%%%%%%%%%%%
%\begin{widetext}
\begin{equation}\label{rhoABcond}
\tilde{\rho}_{A_{\rm out}B_{\rm
out}|E}^{\mathcal{E}}=\mbox{Tr}_{E_{\rho}E_{\chi}}\left[|\Psi^{\mathcal{E}}\rangle\langle\Psi^{\mathcal{E}}|
\Pi_{E_{\rho}}(d_{E_{\rho}})\otimes\tilde{\Pi}_{E_{\chi}}^{\mathcal{E}}({\bf
r}')\right],
\end{equation}
%\end{widetext}
%%%%%%%%%%%%%%%%%%%%%%%%%%%%%%%%%%%%%%%%%%%%%%%%%%%%%%%%%%%%%%%%%%%%
one arrives after some algebra at the conditional state in the
form:
%%%%%%%%%%%%%%%%%%%%%%%%%%%%%%%%%%%%%%%%%%%%%%%%%%%%%%%%%%%%%%%%%%%%%
\begin{widetext}
\begin{equation}\label{tilderhocond}
\tilde{\rho}_{A_{\rm out}B_{\rm
out}|E}^{\mathcal{E}}=\frac{P(d_{E_{\rho}})q({\bf
r}')}{p_{0}}\mbox{Tr}_{AA_{\rm in}BB_{\rm
in}}\left\{\rho_{AB|E_{\rho}}(d_{E_{\rho}})\bigotimes_{j=A,B}\chi_{j_{\rm
in}j_{\rm out}}[(\mathscr{W}\mathbf{r}')_{j_{\rm in}j_{\rm
out}}]\bigotimes_{k=A,B}|\tilde{\{0\}}\rangle_{k k_{\rm
in}}\langle\tilde{\{0\}}|\right\},
\end{equation}
\end{widetext}
%%%%%%%%%%%%%%%%%%%%%%%%%%%%%%%%%%%%%%%%%%%%%%%%%%%%%%%%%%%%%%%%%%%%%
where the state $\rho_{AB|E_{\rho}}(d_{E_{\rho}})$ is defined in
Eq.~(\ref{rhoABErho}). Here and in what follows we do not write
explicitly in some places the dependence of the conditional states
on the measurement outcomes for brevity.

Expressing the operator $|\tilde{\{0\}}\rangle_{jj_{\rm
in}}\langle\tilde{\{0\}}|$ on the RHS of the latter equation using
Eq.~(\ref{beta}) and carrying out the trace over the subsystems
$A$ and $B$ we further get
%%%%%%%%%%%%%%%%%%%%%%%%%%%%%%%%%%%%%%%%%%%%%%%%%%%%%%%%%%%%%%%%%%%%%
\begin{widetext}
\begin{equation}\label{tilderhocond2}
\tilde{\rho}_{A_{\rm out}B_{\rm
out}|E}^{\mathcal{E}}=\frac{P(d_{E_{\rho}})q({\bf
r}')}{p_{0}}\mbox{Tr}_{A_{\rm in}B_{\rm
in}}\left\{\bigotimes_{j=A,B}\chi_{j_{\rm in}j_{\rm
out}}[(\mathscr{W}\mathbf{r}')_{j_{\rm in}j_{\rm
out}}]\rho_{A_{\rm in}B_{\rm
in}|E_{\rho}}^{T}(d_{E_{\rho}})\otimes\openone_{A_{\rm out}B_{\rm
out}}\right\}.
\end{equation}
\end{widetext}
%%%%%%%%%%%%%%%%%%%%%%%%%%%%%%%%%%%%%%%%%%%%%%%%%%%%%%%%%%%%%%%%%%%%%

Let us assume now that the considered separable operation
$\mathcal{E}$ is GLTPOCC, i.e., it can be decomposed into Gaussian
local trace-preserving operations on subsystems $A$ and $B$, and the
addition of classical Gaussian noise. The density matrices
$\chi_{j_{\rm in}j_{\rm out}}$, $j=A,B$ representing the local
operations then satisfy the trace-preservation constraints
(\ref{TPconstraint}), i.e.,
%%%%%%%%%%%%%%%%%%%%%%%%%%%%%%%%%%%%%%%%%%%%%%%%%%%%%%%%%%%%%%%%%%%%%%
\begin{equation}\label{TPchijconstraint}
\mbox{Tr}_{j_{\rm out}}[\chi_{j_{\rm in}j_{\rm
out}}]=\openone_{j_{\rm in}},\quad j=A,B,
\end{equation}
%%%%%%%%%%%%%%%%%%%%%%%%%%%%%%%%%%%%%%%%%%%%%%%%%%%%%%%%%%%%%%%%%%%%%%%
which imply fulfilment of the trace-preservation constraints for
the states (\ref{chisj})
%%%%%%%%%%%%%%%%%%%%%%%%%%%%%%%%%%%%%%%%%%%%%%%%%%%%%%%%%%%%%%%%%%%%%%
\begin{equation}\label{TPchijconstraint}
\mbox{Tr}_{j_{\rm out}}\left\{\chi_{j_{\rm in}j_{\rm
out}}[(\mathscr{W}\mathbf{r}')_{j_{\rm in}j_{\rm
out}}]\right\}=\openone_{j_{\rm in}},\quad j=A,B.
\end{equation}
%%%%%%%%%%%%%%%%%%%%%%%%%%%%%%%%%%%%%%%%%%%%%%%%%%%%%%%%%%%%%%%%%%%%%
As a consequence, one finds the trace of the conditional state
(\ref{tilderhocond2}) to be
%%%%%%%%%%%%%%%%%%%%%%%%%%%%%%%%%%%%%%%%%%%%%%%%%%%%%%%%%%%%%%%%%%%%%%%%
\begin{equation}\label{Tracerhocond}
\mbox{Tr}_{A_{\rm out}B_{\rm out}}\left[\tilde{\rho}_{A_{\rm
out}B_{\rm
out}|E}^{\mathcal{E}}\right]=\tilde{P}(d_{E_{\rho}},{\bf
r}')=\frac{P(d_{E_{\rho}})q({\bf r}')}{p_{0}},
\end{equation}
%%%%%%%%%%%%%%%%%%%%%%%%%%%%%%%%%%%%%%%%%%%%%%%%%%%%%%%%%%%%%%%%%%%%%%%%%
and therefore the normalized conditional state reads as
%%%%%%%%%%%%%%%%%%%%%%%%%%%%%%%%%%%%%%%%%%%%%%%%%%%%%%%%%%%%%%%%%%%%%
\begin{widetext}
\begin{equation}\label{tilderhocondnorm}
\rho_{A_{\rm out}B_{\rm out}|E}^{\mathcal{E}}=\mbox{Tr}_{A_{\rm
in}B_{\rm in}}\left\{\bigotimes_{j=A,B}\chi_{j_{\rm in}j_{\rm
out}}[(\mathscr{W}\mathbf{r}')_{j_{\rm in}j_{\rm
out}}]\rho_{A_{\rm in}B_{\rm
in}|E_{\rho}}^{T}(d_{E_{\rho}})\otimes\openone_{A_{\rm out}B_{\rm
out}}\right\}.
\end{equation}
\end{widetext}
%%%%%%%%%%%%%%%%%%%%%%%%%%%%%%%%%%%%%%%%%%%%%%%%%%%%%%%%%%%%%%%%%%%%%
If we further substitute here for the operators $\chi_{j_{\rm
in}j_{\rm out}}[(\mathscr{W}\mathbf{r}')_{j_{\rm in}j_{\rm out}}]$
from Eq.~(\ref{chisj}) and we use the relation
$D^{T}(d)=D(-\Lambda d)$, where $T$ stands for the transposition
in Fock basis and the diagonal matrix
$\Lambda\equiv\mbox{diag}(1,-1,1,-1,\ldots,1,-1)$ realizes the
transposition operation on the CM level, we get the conditional
state (\ref{tilderhocondnorm}) in the form
%%%%%%%%%%%%%%%%%%%%%%%%%%%%%%%%%%%%%%%%%%%%%%%%%%%%%%%%%%%%%%%%%%%%%
\begin{widetext}
\begin{equation}\label{tilderhocondnorm2}
\rho_{A_{\rm out}B_{\rm out}|E}^{\mathcal{E}}=D_{{A_{\rm
out}B_{\rm out}}}[(\mathscr{W}\mathbf{r}')_{A_{\rm out}B_{\rm
out}}]\left(\mathcal{E}_{A}\otimes\mathcal{E}_{B}\right)(\rho_{A_{\rm
in}B_{\rm in}|E_{\rho}}')D_{{A_{\rm out}B_{\rm
out}}}^{\dag}[(\mathscr{W}\mathbf{r}')_{A_{\rm out}B_{\rm out}}].
\end{equation}
\end{widetext}
%%%%%%%%%%%%%%%%%%%%%%%%%%%%%%%%%%%%%%%%%%%%%%%%%%%%%%%%%%%%%%%%%%%%%
Here
%%%%%%%%%%%%%%%%%%%%%%%%%%%%%%%%%%%%%%%%%%%%%%%%%%%%%%%%%%%%%%%%%%%%%
\begin{widetext}
\begin{equation}\label{rhocondin}
\rho_{A_{\rm in}B_{\rm in}|E_{\rho}}'\equiv D_{{A_{\rm in}B_{\rm
in}}}[-\Lambda(\mathscr{W}\mathbf{r}')_{A_{\rm in}B_{\rm
in}}]\rho_{A_{\rm in}B_{\rm in}|E_{\rho}}(d_{E_{\rho}})D_{{A_{\rm
in}B_{\rm in}}}^{\dag}[-\Lambda(\mathscr{W}\mathbf{r}')_{A_{\rm
in}B_{\rm in}}],
\end{equation}
\end{widetext}
%%%%%%%%%%%%%%%%%%%%%%%%%%%%%%%%%%%%%%%%%%%%%%%%%%%%%%%%%%%%%%%%%%%%%
and $\mathcal{E}_{j}$, $j=A,B$ is the local Gaussian
trace-preserving operation represented by the density matrix
$\chi_{j_{\rm in}j_{\rm out}}$, i.e.,
%%%%%%%%%%%%%%%%%%%%%%%%%%%%%%%%%%%%%%%%%%%%%%%%%%%%%%%%%%%%%%%%%%%%%
\begin{widetext}
\begin{equation}\label{EAEB}
\left(\mathcal{E}_{A}\otimes\mathcal{E}_{B}\right)(\rho_{A_{\rm
in}B_{\rm in}|E_{\rho}}')=\mbox{Tr}_{A_{\rm in}B_{\rm
in}}\left\{\chi_{A_{\rm in}A_{\rm out}}\otimes\chi_{B_{\rm
in}B_{\rm out}}(\rho_{A_{\rm in}B_{\rm in}|E_{\rho}}')^{T_{\rm
in}}\otimes\openone_{A_{\rm out}B_{\rm out}}\right\}.
\end{equation}
\end{widetext}
%%%%%%%%%%%%%%%%%%%%%%%%%%%%%%%%%%%%%%%%%%%%%%%%%%%%%%%%%%%%%%%%%%%%%
We have already said that the RHS of Ineq.~(\ref{Ineq1}) is the
mutual information of the conditional distribution
%%%%%%%%%%%%%%%%%%%%%%%%%%%%%%%%%%%%%%%%%%%%%%%%%%%%%%%%%%%%%%%%%%%%%
\begin{widetext}
\begin{equation}\label{tildePcond1}
\tilde{P}(d_{A},d_{B}|d_{E_{\rho}},{\bf r}')=\mbox{Tr}_{A_{\rm
out}B_{\rm out}}\left[\rho_{A_{\rm out}B_{\rm
out}|E}^{\mathcal{E}}\Pi_{A_{\rm
out}}^{\mathcal{E}}(d_{A})\otimes\Pi_{B_{\rm
out}}^{\mathcal{E}}(d_{B})\right]
\end{equation}
\end{widetext}
%%%%%%%%%%%%%%%%%%%%%%%%%%%%%%%%%%%%%%%%%%%%%%%%%%%%%%%%%%%%%%%%%%%%%
of the outcomes of Gaussian measurements $\Pi_{A_{\rm
out}}^{\mathcal{E}}(d_{A})$ and $\Pi_{B_{\rm
out}}^{\mathcal{E}}(d_{B})$ (characterized by CMs
$\Gamma_{A}^{\mathcal E}$ and $\Gamma_{B}^{\mathcal E}$) on the
conditional state (\ref{tilderhocondnorm2}). Substituting into the
RHS of the latter equation for the conditional state from
Eq.~(\ref{tilderhocondnorm2}) one finds after some algebra that
%%%%%%%%%%%%%%%%%%%%%%%%%%%%%%%%%%%%%%%%%%%%%%%%%%%%%%%%%%%%%%%%%%%%%
\begin{widetext}
\begin{equation}\label{tildePcond2}
\tilde{P}(d_{A},d_{B}|d_{E_{\rho}},{\bf
r}')=\tilde{\mathcal{P}}[d_{A}-(\mathscr{W}\mathbf{r}')_{A_{\rm
out}},d_{B}-(\mathscr{W}\mathbf{r}')_{B_{\rm
out}}|d_{E_{\rho}},{\bf r}'],
\end{equation}
\end{widetext}
%%%%%%%%%%%%%%%%%%%%%%%%%%%%%%%%%%%%%%%%%%%%%%%%%%%%%%%%%%%%%%%%%%%%%
where
%%%%%%%%%%%%%%%%%%%%%%%%%%%%%%%%%%%%%%%%%%%%%%%%%%%%%%%%%%%%%%%%%%%%%
\begin{eqnarray*}\label{Wjout}
(\mathscr{W}\mathbf{r}')_{j_{\rm
out}}\equiv((\mathscr{W}\mathbf{r}')_{j_{\rm
out}1},(\mathscr{W}\mathbf{r}')_{j_{\rm
out}2},\ldots,(\mathscr{W}\mathbf{r}')_{j_{\rm out}2J_{j_{\rm
out}}}))^{T},
\end{eqnarray*}
%%%%%%%%%%%%%%%%%%%%%%%%%%%%%%%%%%%%%%%%%%%%%%%%%%%%%%%%%%%%%%%%%%%%%
where $J_{A_{\rm out}}=N_{\rm out}$ and $J_{B_{\rm out}}=M_{\rm
out}$, and
%%%%%%%%%%%%%%%%%%%%%%%%%%%%%%%%%%%%%%%%%%%%%%%%%%%%%%%%%%%%%%%%%%%%%
\begin{widetext}
\begin{equation}\label{tildecalPcond}
\tilde{\mathcal{P}}(d_{A},d_{B}|d_{E_{\rho}},{\bf
r}')=\mbox{Tr}_{A_{\rm out}B_{\rm
out}}\left[\left(\mathcal{E}_{A}\otimes\mathcal{E}_{B}\right)(\rho_{A_{\rm
in}B_{\rm in}|E_{\rho}}')\Pi_{A_{\rm
out}}^{\mathcal{E}}(d_{A})\otimes\Pi_{B_{\rm
out}}^{\mathcal{E}}(d_{B})\right].
\end{equation}
\end{widetext}
%%%%%%%%%%%%%%%%%%%%%%%%%%%%%%%%%%%%%%%%%%%%%%%%%%%%%%%%%%%%%%%%%%%%%
The mutual information of the distribution in
Eq.~(\ref{tildePcond2}) does not depend on the displacements
$-(\mathscr{W}\mathbf{r}')_{j_{\rm out}}$, $j=A,B$ and hence it is
equal to the mutual information of the distribution
(\ref{tildecalPcond}). The tensor product
$\mathcal{E}_{A}\otimes\mathcal{E}_{B}$ of Gaussian local
trace-preserving operations $\mathcal{E}_{j}$, $j=A,B$ appearing
on the RHS of Eq.~(\ref{tildecalPcond}), transforms the
$(N+M)$-mode Gaussian state $\rho_{A_{\rm in}B_{\rm
in}|E_{\rho}}'$, Eq.~(\ref{rhocondin}), onto an $(N_{\rm
out}+M_{\rm out})$-mode Gaussian state. More precisely, the
operation $\mathcal{E}_{j}$, $j=A,B$, transforms $J_{j_{\rm in}}$
modes $j_{{\rm in}1},j_{{\rm in}2},\ldots,j_{{\rm in}J_{j_{\rm
in}}}$ of the state (\ref{rhocondin}) onto $J_{j_{\rm out}}$
output modes $j_{{\rm out}1},j_{{\rm out}2},\ldots,j_{{\rm
out}J_{j_{\rm out}}}$, where $J_{A_{\rm in}}=N$ and $J_{B_{\rm
in}}=M$. As each operation $\mathcal{E}_{j}$ is Gaussian and
trace-preserving it can be realized in three steps encompassing 1)
a Gaussian unitary interaction $U_{j}$ between the $J_{j_{\rm
in}}$ input modes and $J_{j_{\rm anc}}$ ancillary modes in vacuum
states, where $J_{A_{\rm anc}}=N_{\rm anc}$ and $J_{B_{\rm
anc}}=M_{\rm anc}$, followed by 2) discarding of $J_{j_{\rm
disc}}\equiv J_{j_{\rm in}}+J_{j_{\rm anc}}-J_{j_{\rm out}}$
modes, and 3) addition of classical Gaussian noise
\cite{Eisert_03,Weedbrook_12}. The noise can be created by a
random displacement of the output state in phase space distributed
according to a zero mean Gaussian distribution. The addition of
the zero mean Gaussian noise acts only on the level of the CMs
where it is represented by the addition of a positive-semidefinite
matrix $F_{j}$ to the CM of the output state. Similarly, the
measurement $\Pi_{j_{\rm out}}^{\mathcal{E}}(d_{j})$ on the
subsystem is on the level of the CM represented by the addition of
a CM $\Gamma_{j}^{\mathcal{E}}$ to the CM of the measured state.
Denoting as $\gamma_{2}$ the $2(N_{\rm out}+M_{\rm
out})$-dimensional CM of the state obtained by propagation of the
input state $\rho_{A_{\rm in}B_{\rm in}|E_{\rho}}'$ through steps
1) and 2) of the implementation of the operations
$\mathcal{E}_{A}$ and $\mathcal{E}_{B}$, the CCM of the
distribution (\ref{tildecalPcond}) reads as
%%%%%%%%%%%%%%%%%%%%%%%%%%%%%%%%%%%%%%%%%%%%%%%%%%%%%%%%%%%%%%%%
\begin{equation}\label{gamma3}
\gamma_{2}+F_{A}\oplus
F_{B}+\Gamma_{A}^{\mathcal{E}}\oplus\Gamma_{B}^{\mathcal{E}}=\gamma_{2}+
(\Gamma_{A}^{\mathcal{E}}+F_{A})\oplus(\Gamma_{B}^{\mathcal{E}}+F_{B}).
\end{equation}
%%%%%%%%%%%%%%%%%%%%%%%%%%%%%%%%%%%%%%%%%%%%%%%%%%%%%%%%%%%%%%%%
Therefore, the addition of local classical Gaussian noise into
subsystems $A_{\rm out}$ and $B_{\rm out}$ followed by the local
Gaussian measurements $\Pi_{A_{\rm out}}^{\mathcal{E}}(d_{A})$ and
$\Pi_{B_{\rm out}}^{\mathcal{E}}(d_{B})$ on the subsystems can be
viewed just as more noisy local Gaussian measurements
$\bar{\Pi}_{A_{\rm out}}^{\mathcal{E}}(d_{A})$ and
$\bar{\Pi}_{B_{\rm out}}^{\mathcal{E}}(d_{B})$ characterized by
the CM
$\bar{\Gamma}_{A}^{\mathcal{E}}\equiv\Gamma_{A}^{\mathcal{E}}+F_{A}$
and
$\bar{\Gamma}_{B}^{\mathcal{E}}\equiv\Gamma_{B}^{\mathcal{E}}+F_{B}$.
Consequently, the conditional distribution (\ref{tildecalPcond})
can be expressed as
%%%%%%%%%%%%%%%%%%%%%%%%%%%%%%%%%%%%%%%%%%%%%%%%%%%%%%%%%%%%%%%%%%%%%
\begin{widetext}
\begin{equation}\label{tildecalPcond1}
\tilde{\mathcal{P}}(d_{A},d_{B}|d_{E_{\rho}},{\bf
r}')=\mbox{Tr}_{A_{\rm out}B_{\rm
out}}\left[\left(\bar{\mathcal{E}}_{A}\otimes\bar{\mathcal{E}}_{B}\right)(\rho_{A_{\rm
in}B_{\rm in}|E_{\rho}}')\bar{\Pi}_{A_{\rm
out}}^{\mathcal{E}}(d_{A})\otimes\bar{\Pi}_{B_{\rm
out}}^{\mathcal{E}}(d_{B})\right],
\end{equation}
\end{widetext}
%%%%%%%%%%%%%%%%%%%%%%%%%%%%%%%%%%%%%%%%%%%%%%%%%%%%%%%%%%%%%%%%%%%%%
where $\bar{\mathcal{E}}_{A}$ and $\bar{\mathcal{E}}_{B}$ are
local Gaussian trace-preserving operations which can be
implemented using steps 1) and 2) but which do not require
addition of classical noise. If we now express the latter two
operations via local Gaussian unitary transformations $U_A$ and
$U_B$ on a larger system consisting of $N$-mode subsystem $A_{\rm
in}$, $M$-mode subsystem $B_{\rm in}$, $N_{\rm anc}$ auxiliary
vacuum modes denoted as a subsystem $A_{\rm anc}$ and $M_{\rm
anc}$ auxiliary vacuum modes denoted as a subsystem $B_{\rm anc}$,
the distribution (\ref{tildecalPcond1}) attains the form
%%%%%%%%%%%%%%%%%%%%%%%%%%%%%%%%%%%%%%%%%%%%%%%%%%%%%%%%%%%%%%%%%%%%%
\begin{widetext}
\begin{eqnarray}\label{tildecalPcond2}
\tilde{\mathcal{P}}(d_{A},d_{B})&=&\mbox{Tr}_{A_{\rm out}B_{\rm
out}}\mbox{Tr}_{A_{\rm disc}B_{\rm disc}}\left[\left(U_{A}\otimes
U_{B}\right)\rho_{A_{\rm in}B_{\rm
in}|E_{\rho}}'\otimes|\{0\}\rangle_{A_{\rm
anc}}\langle\{0\}|\otimes|\{0\}\rangle_{B_{\rm
anc}}\langle\{0\}|\left(U_{A}^{\dag}\otimes
U_{B}^{\dag}\right)\right.\nonumber\\
&&\left.\bar{\Pi}_{A_{\rm
out}}^{\mathcal{E}}(d_{A})\otimes\bar{\Pi}_{B_{\rm
out}}^{\mathcal{E}}(d_{B})\otimes\openone_{A_{\rm
disc}}\otimes\openone_{B_{\rm disc}}\right],
\end{eqnarray}
\end{widetext}
%%%%%%%%%%%%%%%%%%%%%%%%%%%%%%%%%%%%%%%%%%%%%%%%%%%%%%%%%%%%%%%%%%%%%
where here and in what follows we omit the dependence of the
distribution on the variables $d_{E_{\rho}}$ and ${\bf r}'$ for
brevity. Here $\mbox{Tr}_{j_{\rm disc}}$, $j=A,B$, is the trace
over the discarded $J_{j_{\rm disc}}$-mode subsystem $j_{\rm
disc}$ ($J_{A_{\rm disc}}= N+N_{\rm anc}-N_{\rm out}$ and
$J_{B_{\rm disc}}= M+M_{\rm anc}-M_{\rm out}$),
$|\{0\}\rangle_{j_{\rm anc}}$ is the tensor product of $J_{j_{\rm
anc}}$ vacuum states, and $\openone_{j_{\rm disc}}$ is the
identity operator on the space of the discarded subsystem $j_{\rm
disc}$. Next, the linearity of the Gaussian unitary transformation
$U_{A}\otimes U_{B}$ allows us to transform the displacement
$D_{{A_{\rm in}B_{\rm
in}}}[-\Lambda(\mathscr{W}\mathbf{r}')_{A_{\rm in}B_{\rm in}}]$ in
Eq.~(\ref{rhocondin}) through the transformation which will
result, together with utilization of the invariance of the trace
under cyclic permutations, in a displacement of the measurement
outcomes $d_{A}$ and $d_{B}$. However, as we have already said
such a displacement is irrelevant from the point of view of mutual
information and hence we can replace in what follows the displaced
state $\rho_{A_{\rm in}B_{\rm in}|E_{\rho}}'$ on the RHS of
Eq.(\ref{tildecalPcond2}) with the undisplaced state $\rho_{A_{\rm
in}B_{\rm in}|E_{\rho}}(d_{E_{\rho}})$ defined in
Eq.~(\ref{rhoABErho}). Further, the distribution
(\ref{tildecalPcond2}) can be seen as the reduction
%%%%%%%%%%%%%%%%%%%%%%%%%%%%%%%%%%%%%%%%%%%%%%%%%%%%%%%%%%%%%%%%%%%%%
\begin{equation}\label{tildePred}
\tilde{\mathcal{P}}(d_{A},d_{B})=\int\mathfrak{P}(d_{A},d_{A}',d_{B},
d_{B}')d^{2J_{A_{\rm disc}}}d_{A}'d^{2J_{B_{\rm disc}}}d_{B}'
\end{equation}
%%%%%%%%%%%%%%%%%%%%%%%%%%%%%%%%%%%%%%%%%%%%%%%%%%%%%%%%%%%%%%%%%%%%%
of the following distribution
%%%%%%%%%%%%%%%%%%%%%%%%%%%%%%%%%%%%%%%%%%%%%%%%%%%%%%%%%%%%%%%%%%%%%
\begin{widetext}
\begin{eqnarray}\label{tildecalPcond3}
\mathfrak{P}(d_{A},d_{A}',d_{B}, d_{B}')&=&\mbox{Tr}_{A_{\rm
out}B_{\rm out}}\mbox{Tr}_{A_{\rm disc}B_{\rm
disc}}\left[\left(U_{A}\otimes U_{B}\right)\rho_{A_{\rm in}B_{\rm
in}|E_{\rho}}\otimes|\{0\}\rangle_{A_{\rm
anc}}\langle\{0\}|\otimes|\{0\}\rangle_{B_{\rm
anc}}\langle\{0\}|\right.\nonumber\\
&&\left.\times\left(U_{A}^{\dag}\otimes
U_{B}^{\dag}\right)\bar{\Pi}_{A_{\rm
out}}^{\mathcal{E}}(d_{A})\otimes\bar{\Pi}_{B_{\rm
out}}^{\mathcal{E}}(d_{B})\otimes\Pi_{A_{\rm dsic}}(d_{A}')\otimes
\Pi_{B_{\rm disc}}(d_{B}')\right],
\end{eqnarray}
\end{widetext}
%%%%%%%%%%%%%%%%%%%%%%%%%%%%%%%%%%%%%%%%%%%%%%%%%%%%%%%%%%%%%%%%%%%%%
where $\Pi_{j_{\rm dsic}}(d_{j}')$, $j=A,B$ is a Gaussian
measurement on the discarded subsystem $j_{\rm disc}$ with the
measurement outcome $d_{j}'$ and where we have omitted the
dependence of the state $\rho_{A_{\rm in}B_{\rm in}|E_{\rho}}$ on
the measurement outcome $d_{E_{\rho}}$ for simplicity. As
discarding variables cannot increase the mutual information
\cite{Cover_06}, one obtains that the mutual information $I(A;B)$
of the distribution (\ref{tildecalPcond2}) and the mutual
information $I(A,A';B,B')$ of the distribution
(\ref{tildecalPcond3}) satisfy the inequality $I(A;B)\leq
I(A,A';B,B')$. Now, making use of the invariance of the trace
under cyclic permutations and the equality
$U^{\dag}(\mathscr{S})\Pi(d)U(\mathscr{S})=\Pi_{\mathscr{S}}(\mathscr{S}^{-1}d)$,
where $\Pi_{\mathscr{S}}(d)$ is a component of a Gaussian POVM
with the seed element $U^{\dag}(\mathscr{S})\Pi_{0}U(\mathscr{S})$
and $U(\mathscr{S})$ is a Gaussian unitary transformation
corresponding to the symplectic matrix $\mathscr{S}$, we can write
down the distribution (\ref{tildecalPcond3}) as
%%%%%%%%%%%%%%%%%%%%%%%%%%%%%%%%%%%%%%%%%%%%%%%%%%%%%%%%%%%%%%%%%%%%%%%
\begin{equation}\label{frakPeqscrP}
\mathfrak{P}(d_{A},d_{A}',d_{B},
d_{B}')=\mathscr{P}\left[\left(\mathscr{S}_{A}^{-1}\Delta_{A}\right)^{T},\left(\mathscr{S}_{B}^{-1}\Delta_{B}\right)^{T}\right].\nonumber\\
\end{equation}
%%%%%%%%%%%%%%%%%%%%%%%%%%%%%%%%%%%%%%%%%%%%%%%%%%%%%%%%%%%%%%%%%%%%%%%
%%%%%%%%%%%%%%%%%%%%%%%%%%%%%%%%%%%%%%%%%%%%%%%%%%%%%%%%%%%%%%%%%%%%%%%
%\begin{equation}\label{frakP}
%\mathfrak{P}(d_{A},d_{A}',d_{B},
%d_{B}')=\mathscr{P}\left\{\left[\mathscr{S}_{A}^{-1}\left(\begin{array}{c}
%d_{A} \\
%d_{A}' \\
%\end{array}\right)\right]^{T},\left[\mathscr{S}_{B}^{-1}\left(\begin{array}{c}
%d_{B} \\
%d_{B}'\\
%\end{array}\right)\right]^{T}\right\}
%\end{equation}
%%%%%%%%%%%%%%%%%%%%%%%%%%%%%%%%%%%%%%%%%%%%%%%%%%%%%%%%%%%%%%%%%%%%%%%
Here $\mathscr{S}_{A}$ and $\mathscr{S}_{B}$ denote the symplectic
matrices corresponding to the local Gaussian unitaries $U_{A}$ and
$U_{B}$, respectively, $\Delta_{j}=(d_{j}^{T},d_{j}'^{T})^{T}$,
$j=A,B$, and
%%%%%%%%%%%%%%%%%%%%%%%%%%%%%%%%%%%%%%%%%%%%%%%%%%%%%%%%%%%%%%%%%%%%%
\begin{widetext}
\begin{eqnarray}\label{scrP}
\mathscr{P}\left(d_{A},d_{A}',d_{B},
d_{B}'\right)&=&\mbox{Tr}_{A_{\rm in}A_{\rm anc}B_{\rm in}B_{\rm
anc}}\left[\rho_{A_{\rm in}B_{\rm
in}|E_{\rho}}\otimes|\{0\}\rangle_{A_{\rm anc}B_{\rm
anc}}\langle\{0\}|\Pi_{A_{\rm in}A_{\rm
anc}}'(d_{A},d_{A}')\otimes\Pi_{B_{\rm in}B_{\rm
anc}}'(d_{B},d_{B}')\right],\nonumber\\
\end{eqnarray}
\end{widetext}
%%%%%%%%%%%%%%%%%%%%%%%%%%%%%%%%%%%%%%%%%%%%%%%%%%%%%%%%%%%%%%%%%%%%%
where $|\{0\}\rangle_{A_{\rm anc}B_{\rm
anc}}=|\{0\}\rangle_{A_{\rm anc}}\otimes|\{0\}\rangle_{B_{\rm
anc}}$ and $\Pi_{j_{\rm in}j_{\rm anc}}'(d_{j},d_{j}')$, $j=A,B$,
is the Gaussian measurement on the subsystem $(j_{\rm in},j_{\rm
anc})$ with the seed element $\Pi_{0 j_{\rm in}j_{\rm anc}}'\equiv
U_{j}^{\dag}\bar{\Pi}_{0 j_{\rm out}}^{\mathcal{E}}\otimes\Pi_{0
j_{\rm disc}}U_{j}$, where $\bar{\Pi}_{0 j_{\rm
out}}^{\mathcal{E}}$ and $\Pi_{0 j_{\rm disc}}$ are the seed
elements of the Gaussian measurements $\bar{\Pi}_{j_{\rm
out}}^{\mathcal{E}}(d_{j})$ and $\Pi_{j_{\rm dsic}}(d_{j}')$,
respectively, which appear on the RHS of
Eq.~(\ref{tildecalPcond3}). From the invariance of the mutual
information under local symplectic transformations it then follows
that the mutual information of the distribution
(\ref{tildecalPcond3}) and the distribution (\ref{scrP}) are equal
and hence we can further work with the distribution (\ref{scrP}).

Let us denote now the CM of the conditional state $\rho_{A_{\rm
in}B_{\rm in}|E_{\rho}}$ as $\gamma_{AB}^{c}$ and
the CMs of the measurements $\Pi_{A_{\rm in}A_{\rm
anc}}'(d_{A},d_{A}')$ and $\Pi_{B_{\rm in}B_{\rm
anc}}'(d_{B},d_{B}')$ as $\Gamma_{A}'$ and
$\Gamma_{B}'$, respectively. The mutual
information of the distribution (\ref{scrP}) then attains the form \cite{Gelfand_57}
%%%%%%%%%%%%%%%%%%%%%%%%%%%%%%%%%%%%%%%%%%%%%%%%%%%%%%%%%%%%%%%%%%%%
\begin{eqnarray}\label{I}
I(A,A';B,B')=\frac{1}{2}\ln\left(\frac{\mbox{det}\sigma_{A}'\mbox{det}\sigma_{B}'}{\mbox{det}\sigma_{AB}'}\right),
\end{eqnarray}
%%%%%%%%%%%%%%%%%%%%%%%%%%%%%%%%%%%%%%%%%%%%%%%%%%%%%%%%%%%%%%%%%%%%
where
%%%%%%%%%%%%%%%%%%%%%%%%%%%%%%%%%%%%%%%%%%%%%%%%%%%%%%%%%%%%%%%%%%%%
\begin{equation}\label{sigmaprimed}
\sigma_{AB}'=\gamma_{AB}^{c}\oplus\openone_{\rm anc}+\Gamma_{A}'\oplus\Gamma_{B}',
\end{equation}
%%%%%%%%%%%%%%%%%%%%%%%%%%%%%%%%%%%%%%%%%%%%%%%%%%%%%%%%%%%%%%%%%%%%
with $\sigma_{j}'$ being the CM of the reduced state of the
subsystem $(j_{\rm in},j_{\rm anc})$, $j=A,B$, and $\openone_{\rm
anc}$ is the $2(N_{\rm anc}+M_{\rm anc})$-dimensional identity
matrix describing the CM of the vacuum state
$|\{0\}\rangle_{A_{\rm anc}B_{\rm anc}}$. Further, it is
convenient to express the CMs $\Gamma_{A}'$ and $\Gamma_{B}'$ with
respect to $\mbox{in}|\mbox{anc}$ splitting as
%%%%%%%%%%%%%%%%%%%%%%%%%%%%%%%%%%%%%%%%%%%%%%%%%%%%%%%%%%%%%%%%%%%%%%%%%%%%%%%%%%%%%%%%%%%%%%%%%%%%
\begin{eqnarray}\label{Gammaprimed}
\Gamma_{A}'=\left(\begin{array}{cc}
A_{\rm in} & C_{A}\\
C_{A}^{T} & A_{\rm anc}\\
\end{array}\right),\,\, \Gamma_{B}'=\left(\begin{array}{cc}
B_{\rm in} & C_{B}\\
C_{B}^{T} & B_{\rm anc}\\
\end{array}\right).
\end{eqnarray}
%%%%%%%%%%%%%%%%%%%%%%%%%%%%%%%%%%%%%%%%%%%%%%%%%%%%%%%%%%%%%%%%%%%%%%%%%%%%%%%%%%%%%%%%%%%%%%%%%%%%
%%%%%%%%%%%%%%%%%%%%%%%%%%%%%%%%%%%%%%%%%%%%%%%%%%%%%%%%%%%%%%%%%%%%%%%%%%%%%%%%%%%%%%%%%%%%%%%%%%%%
%\begin{eqnarray}\label{gammaABc}
%\gamma_{AB}^{c}=\left(\begin{array}{cc}
%A & C\\
%C^{T} & B\\
%\end{array}\right).
%\end{eqnarray}
%%%%%%%%%%%%%%%%%%%%%%%%%%%%%%%%%%%%%%%%%%%%%%%%%%%%%%%%%%%%%%%%%%%%%%%%%%%%%%%%%%%%%%%%%%%%%%%%%%%%
Consider now the determinant formula \cite{Horn_85}
%%%%%%%%%%%%%%%%%%%%%%%%%%%%%%%%%%%%%%%%%%%%%%%%%%%%%%%%%%%%%%%%%%%%%%%%%%%%%%%%%%%%%%%%%%%%%%%%%%%%%
\begin{equation}\label{det}
\mbox{det}(M)=\mbox{det}(\mathscr{D})\mbox{det}(\mathscr{A}-\mathscr{B}\mathscr{D}^{-1}\mathscr{C}),
\end{equation}
%%%%%%%%%%%%%%%%%%%%%%%%%%%%%%%%%%%%%%%%%%%%%%%%%%%%%%%%%%%%%%%%%%%%%%%%%%%%%%%%%%%%%%%%%%%%%%%%%%%%%
which is valid for any $(n+m)\times(n+m)$ matrix
%%%%%%%%%%%%%%%%%%%%%%%%%%%%%%%%%%%%%%%%%%%%%%%%%%%%%%%%%%%%%%%%%%%%%%%%%%%%%%%%%%%%%%%%%%%%%%%%%%%%
\begin{eqnarray}\label{M}
M=\left(\begin{array}{cc}
\mathscr{A} & \mathscr{B}\\
\mathscr{C} & \mathscr{D}\\
\end{array}\right),
\end{eqnarray}
%%%%%%%%%%%%%%%%%%%%%%%%%%%%%%%%%%%%%%%%%%%%%%%%%%%%%%%%%%%%%%%%%%%%%%%%%%%%%%%%%%%%%%%%%%%%%%%%%%%%
where $\mathscr{A}$, $\mathscr{B}$ and $\mathscr{C}$ are
respectively $n\times n$, $n\times m$ and $m\times n$ matrices and
$\mathscr{D}$ is an $m\times m$ invertible matrix. Applying the
formula to the RHS of Eq.~(\ref{I}) we can bring it after some
algebra into the form
%%%%%%%%%%%%%%%%%%%%%%%%%%%%%%%%%%%%%%%%%%%%%%%%%%%%%%%%%%%%%%%%%%%%
\begin{eqnarray}\label{I2}
I(A,A';B,B')=\frac{1}{2}\ln\left(\frac{\mbox{det}\mu_{A}\mbox{det}\mu_{B}}{\mbox{det}\mu_{AB}}\right),
\end{eqnarray}
%%%%%%%%%%%%%%%%%%%%%%%%%%%%%%%%%%%%%%%%%%%%%%%%%%%%%%%%%%%%%%%%%%%%
where
%%%%%%%%%%%%%%%%%%%%%%%%%%%%%%%%%%%%%%%%%%%%%%%%%%%%%%%%%%%%%%%%%%%%
\begin{equation}\label{mu}
\mu_{AB}=\gamma_{AB}^{c}+\tilde{\Gamma}_{A}\oplus\tilde{\Gamma}_{B}
\end{equation}
%%%%%%%%%%%%%%%%%%%%%%%%%%%%%%%%%%%%%%%%%%%%%%%%%%%%%%%%%%%%%%%%%%%%
and $\mu_{A,B}$ are CMs of the reduced states of the subsystems
$A$ and $B$. Here
%%%%%%%%%%%%%%%%%%%%%%%%%%%%%%%%%%%%%%%%%%%%%%%%%%%%%%%%%%%%%%%%%%%%
\begin{equation}\label{tildeGammaA}
\tilde{\Gamma}_{A}=A_{\rm in}-C_{A}(A_{\rm anc}+\openone_{A_{\rm
anc}})^{-1}C_{A}^{T}
\end{equation}
%%%%%%%%%%%%%%%%%%%%%%%%%%%%%%%%%%%%%%%%%%%%%%%%%%%%%%%%%%%%%%%%%%%
is the $N$-mode CM,
%%%%%%%%%%%%%%%%%%%%%%%%%%%%%%%%%%%%%%%%%%%%%%%%%%%%%%%%%%%%%%%%%%%%
\begin{equation}\label{tildeGammaB}
\tilde{\Gamma}_{B}=B_{\rm in}-C_{B}(B_{\rm anc}+\openone_{B_{\rm
anc}})^{-1}C_{B}^{T}
\end{equation}
%%%%%%%%%%%%%%%%%%%%%%%%%%%%%%%%%%%%%%%%%%%%%%%%%%%%%%%%%%%%%%%%%%%
is the $M$-mode CM, $\openone_{A_{\rm anc}}$ is the $2N_{\rm
anc}\times2N_{\rm anc}$ identity matrix, and $\openone_{B_{\rm
anc}}$ is the $2M_{\rm anc}\times2M_{\rm anc}$ identity matrix.
Hence, we can interpret the mutual information $I(A,A';B,B')$ as
the mutual information of a new conditional Gaussian probability
density $\tilde{p}(\tilde{d}_{A},\tilde{d}_{B}|d_{E_{\rho}})$
given in Eq.~(\ref{Psuboptimal}), which is obtained by the
Gaussian measurements $\tilde{\Pi}_{A}(\tilde{d}_{A})$ and
$\tilde{\Pi}_{B}(\tilde{d}_{B})$ with CMs $\tilde{\Gamma}_{A}$ and
$\tilde{\Gamma}_{B}$ on the conditional state $\rho_{A_{\rm
in}B_{\rm in}|E_{\rho}}$ defined in Eq.~(\ref{rhoABErho}). If we
now take into account the fact that the CM $\gamma_{AB}^{c}$ of
the state reads as
%%%%%%%%%%%%%%%%%%%%%%%%%%%%%%%%%%%%%%%%%%%%%%%%%%%%%%%%%%%%%%
\begin{equation}\label{gammaABc}
\gamma_{AB}^{c}=\gamma_{AB}-\gamma_{ABE}^{\cal
I}\left(\gamma_{E}^{\cal I}+\Gamma_{E}^{\cal
I}\right)^{-1}(\gamma_{ABE}^{\cal I})^{T},
\end{equation}
%%%%%%%%%%%%%%%%%%%%%%%%%%%%%%%%%%%%%%%%%%%%%%%%%%%%%%%%%%%%%%%%
where $\gamma_{ABE}^{\cal I}$ and $\gamma_{E}^{\cal I}$ are the
respective blocks of the CM $\gamma_{\pi}^{\cal I}$, we find that
the mutual information (\ref{I2}) is equal to
%%%%%%%%%%%%%%%%%%%%%%%%%%%%%%%%%%%%%%%%%%%%%%%%%%%%%%%%%%%%%%%%%%%%
\begin{eqnarray}\label{I3}
I(A,A';B,B')=f\left(\gamma_{\pi}^{\cal
I},\tilde{\Gamma}_{A},\tilde{\Gamma}_{B},\Gamma_{E}^{\cal
I}\right),
\end{eqnarray}
%%%%%%%%%%%%%%%%%%%%%%%%%%%%%%%%%%%%%%%%%%%%%%%%%%%%%%%%%%%%%%%%%%%%
and thus the inequality $I(A;B)\leq I(A,A';B,B')$ translates into
the inequality (\ref{Ineq2}) as we wanted to prove.

Finally, as at given CMs $\gamma_{\pi}^{\cal I}$ and
$\Gamma_{E}^{\cal I}$, the CMs $\tilde{\Gamma}_{A}$ and
$\tilde{\Gamma}_{B}$ given in Eqs.~(\ref{tildeGammaA}) and
(\ref{tildeGammaB}) generally do not maximize the function
$f\left(\gamma_{\pi}^{\cal
I},\Gamma_{A},\Gamma_{B},\Gamma_{E}^{\cal I}\right)$ with respect
to CMs $\Gamma_{A}$ and $\Gamma_{B}$ one gets
%%%%%%%%%%%%%%%%%%%%%%%%%%%%%%%%%%%%%%%%%%%%%%%%%%%%%%%%%%%%%%%%%%%%%%%%%%%%%%%%%%%%%%%%%%%%%%%%%%%%%%%%%%%
\begin{widetext}
\begin{eqnarray}\label{Ineq3}
f\left(\gamma_{\pi}^{\cal
I},\tilde{\Gamma}_{A},\tilde{\Gamma}_{B},\Gamma_{E}^{\cal
I}\right)\leq f\left(\gamma_{\pi}^{\cal I},\Gamma_{A}^{\cal
I},\Gamma_{B}^{\cal I},\Gamma_{E}^{\cal
I}\right)=E_{\downarrow}^{G}\left(\rho_{AB}^{\cal I}\right),
\end{eqnarray}
\end{widetext}
%%%%%%%%%%%%%%%%%%%%%%%%%%%%%%%%%%%%%%%%%%%%%%%%%%%%%%%%%%%%%%%%%%%%%%%%%%%%%%%%%%%%%%%%%%%%%%%%%%%%%%%%%%%
where $\Gamma_{A}^{\cal I}$ and $\Gamma_{B}^{\cal I}$ are CMs of
the optimal measurements $\Pi_{A}(d_{A})$ and $\Pi_{B}(d_{B})$
which maximize $f$ and the equality follows from Eq.~(\ref{Eq10}).

In summary, combining inequalities (\ref{Ineq1}), (\ref{Ineq2})
and (\ref{Ineq3}) the monotonicity of GIE, Eq.~(\ref{Gaussianmu}),
under GLTPOCC can be expressed by the following chain of
inequalities:
%%%%%%%%%%%%%%%%%%%%%%%%%%%%%%%%%%%%%%%%%%%%%%%%%%%%%%%%%%%%%%%%%%%%%
\begin{widetext}
\begin{equation}\label{monotonicity}
E_{\downarrow}^{G}\left(\rho_{AB}^{\mathcal{E}}\right)\leq
f\left(\gamma_{\pi}^{\mathcal{E}},\Gamma_{A}^{\mathcal{E}},\Gamma_{B}^{\mathcal{E}},\Gamma_{E}^{\cal
I}\oplus\tilde{\Gamma}_{E_{\chi}}^{\mathcal{E}}\right) \leq
f\left(\gamma_{\pi}^{\cal
I},\tilde{\Gamma}_{A},\tilde{\Gamma}_{B},\Gamma_{E}^{\cal
I}\right)\leq E_{\downarrow}^{G}\left(\rho_{AB}^{\cal I}\right),
\end{equation}
\end{widetext}
%%%%%%%%%%%%%%%%%%%%%%%%%%%%%%%%%%%%%%%%%%%%%%%%%%%%%%%%%%%%%%%%%%%%%
which accomplishes the monotonicity proof.

%%%%%%%%%%%%%%%%%%%%%%%%%%%%%%%%%%%%%%%%%%%%%%%%%%%%%%%%%%%%%%%%%%%%%%%%%%%%%%%%%%%%%%%%%%%%%%%%%%%%%
Before moving to an explicit evaluation of GIE, let us note that an important subset of GLTPOCC operations is the class of Gaussian local
unitary operations ($\equiv U_{A}\otimes U_{B}$) which transform the input Gaussian state
$\rho_{AB}^{\cal I}$ to $\rho_{AB}^{\mathcal{U}}\equiv (U_{A}\otimes U_{B})\rho_{AB}^{\cal I}(U_{A}^{\dag}\otimes U_{B}^{\dag})$.
Inequality (\ref{monotonicity}) and the reversibility of unitary operations then implies the invariance of
GIE with respect to the local Gaussian unitary operations,
$E_{\downarrow}^{G}\left(\rho_{AB}^{\mathcal{U}}\right)=E_{\downarrow}^{G}\left(\rho_{AB}^{\cal I}\right)$.
When calculating GIE we can therefore assume without any loss of generality that the CM $\gamma_{AB}$ of the
considered state is in the standard form \cite{Simon_00}
%%%%%%%%%%%%%%%%%%%%%%%%%%%%%%%%%%%%%%%%%%%%%%%%%%%%%%%%%%%%%%%%%%%%%%%%%%%%%%%%%%%%%%%%%%%%%%%%%%
\begin{equation}\label{gammaABadd}
\gamma_{AB}=\left(
\begin{array}{cccc}
 a & 0 & c_1 & 0 \\
 0 & a & 0 & c_2 \\
 c_1 & 0 & b & 0 \\
 0 & c_2 & 0 & b
\end{array}
\right),
\end{equation}
%%%%%%%%%%%%%%%%%%%%%%%%%%%%%%%%%%%%%%%%%%%%%%%%%%%%%%%%%%%%%%%%%%%%%%%%%%%%%%%%%%%%%%%%%%%%%%%%%%%
where $c_1 \ge |c_2| \ge 0$, which can greatly simplify our calculations.
%%%%%%%%%%%%%%%%%%%%%%%%%%%%%%%%%%%%%%%%%%%%%%%%%%%%%%%%%%%%%%%%%%%%%%%%%%%%%%%%%%%%%%%%%%%%%%%%%%%%%

\section{GIE for pure states}\label{sec_5}

\begin{figure}[ht]
\includegraphics[width=0.9\columnwidth]{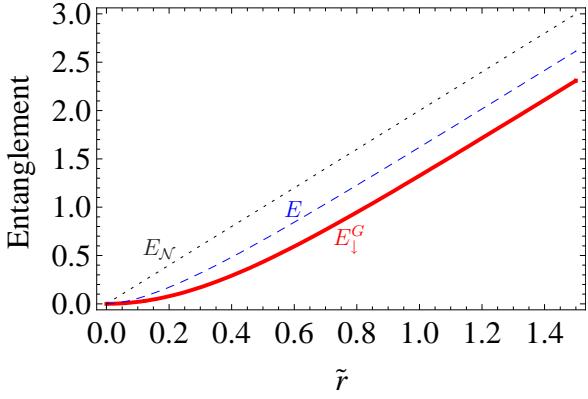}
\caption{(Color online) GIE $E_{\downarrow}^{G}$ (solid red
curve), entropy of entanglement $E$ (dashed blue curve), and
logarithmic negativity $E_{\cal{N}}$ (dotted black curve) for pure
Gaussian states versus the squeezing parameter $\tilde{r}$.}
\label{fig_GIEpure}
\end{figure}

As a first example we calculate GIE for the class of pure Gaussian
states $\rho_{\rm p}$ with CM $\gamma_{AB}^{\rm p}$. For these
states any purification is a product state with respect to the
$AB|E$ splitting and therefore the block $\gamma_{ABE}$ in CCM
(\ref{gammaABEsplitting}) is a matrix of zeros. This implies that
the Schur complement (\ref{sigma}) reads as
$\sigma_{AB}=\gamma_{AB}^{\rm p}+\Gamma_{A}\oplus\Gamma_{B}$ and
the GIE coincides with the Gaussian classical mutual information
($\equiv\mathcal{I}_{c}^{G}$) of a quantum state $\rho_{\rm p}$
\cite{Terhal_02,Mista_11},
%%%%%%%%%%%%%%%%%%%%%%%%%%%%%%%%%%%%%%%%%%%%%%%%%%%%%%%%%%%%%%%%%%%%
\begin{equation}\label{IcGadd}
E_{\downarrow}^{G}\left(\rho_{\rm
p}\right)=\mathcal{I}_{c}^{G}\left(\rho_{\rm p}\right)\equiv\mathop{\mbox{sup}}_{\Gamma_{A},\Gamma_{B}}\frac{1}{2}\ln\left(\frac{\mbox{det}\sigma_{A}\mbox{det}\sigma_{B}}{\mbox{det}\sigma_{AB}}\right).
\end{equation}
%%%%%%%%%%%%%%%%%%%%%%%%%%%%%%%%%%%%%%%%%%%%%%%%%%%%%%%%%%%%%%%%%%%%%
From the results of Ref.~\cite{Mista_11} it then follows
that the supremum is attained by double homodyne detection which gives \cite{Mista_14}
%%%%%%%%%%%%%%%%%%%%%%%%%%%%%%%%%%%%%%%%%%%%%%%%%%%%%%%%%%%%%%%%%%%%
\begin{equation}\label{GIEpure}
E_{\downarrow}^{G}\left(\rho_{\rm
p}\right)=\frac{1}{2}\ln\left(\mbox{det}\gamma_{A}\right)=\ln[\cosh(2\tilde{r})],
\end{equation}
%%%%%%%%%%%%%%%%%%%%%%%%%%%%%%%%%%%%%%%%%%%%%%%%%%%%%%%%%%%%%%%%%%%%
where $\gamma_{A}$ is the CM of the reduced state $\rho_{A}$ of
mode $A$ of the state $\rho_{\rm p}$ and $\tilde{r}\geq0$ is the
squeezing parameter characterizing the latter state, which is
defined by the equation
$\cosh(2\tilde{r})=\sqrt{\mbox{det}\gamma_{A}}$. Interestingly,
the RHS of Eq.~(\ref{GIEpure}) is equal to the Gaussian R\'enyi-2
(GR2) entropy $\mathcal{S}_{2}(\rho_{A})$ which is nothing but the
GR2 entanglement $E_{2}^{G}(\rho_{\rm p})$ \cite{Adesso_12}. This
means that for all pure Gaussian states it holds that
$E_{\downarrow}^{G}=E_{2}^{G}$. Comparing, on the other hand, GIE
with the entropy of entanglement $E(\rho_{\rm
p})=\mathcal{S}(\rho_{A})$ \cite{Bennett_96,Bennett_96a}, where
\cite{Parker_00}
%%%%%%%%%%%%%%%%%%%%%%%%%%%%%%%%%%%%%%%%%%%%%%%%%%%%%%%%%%%%%%%%%%
\begin{eqnarray}\label{S}
\mathcal{S}(\rho_{A})=\cosh^2(\tilde{r})\ln[\cosh^2(\tilde{r})]-\sinh^2(\tilde{r})\ln[\sinh^2(\tilde{r})]\nonumber\\
\end{eqnarray}
%%%%%%%%%%%%%%%%%%%%%%%%%%%%%%%%%%%%%%%%%%%%%%%%%%%%%%%%%%%%%%%%%%
is the marginal von Neumann entropy, one finds that the inequality
$E\geq E_{\downarrow}^{G}$ is satisfied for all pure Gaussian
states \cite{Mista_14}. However, the equality to the entropy of
entanglement is restored for true IE $E_{\downarrow}$,
Eq.~(\ref{Edownarrow}), which admits also non-Gaussian
measurements on modes $A$ and $B$. Namely,
$E_{\downarrow}\left(\rho_{\rm
p}\right)=\mathcal{I}_{c}\left(\rho_{\rm
p}\right)\equiv\mathop{\mbox{sup}}_{\Pi_{A}\otimes\Pi_{B}}
I\left(A; B\right)$, where the RHS is the classical mutual
information of a quantum state $\rho_{\rm p}$ \cite{Terhal_02}
with $I(A;B)$ being the mutual information of a distribution of
outcomes of generally non-Gaussian measurements $\Pi_{A}$ and
$\Pi_{B}$ on modes $A$ and $B$ of the state $\rho_{\rm p}$. The
quantity $\mathcal{I}_{c}\left(\rho_{\rm p}\right)$ is invariant
with respect to local unitaries and thus $\rho_{\rm p}$ can be
replaced by the locally unitarily equivalent two-mode squeezed
vacuum state $\rho_{\rm
p}(\lambda)=|\psi(\lambda)\rangle\langle\psi(\lambda)|$, where
%%%%%%%%%%%%%%%%%%%%%%%%%%%%%%%%%%%%%%%%%%%%%%%%%%%%%%%%%%%%%%%%%%%%%%%%%
\begin{equation}\label{TMSV}
|\psi(\lambda)\rangle=\sqrt{1-\lambda^2}\sum_{n=0}^{\infty}\lambda^{n}|n,n\rangle_{AB}
\end{equation}
%%%%%%%%%%%%%%%%%%%%%%%%%%%%%%%%%%%%%%%%%%%%%%%%%%%%%%%%%%%%%%%%%%%%%%%%
with $\lambda=\tanh\tilde{r}$. Non-Gaussian local photon counting
on modes $A$ and $B$ of the state $|\psi(\lambda)\rangle$ then
yields a probability distribution with $I(A;B)={\cal S}(\rho_{A})$
\cite{Mista_11}, which is the highest mutual information one can
achieve \cite{Wu_09}. Thus we find that $E_{\downarrow}=E$ holds
for all pure Gaussian states as required. A comparison of GIE
(\ref{GIEpure}), entropy of entanglement (\ref{S}) and logarithmic
negativity \cite{Vidal_02,Eisert_PhD} $E_{\cal N}(\rho_{\rm
p})=2\tilde{r}$ \cite{Ohliger_10} as functions of the squeezing
parameter $\tilde{r}$ is depicted in Fig.~\ref{fig_GIEpure}.

%%%%%%%%%%%%%%%%%%%%%%%%%%%%%%%%%%%%%%%%%%%%%%%%%%%%%%%%%%%%%%%%%%%%%%%%%%%%%%%%%%%%%%%%%%%%%%%%%%
%%%%%%%%%%%%%%%%%%%%%%%%%%%%%%%%%%%%%%%%%%%%%%%%%%%%%%%%%%%%%%%%%%%%%%%%%%%%%%%%%%%%%%%%%%%%%%%%%%%

\section{GIE for a two-mode reduction of the three-mode CV GHZ state}\label{sec_6}

%In this Section we calculate GIE for Gaussian state $\rho_{AB}^{GHZ}$ with CM given in Eq.~(19) of the main text in the
%interval $r\in[0,0.684]$.

Despite the complexity of optimization in Eq.~(\ref{Gaussianmu})
it is possible to calculate GIE analytically for some mixed
two-mode Gaussian states. In what follows we illustrate this by
calculating GIE for a two-mode Gaussian state
$(\equiv\rho_{AB}^{GHZ})$ with CM
%%%%%%%%%%%%%%%%%%%%%%%%%%%%%%%%%%%%%%%%%%%%%%%%%%%%%%%%%%%%%%%%%%%%%%%%%%%%%%%%%%%%%%%%%%%%%%%%%%%%
\begin{eqnarray}\label{gammaABGHZ}
\gamma_{AB}^{GHZ}=\left(\begin{array}{cc}
\alpha & \kappa \\
\kappa & \alpha \\
\end{array}\right),
\end{eqnarray}
%%%%%%%%%%%%%%%%%%%%%%%%%%%%%%%%%%%%%%%%%%%%%%%%%%%%%%%%%%%%%%%%%%%%%%%%%%%%%%%%%%%%%%%%%%%%%%%%%%%%%%
which is a reduction of the three-mode CV GHZ state \cite{Loock_00} having CM
%%%%%%%%%%%%%%%%%%%%%%%%%%%%%%%%%%%%%%%%%%%%%%%%%%%%%%%%%%%%%%%%%%%%%%%%%%%%%%%%%%%%%%%%%%%%%%%%%%%%
\begin{eqnarray}\label{gammaGHZ}
\gamma_{ABE}^{GHZ}=\left(\begin{array}{ccc}
\alpha & \kappa & \kappa\\
\kappa & \alpha & \kappa\\
\kappa & \kappa & \alpha\\
\end{array}\right).
\end{eqnarray}
%%%%%%%%%%%%%%%%%%%%%%%%%%%%%%%%%%%%%%%%%%%%%%%%%%%%%%%%%%%%%%%%%%%%%%%%%%%%%%%%%%%%%%%%%%%%%%%%%%%%%%
Here $\alpha=\mbox{diag}\left(x_{+},x_{-}\right)$ and
$\kappa=(x_{-}-x_{+})\sigma_{z}$, where $x_{\pm}=(e^{\pm
2r}+2e^{\mp 2r})/3$ and $r\geq0$ is a squeezing parameter.
This calculation will be accomplished in two steps. First, we will calculate an easier computable upper bound
$(\equiv U\left(\rho_{AB}^{GHZ}\right))$ on $E_{\downarrow}^{G}\left(\rho_{AB}^{GHZ}\right)$. In the second step we will show,
that for homodyne detections on modes $A,B$ with CMs $\Gamma_{A}^{x'}$ and $\Gamma_{B}^{x'}$ homodyne detection on mode
$E$ with CM $\Gamma_{E}^{x'}$ minimizes the mutual information (\ref{f}), i.e.
$f(\gamma_{\pi},\Gamma_{A}^{x'},\Gamma_{B}^{x'},\Gamma_{E}^{x'})=
\mathop{\mbox{inf}}_{\Gamma_{E}}f\left(\gamma_{\pi},\Gamma_{A}^{x'},\Gamma_{B}^{x'},\Gamma_{E}\right)$,
and simultaneously the upper bound $U\left(\rho_{AB}^{GHZ}\right)$ is saturated, i.e.
%%%%%%%%%%%%%%%%%%%%%%%%%%%%%%%%%%%%%%%%%%%%%%%%%%%%%%%%%%%%%%%%%%%%%%%%%%%%
\begin{equation}\label{saturation}
U\left(\rho_{AB}^{GHZ}\right)=f(\gamma_{\pi},\Gamma_{A}^{x'},\Gamma_{B}^{x'},\Gamma_{E}^{x'}),
\end{equation}
%%%%%%%%%%%%%%%%%%%%%%%%%%%%%%%%%%%%%%%%%%%%%%%%%%%%%%%%%%%%%%%%%%%%%%%%%%%%%
where $\gamma_{\pi}$ denotes the CM of the purification of the state
$\rho_{AB}^{GHZ}$. The quantity
$f(\gamma_{\pi},\Gamma_{A}^{x'},\Gamma_{B}^{x'},\Gamma_{E}^{x'})$ is
thus the largest possible minimal mutual information with respect
to all Gaussian measurements on mode $E$, which finally yields
%%%%%%%%%%%%%%%%%%%%%%%%%%%%%%%%%%%%%%%%%%%%%%%%%%%%%%%%%%%%%%%%%%%%%%%%%%%%
\begin{equation}\label{GIEGHZ}
E_{\downarrow}^{G}\left(\rho_{AB}^{GHZ}\right)=f(\gamma_{\pi},\Gamma_{A}^{x'},\Gamma_{B}^{x'},\Gamma_{E}^{x'}).
\end{equation}
%%%%%%%%%%%%%%%%%%%%%%%%%%%%%%%%%%%%%%%%%%%%%%%%%%%%%%%%%%%%%%%%%%%%%%%%%%%%%

Let us start by noting that from the max-min inequality \cite{Boyd_04} it follows that GIE satisfies
inequality $E_{\downarrow}^{G}\left(\rho_{AB}^{GHZ}\right)\leq U\left(\rho_{AB}^{GHZ}\right)$, where
%%%%%%%%%%%%%%%%%%%%%%%%%%%%%%%%%%%%%%%%%%%%%%%%%%%%%%%%%%%%%%%%%%%%
\begin{equation}\label{U}
U\left(\rho_{AB}^{GHZ}\right)\equiv\mathop{\mbox{inf}}_{\Gamma_{E}}\mathop{\mbox{sup}}_{\Gamma_{A},\Gamma_{B}}
f(\gamma_{\pi},\Gamma_{A},\Gamma_{B},\Gamma_{E}).
\end{equation}
%%%%%%%%%%%%%%%%%%%%%%%%%%%%%%%%%%%%%%%%%%%%%%%%%%%%%%%%%%%%%%%%%%%%%
Next, consider the quantity
%%%%%%%%%%%%%%%%%%%%%%%%%%%%%%%%%%%%%%%%%%%%%%%%%%%%%%%%%%%%%%%%%%%%
\begin{equation}\label{IcG}
\mathcal{I}_{c}^{G}\left(\rho_{AB|E}\right)=\mathop{\mbox{sup}}_{\Gamma_{A},\Gamma_{B}}
f(\gamma_{\pi},\Gamma_{A},\Gamma_{B},\Gamma_{E}),
\end{equation}
%%%%%%%%%%%%%%%%%%%%%%%%%%%%%%%%%%%%%%%%%%%%%%%%%%%%%%%%%%%%%%%%%%%%%
which is the Gaussian classical mutual information of the
conditional quantum state $\rho_{AB|E}$ of modes $A$ and $B$ after
a measurement with CM $\Gamma_{E}$ on mode $E$ of the purification
with CM $\gamma_{\pi}$ \cite{Mista_11}. Let us take as the CM $\gamma_{\pi}$ the CM (\ref{gammaGHZ}),
$\gamma_{\pi}=\gamma_{ABE}^{GHZ}$, and denote as $\gamma_{AB|E}$ the CM of the conditional
state $\rho_{AB|E}$. As the CM $\gamma_{\pi}$ is symmetric under exchange of any pair of modes,
the CM $\gamma_{AB|E}$ is also symmetric for any CM $\Gamma_{E}$. To calculate the expression on the
RHS of Eq.~(\ref{IcG}) it is convenient first to express the CM $\gamma_{AB|E}$ in the standard
form (\ref{gammaABadd}) where $a=b$ due to the symmetry, i.e.
%%%%%%%%%%%%%%%%%%%%%%%%%%%%%%%%%%%%%%%%%%%%%%%%%%%%%%%%%%%%%%%%%%%%%%%%%%%%%%%%%%%%%%%%%%%%%%%%%%%%
\begin{eqnarray}\label{gammaABcondE}
\gamma_{AB|E}^{\rm st}=\left(\begin{array}{cccc}
a & 0 & c_{1} & 0\\
0 & a & 0 & c_{2}\\
c_{1} & 0 & a & 0\\
0 & c_{2} & 0 & a\\
\end{array}\right).
\end{eqnarray}
%%%%%%%%%%%%%%%%%%%%%%%%%%%%%%%%%%%%%%%%%%%%%%%%%%%%%%%%%%%%%%%%%%%%%%%%%%%%%%%%%%%%%%%%%%%%%%%%%%%%%%
The mutual information $f(\gamma_{\pi},\Gamma_{A},\Gamma_{B},\Gamma_{E})$ is then given
by Eq.~(\ref{f}) where $\sigma_{AB}=\gamma_{AB|E}^{\rm
st}+\Gamma_{A}\oplus\Gamma_{B}$. Further, in Ref.~\cite{Mista_11}
it was shown that for symmetric states with CM
(\ref{gammaABcondE}) the optimal measurements on modes $A$ and $B$
are always symmetric with CMs of the form
$\Gamma_{A}=\Gamma_{B}=\mbox{diag}(e^{-2t},e^{2t})$, $t\geq 0$.
From Eqs.~(\ref{f}) and (\ref{gammaABcondE}) it then follows that
%%%%%%%%%%%%%%%%%%%%%%%%%%%%%%%%%%%%%%%%%%%%%%%%%%%%%%%%%%%%%%%%%%%%
\begin{equation}\label{fsym}
f(\gamma_{\pi},\Gamma_{A},\Gamma_{B},\Gamma_{E})=-\ln\sqrt{h},
\end{equation}
%%%%%%%%%%%%%%%%%%%%%%%%%%%%%%%%%%%%%%%%%%%%%%%%%%%%%%%%%%%%%%%%%%%%%
where
%%%%%%%%%%%%%%%%%%%%%%%%%%%%%%%%%%%%%%%%%%%%%%%%%%%%%%%%%%%%%%%%%%%%%
\begin{equation}\label{oneminh}
h=\left[1-\frac{c_{1}^2}{(a+e^{-2t})^2}\right]\left[1-\frac{c_{2}^2}{(a+e^{2t})^2}\right].
\end{equation}
%%%%%%%%%%%%%%%%%%%%%%%%%%%%%%%%%%%%%%%%%%%%%%%%%%%%%%%%%%%%%%%%%%%%%
In order to maximize the function (\ref{fsym}) with respect to CMs
$\Gamma_{A}$ and $\Gamma_{B}$, we have to minimize the function on
the RHS of Eq.~(\ref{oneminh}) with respect to $t\geq0$. This can
be done by the following chain of inequalities:
%%%%%%%%%%%%%%%%%%%%%%%%%%%%%%%%%%%%%%%%%%%%%%%%%%%%%%%%%%%%%%%%%%%%%
\begin{widetext}
\begin{eqnarray}\label{chainineq}
\left[1-\frac{c_{1}^2}{(a+e^{-2t})^2}\right]\left[1-\frac{c_{2}^2}{(a+e^{2t})^2}\right]&\geq&\left[1-\frac{c_{1}^2}{(a+e^{-2t})^2}\right]\left[1-\frac{c_{1}^2}{(a+e^{2t})^2}\right]
\nonumber\\
&=&1-\frac{c_{1}^2}{a^2}+\frac{c_{1}^{2}}{a^2\left[1+a^2+2a\cosh(2t)\right]}\left[2+\frac{a^{2}c_{1}^2-(a^2-1)^2}{1+a^2+2a\cosh(2t)}\right]
\nonumber\\
&\geq& 1-\frac{c_{1}^2}{a^2}.
\end{eqnarray}
\end{widetext}
%%%%%%%%%%%%%%%%%%%%%%%%%%%%%%%%%%%%%%%%%%%%%%%%%%%%%%%%%%%%%%%%%%%%%
Here, the first inequality is a consequence of inequality
$c_{1}\geq |c_{2}|$ and the second inequality is fulfilled if
%%%%%%%%%%%%%%%%%%%%%%%%%%%%%%%%%%%%%%%%%%%%%%%%%%%%%%%%%%%%%%%%%%%%%
\begin{equation}\label{IcGcondition}
(2a+1)^2\geq a^2(a^2-c_{1}^2).
\end{equation}
%%%%%%%%%%%%%%%%%%%%%%%%%%%%%%%%%%%%%%%%%%%%%%%%%%%%%%%%%%%%%%%%%%%%%
Importantly, the lower bound $1-c_{1}^2/a^2$ in inequalities
(\ref{chainineq}) is tight because it can be achieved in the limit
$t\rightarrow+\infty$ which corresponds to the homodyne detection
of $x$-quadratures on both modes $A$ and $B$. We have thus arrived
to the finding that, for all symmetric states with CM
(\ref{gammaABcondE}) for which the parameters $a$ and $c_{1}$
satisfy inequality (\ref{IcGcondition}), the optimal measurement
in Gaussian classical mutual information (\ref{IcG}) is double
homodyne detection of $x$-quadratures. Hence, one gets
%%%%%%%%%%%%%%%%%%%%%%%%%%%%%%%%%%%%%%%%%%%%%%%%%%%%%%%%%%%%%%%%%%%%
\begin{equation}\label{IcGsym}
\mathcal{I}_{c}^{G}\left(\rho_{AB|E}\right)=\frac{1}{2}\ln\frac{a^2}{a^2-c_{1}^2}.
\end{equation}
%%%%%%%%%%%%%%%%%%%%%%%%%%%%%%%%%%%%%%%%%%%%%%%%%%%%%%%%%%%%%%%%%%%%%
Before going further let us note that the inequality
(\ref{IcGcondition}) has been derived in Ref.~\cite{Mista_11} as a
condition under which, for two-mode squeezed thermal states which possess
CMs (\ref{gammaABcondE}) with $c_{2}=-c_{1}$, the
optimal measurement in (\ref{IcG}) is double homodyne detection. The
present analysis thus extends the result of Ref.~\cite{Mista_11}
to all symmetric states satisfying condition
(\ref{IcGcondition}).

Moving to the derivation of the upper bound (\ref{U}) it is first
convenient to find a simpler condition under which the state
$\rho_{AB}^{GHZ}$ with CM (\ref{gammaABcondE}) satisfies
inequality (\ref{IcGcondition}). For this purpose we first rewrite
inequality (\ref{IcGcondition}) into an equivalent form
%%%%%%%%%%%%%%%%%%%%%%%%%%%%%%%%%%%%%%%%%%%%%%%%%%%%%%%%%%%%%%%%%%%%%
\begin{equation}\label{IcGcondition2}
2+\frac{1}{a}-s\geq 0,
\end{equation}
%%%%%%%%%%%%%%%%%%%%%%%%%%%%%%%%%%%%%%%%%%%%%%%%%%%%%%%%%%%%%%%%%%%%%
where we have introduced $s\equiv\sqrt{a^2-c_{1}^2}$. Since $a$ is
a symplectic eigenvalue of the local state of mode $A$, it
satisfies the inequality $a\geq1>0$ and therefore $1/a>0$.
Consequently, for CMs (\ref{gammaABcondE}) for which $s\leq2$ the
inequality (\ref{IcGcondition2}) is always satisfied. Let us
denote now as $a_{\rm max}$ the maximal value of the parameter $a$
of the CM (\ref{gammaABcondE}) over all CMs $\Gamma_{E}$ of Eve's
measurements. From the obvious inequality $a\geq s$ it then follows
that if
%%%%%%%%%%%%%%%%%%%%%%%%%%%%%%%%%%%%%%%%%%%%%%%%%%%%%%%%%%%%%%%%%%%%
\begin{equation}\label{amax}
a_{\rm max}\leq 2,
\end{equation}
%%%%%%%%%%%%%%%%%%%%%%%%%%%%%%%%%%%%%%%%%%%%%%%%%%%%%%%%%%%%%%%%%%%%%%
then $s\leq a\leq a_{\rm max}\leq 2$, and inequality
(\ref{IcGcondition2}) is therefore always satisfied. By
calculating $a_{\rm max}$ for the state $\rho_{AB}^{GHZ}$ and
using inequality (\ref{amax}), we can find easily a region of
the squeezing parameter $r$ for which the Gaussian classical
mutual information (\ref{IcG}) is given by formula
(\ref{IcGsym}).

To calculate the quantity $a_{\rm max}$ we first calculate the
local symplectic eigenvalue $a$ of CM (\ref{gammaABcondE}).
The CM describes a conditional quantum state obtained by a Gaussian
measurement with CM $\Gamma_{E}$ on mode $E$ of the purification
of the state $\rho_{AB}^{GHZ}$ with CM (\ref{gammaGHZ}). We further decompose the latter CM as
%%%%%%%%%%%%%%%%%%%%%%%%%%%%%%%%%%%%%%%%%%%%%%%%%%%%%%%%%%%%%%%%%%%%%%%%%%%%%%%%%%%%%%%%%%%%%%%%%%%%%%
\begin{equation}\label{decomposition}
\gamma_{ABE}^{GHZ}=S_{ABE}\left(\gamma_{AE}^{TMSV}\oplus\gamma_{B}^{\rm
sq}\right)S_{ABE}^{T},
\end{equation}
%%%%%%%%%%%%%%%%%%%%%%%%%%%%%%%%%%%%%%%%%%%%%%%%%%%%%%%%%%%%%%%%%%%%%%%%%%%%%%%%%%%%%%%%%%%%%%%%%%%%%%
where
%%%%%%%%%%%%%%%%%%%%%%%%%%%%%%%%%%%%%%%%%%%%%%%%%%%%%%%%%%%%%%%%%%%%%%%%%%%%%%%%%%%%%%%%%%%%%%%%%%%%
\begin{eqnarray}\label{gammaTMSV}
\gamma_{AE}^{TMSV}=\left(\begin{array}{cc}
\nu\openone_{2} & \sqrt{\nu^{2}-1}\sigma_{z}\\
\sqrt{\nu^{2}-1}\sigma_{z} & \nu\openone_{2}\\
\end{array}\right),
\end{eqnarray}
%%%%%%%%%%%%%%%%%%%%%%%%%%%%%%%%%%%%%%%%%%%%%%%%%%%%%%%%%%%%%%%%%%%%%%%%%%%%%%%%%%%%%%%%%%%%%%%%%%%%%%
is the CM of pure two-mode squeezed vacuum state with
%%%%%%%%%%%%%%%%%%%%%%%%%%%%%%%%%%%%%%%%%%%%%%%%%%%%%%%%%%%%%%%%%%%%%%%%%%%%%%%%%%%%%%%%%%%%%%%%%%%%%%
\begin{eqnarray}\label{nu}
\nu=\sqrt{x_{+}x_{-}}=\frac{1}{3}\sqrt{5+4\cosh(4r)},
\end{eqnarray}
%%%%%%%%%%%%%%%%%%%%%%%%%%%%%%%%%%%%%%%%%%%%%%%%%%%%%%%%%%%%%%%%%%%%%%%%%%%%%%%%%%%%%%%%%%%%%%%%%%%%%%
$\gamma_{B}^{\rm sq}=\mbox{diag}(e^{-2r},e^{2r})$, and
$S_{ABE}=(U_{AB}\oplus\openone_{E})(S_{A}\oplus\openone_{B}\oplus
S_{E})$, where
$S_{A}=S_{E}^{-1}=\mbox{diag}(\sqrt[4]{x_{-}/x_{+}},\sqrt[4]{x_{+}/x_{-}})$
and
%%%%%%%%%%%%%%%%%%%%%%%%%%%%%%%%%%%%%%%%%%%%%%%%%%%%%%%%%%%%%%%%%%%%%%%%%%%%%%%%%%%%%%%%%%%%%%%%%%%%
\begin{eqnarray}\label{UBS}
U_{AB}=\frac{1}{\sqrt{2}}\left(\begin{array}{cc}
\openone_{2} & \openone_{2}\\
\openone_{2} & -\openone_{2}\\
\end{array}\right).
\end{eqnarray}
%%%%%%%%%%%%%%%%%%%%%%%%%%%%%%%%%%%%%%%%%%%%%%%%%%%%%%%%%%%%%%%%%%%%%%%%%%%%%%%%%%%%%%%%%%%%%%%%%%%%%%
The decomposition (\ref{decomposition}) expresses the simple fact
that the CV GHZ state can be obtained by the mixing of mode $A$ of the
TMSV state with CM (\ref{gammaTMSV}) transformed by the squeezing
operation described by the matrix $S_{A}\oplus S_{E}$ with the
squeezed state in mode $B$ with CM $\gamma_{B}^{\rm sq}$ on a
balanced beam splitter described by the matrix $U_{AB}$
\cite{Giedke_03}. The conditional state $\rho_{AB|E}$ is then
obtained by performing a Gaussian measurement with CM $\Gamma_{E}$
on mode $E$ of the purification. Since the maximization of $a$ is
carried out over all CMs $\Gamma_{E}$, we can integrate the
squeezing transformation $S_{E}$ into the CM $\Gamma_{E}$ and can therefore drop the matrix $S_{E}$ from any further considerations. Let
us express now the CM of Eve's measurement as
$\Gamma_E=U(\varphi)\mbox{diag}(V_{x},V_{p})U^{T}(\varphi)$, where
%%%%%%%%%%%%%%%%%%%%%%%%%%%%%%%%%%%%%%%%%%%%%%%%%%%%%%%%%%%%%%%%%%%%%%%%%%%%%%%%%%%%%%%%%%%%%%%%%%%%%%%%%%%%%%%%%%%%%%%%%%%%%%%%%%%%%%%%
\begin{equation}\label{UV}
U(\varphi)= \left(
\begin{array}{cc}
\cos\varphi  & -\sin \varphi \\
\sin\varphi & \cos\varphi \\
\end{array}
\right),
\end{equation}
%%%%%%%%%%%%%%%%%%%%%%%%%%%%%%%%%%%%%%%%%%%%%%%%%%%%%%%%%%%%%%%%%%%%%%%%%%%%%%%%%%%%%%%%%%%%%%%%%%%%%%%%%%%%%%%%%%%%%%%%%%%%%%%%%%%%%%%%%%%
where $\varphi\in[0,\pi)$, $V_{x}\geq V_{p}\geq0$, and
$V_{x}V_{p}\geq 1$. By performing the Gaussian measurement with CM
$\Gamma_{E}$ on mode $E$ of the TMSV state with CM
(\ref{gammaTMSV}), mode $A$ collapses into the Gaussian state with
CM $\gamma_{A}^{\rm
cond}=U^{T}(\varphi)\mbox{diag}(\mathcal{V}_{x},\mathcal{V}_{p})U(\varphi)$,
where
%%%%%%%%%%%%%%%%%%%%%%%%%%%%%%%%%%%%%%%%%%%%%%%%%%%%%%%%%%%%%%%%%%%%%%%%%%%%%%%%%%%%%%%%%%%%%%%%%%%%
\begin{equation}\label{calVxVp}
\mathcal{V}_{x}=\frac{\nu V_{x}+1}{\nu+V_{x}},\quad
\mathcal{V}_{p}=\frac{\nu V_{p}+1}{\nu+V_{p}}.
\end{equation}
%%%%%%%%%%%%%%%%%%%%%%%%%%%%%%%%%%%%%%%%%%%%%%%%%%%%%%%%%%%%%%%%%%%%%%%%%%%%%%%%%%%%%%%%%%%%%%%%%%%%
Hence, at given $\nu$ the quantities $\mathcal{V}_{x}$ and
$\mathcal{V}_{p}$ will lie in the subset $\mathscr{M}$ of the
$(\mathcal{V}_{p},\mathcal{V}_{x})$-plane characterized by the
inequalities $1/\nu\leq\mathcal{V}_{p}\leq\nu$,
$1/\mathcal{V}_{p}\leq\mathcal{V}_{x}\leq\nu$ and
$\mathcal{V}_{x}\geq\mathcal{V}_{p}$. In other words, if
$\mathcal{V}_{p}\in[1/\nu,1]$ then
$\mathcal{V}_{x}\in[1/\mathcal{V}_{p},\nu]$, whereas if
$\mathcal{V}_{p}\in(1,\nu]$ then
$\mathcal{V}_{x}\in[\mathcal{V}_{p},\nu]$.
%%%%%%%%%%%%%%%%%%%%%%%%%%%%%%%%%%%%%%%%%%%%%%%%%%%%%%%%%%%%%%%%%%%%%%%%%%%%%%%%%%%%%%%%%%%%%%%%%%%%

Let us now return back to the derivation of the local symplectic
eigenvalue $a$. After the measurement on mode $E$ of the TMSV
state, mode $A$ collapses into a
Gaussian state with CM $\gamma_{A}^{\rm cond}$ which is
subsequently transformed by the squeezing operation described by
the matrix $S_{A}$ and then mixed with the squeezed state with CM
$\gamma_{B}^{\rm sq}$ on a balanced beam splitter characterized by
the matrix $U_{AB}$. This gives the conditional state
$\rho_{AB|E}$ with CM
%%%%%%%%%%%%%%%%%%%%%%%%%%%%%%%%%%%%%%%%%%%%%%%%%%%%%%%%%%%%%%%%%%%%%%%%%%%%%%%%%%%%%%%%%%%%%%%%%%%%%
\begin{eqnarray}\label{gammaABcondEexp}
\gamma_{AB|E}=U_{AB}(S_{A}\gamma_{A}^{\rm
cond}S_{A}^{T}\oplus\gamma_{B}^{\rm sq})U_{AB}^{T}.
\end{eqnarray}
%%%%%%%%%%%%%%%%%%%%%%%%%%%%%%%%%%%%%%%%%%%%%%%%%%%%%%%%%%%%%%%%%%%%%%%%%%%%%%%%%%%%%%%%%%%%%%%%%%%%%
Expressing further the latter CM in block form with respect to
$A|B$ splitting,
%%%%%%%%%%%%%%%%%%%%%%%%%%%%%%%%%%%%%%%%%%%%%%%%%%%%%%%%%%%%%%%%%%%%%%%%%%%%%%%%%%%%%%%%%%%%%%%%%%%%%%%%%%%%%%%%%%%%%%%%%%%%%%%%%%%%%%%%
\begin{equation}\label{gammaABcondEblock}
\gamma_{AB|E}=\left(
\begin{array}{cc}
A  & C \\
C & A \\
\end{array}
\right),
\end{equation}
%%%%%%%%%%%%%%%%%%%%%%%%%%%%%%%%%%%%%%%%%%%%%%%%%%%%%%%%%%%%%%%%%%%%%%%%%%%%%%%%%%%%%%%%%%%%%%%%%%%%%%%%%%%%%%%%%%%%%%%%%%%%%%%%%%%%%%%%%%%
one can calculate the entry $a$ of the CM (\ref{gammaABcondE})
from the formula $a=\sqrt{\mbox{det}A}$ in the form
%%%%%%%%%%%%%%%%%%%%%%%%%%%%%%%%%%%%%%%%%%%%%%%%%%%%%%%%%%%%%%%%%%%%%%%%%%%%%%%%%%%%%%%%%%%%%%%%%%%%%%%%%
\begin{eqnarray}\label{a}
a=\frac{\sqrt{1+\mathcal{V}_{x}\mathcal{V}_{p}+2[\mathcal{V}_{+}\cosh(2q)+\mathcal{V}_{-}\sinh(2q)\cos(2\varphi)]}}{2},\nonumber\\
\end{eqnarray}
%%%%%%%%%%%%%%%%%%%%%%%%%%%%%%%%%%%%%%%%%%%%%%%%%%%%%%%%%%%%%%%%%%%%%%%%%%%%%%%%%%
where $\mathcal{V}_{\pm}=(\mathcal{V}_{x}\pm\mathcal{V}_{p})/2$
and $q=r+\ln(\sqrt{x_{-}/x_{+}})/2$. As the inequality $\mathcal{V}_{-}\geq0$ holds $a$ is maximized if $\varphi=0$.
Further, the extremal equations $\partial a/\partial\mathcal{V}_{x}=0$ and $\partial a/\partial\mathcal{V}_{p}=0$ have no solution
in the interior of the set $\mathscr{M}$ and therefore the maximum lies on the boundary of the set. On the boundary the
local symplectic eigenvalue $a$ attains the maximum
%%%%%%%%%%%%%%%%%%%%%%%%%%%%%%%%%%%%%%%%%%%%%%%%%%%%%%%%%%%%%%%%%%%%%%%%%%%%%%%%%%%%%%%%%%%%%%%%%%%%%%%%%
\begin{eqnarray}\label{amax2}
a_{\rm max}=\frac{\sqrt{1+\nu^2+2\nu\cosh(2q)}}{2}=\nu
\end{eqnarray}
%%%%%%%%%%%%%%%%%%%%%%%%%%%%%%%%%%%%%%%%%%%%%%%%%%%%%%%%%%%%%%%%%%%%%%%%%%%%%%%%%%
for $\mathcal{V}_{x}=\mathcal{V}_{p}=\nu$. Next, making use of the explicit expression for the symplectic
eigenvalue $\nu$, Eq.~(\ref{nu}), and the inequality
(\ref{amax}), one finds after some algebra that the inequality (\ref{amax}) is fulfilled
if the squeezing parameter $r$ satisfies the inequality
%%%%%%%%%%%%%%%%%%%%%%%%%%%%%%%%%%%%%%%%%%%%%%%%%%%%%%%%%%%%%%%%%%%%%%%%%%%%%%%%%%
\begin{eqnarray}\label{rmax}
r\leq r_{\rm
th}\equiv\frac{1}{4}\arccosh\left(\frac{31}{4}\right)\doteq0.684.
\end{eqnarray}
%%%%%%%%%%%%%%%%%%%%%%%%%%%%%%%%%%%%%%%%%%%%%%%%%%%%%%%%%%%%%%%%%%%%%%%%%%%%%%%%%
Consequently, for the class of two-mode Gaussian states
$\rho_{AB}^{GHZ}$ for which $r$ satisfies inequality (\ref{rmax})
the Gaussian classical mutual information (\ref{IcG}) of the
conditional state $\rho_{AB|E}$ is for any Gaussian measurement on
mode $E$ given by the formula (\ref{IcGsym}). Later in this
section we show explicitly that the latter statement in fact holds
for all $r\geq0$. This is because for derivation of the inequality
(\ref{rmax}) we used the inequality (\ref{amax}) which is stronger
than the original inequality (\ref{IcGcondition2}), and therefore
the threshold squeezing for which the latter inequality is
satisfied is larger than $r_{\rm th}$. By minimizing the left-hand
side (LHS) of inequality (\ref{IcGcondition2}) over all CMs
$\Gamma_{E}$ one finds that the LHS has a lower bound of the form
%%%%%%%%%%%%%%%%%%%%%%%%%%%%%%%%%%%%%%%%%%%%%%%%%%%%%%%%%%%%%%%%%%%%%%%%%%%%%%%%%%%%%%
\begin{eqnarray}\label{lowerbound1}
2+\frac{1}{a}-s\geq
2+\frac{1}{\sqrt{x_{+}x_{-}}}-\frac{x_{-}}{e^{r}\sqrt{x_{+}}},
\end{eqnarray}
%%%%%%%%%%%%%%%%%%%%%%%%%%%%%%%%%%%%%%%%%%%%%%%%%%%%%%%%%%%%%%%%%%%%%%%%%%%%%%%%%%%%%%
where the parameters $x_{\pm}$ are defined below
Eq.~(\ref{gammaGHZ}). Further, the RHS of the latter inequality is
a monotonously decreasing function of the squeezing parameter $r$
which approaches the value $2-2/\sqrt{3}$ in the limit of
$r\rightarrow+\infty$. Hence, one finally gets the following lower
bound
%%%%%%%%%%%%%%%%%%%%%%%%%%%%%%%%%%%%%%%%%%%%%%%%%%%%%%%%%%%%%%%%%%%%%%%%%%%%%%%%%%%%%%
\begin{eqnarray}\label{lowerbound2}
2+\frac{1}{a}-s\geq 2-\frac{2}{\sqrt{3}}\doteq 0.845
\end{eqnarray}
%%%%%%%%%%%%%%%%%%%%%%%%%%%%%%%%%%%%%%%%%%%%%%%%%%%%%%%%%%%%%%%%%%%%%%%%%%%%%%%%%%%%%%
for the LHS of the inequality (\ref{IcGcondition2}) and therefore
the inequality is indeed satisfied for any $r\geq0$. Since the
minimization of the LHS of the inequality (\ref{IcGcondition2}) is
very similar to the minimization needed for calculation of the
upper bound (\ref{U}), it is more convenient first to carry out
the latter minimization. Explicit minimization of the LHS of the
inequality (\ref{IcGcondition2}) is postponed until near the end
of the present section.

In the last step of the calculation of the upper bound
$U\left(\rho_{AB}^{GHZ}\right)$, Eq.~(\ref{U}), we perform
minimization on the RHS of the following equation
%%%%%%%%%%%%%%%%%%%%%%%%%%%%%%%%%%%%%%%%%%%%%%%%%%%%%%%%%%%%%%%%%%%%
\begin{equation}\label{infIcG}
U\left(\rho_{AB}^{GHZ}\right)=\mathop{\mbox{inf}}_{\Gamma_{E}}\left[\frac{1}{2}\ln\left(\frac{a^2}{a^2-c_{1}^2}\right)\right]
\end{equation}
%%%%%%%%%%%%%%%%%%%%%%%%%%%%%%%%%%%%%%%%%%%%%%%%%%%%%%%%%%%%%%%%%%%%%
over all single-mode CMs $\Gamma_{E}$. This amounts to the
minimization of the ratio $c_{1}/a$, where $a$ is given in
Eq.~(\ref{a}). The parameter $c_{1}$ appearing in CM (\ref{gammaABcondE})
can be calculated as a larger eigenvalue of the matrix $QCQ^{T}$,
%%%%%%%%%%%%%%%%%%%%%%%%%%%%%%%%%%%%%%%%%%%%%%%%%%%%%%%%%%%%%%%%%%%%%%
\begin{eqnarray}\label{c1}
c_{1}=\frac{\mbox{Tr}(QCQ^{T})+\sqrt{[\mbox{Tr}(QCQ^{T})]^{2}-4\mbox{det}C}}{2},
\end{eqnarray}
%%%%%%%%%%%%%%%%%%%%%%%%%%%%%%%%%%%%%%%%%%%%%%%%%%%%%%%%%%%%%%%%%%%%%%
where $Q$ symplectically diagonalizes the matrix $A$, i.e.
$QAQ^{T}=a\openone_{2}$, and where we have used the equality
$\mbox{det}(QCQ^{T})=\mbox{det}C$. If we calculate explicitly
the CM (\ref{gammaABcondEexp}) we get after some algebra
%%%%%%%%%%%%%%%%%%%%%%%%%%%%%%%%%%%%%%%%%%%%%%%%%%%%%%%%%%%%%%%%%%%%%%
\begin{eqnarray}\label{detC}
\mbox{det}C=\frac{1}{2}(1+\mathcal{V}_{x}\mathcal{V}_{p})-a^2,
\end{eqnarray}
%%%%%%%%%%%%%%%%%%%%%%%%%%%%%%%%%%%%%%%%%%%%%%%%%%%%%%%%%%%%%%%%%%%%%%
and the utilization of the expression
$Q=\mbox{diag}(\sqrt[4]{\lambda_{2}/\lambda_{1}},\sqrt[4]{\lambda_{1}/\lambda_{2}})U(\theta)S_{A}^{-1}$, where
$U(\theta)S_{A}^{-1}A(S_{A}^{T})^{-1}U^{T}(\theta)=\mbox{diag}(\lambda_{1},\lambda_{2})$, $\lambda_{1}\geq\lambda_{2}$,
yields
%%%%%%%%%%%%%%%%%%%%%%%%%%%%%%%%%%%%%%%%%%%%%%%%%%%%%%%%%%%%%%%%%%%%%%
\begin{eqnarray}\label{TrQCQT}
\mbox{Tr}(QCQ^{T})=a\mbox{Tr}(CA^{-1})=\frac{(\mathcal{V}_{x}\mathcal{V}_{p}-1)}{2a}.
\end{eqnarray}
%%%%%%%%%%%%%%%%%%%%%%%%%%%%%%%%%%%%%%%%%%%%%%%%%%%%%%%%%%%%%%%%%%%%%%
Substituting now from Eqs.~(\ref{detC}) and (\ref{TrQCQT}) into Eq.~(\ref{c1})
one finds the ratio $c_{1}/a$ to be minimized in the form
%%%%%%%%%%%%%%%%%%%%%%%%%%%%%%%%%%%%%%%%%%%%%%%%%%%%%%%%%%%%%%%%%%%%%%
\begin{eqnarray}\label{c1a}
\frac{c_{1}}{a}=\frac{\mathcal{K}}{a^2}+\sqrt{\left(\frac{\mathcal{K}}{a^2}-1\right)^2-\frac{1}{a^2}}\equiv
g
\end{eqnarray}
%%%%%%%%%%%%%%%%%%%%%%%%%%%%%%%%%%%%%%%%%%%%%%%%%%%%%%%%%%%%%%%%%%%%%%
with $\mathcal{K}=(\mathcal{V}_{x}\mathcal{V}_{p}-1)/4$.

The minimal value of the ratio (\ref{c1a}) is easily found by a
direct substitution for $r=0$ which corresponds to the vacuum
density matrix $\rho_{AB}^{GHZ}$. In this case one has $\nu=1$
which implies $\mathcal{V}_{x}=\mathcal{V}_{p}=1$ and therefore $\mathcal{K}=0$ which gives
$g=\sqrt{(a^2-1)/a^2}$.
As for $r=0$ one further gets $q=0$ and we see from Eq.~(\ref{a})
that $a=1$ and thus $g=0$. Consequently, for $r=0$ the upper bound
(\ref{U}) vanishes, $U\left(\rho_{AB}^{GHZ}\right)=0$, and
therefore $E_{\downarrow}^{G}\left(\rho_{AB}^{GHZ}\right)=0$ which
is in accordance with our previous finding that GIE vanishes on
all separable states.

For $r>0$ the minimization of $g$, Eq.~(\ref{c1a}), with respect
to the variables $\varphi,\mathcal{V}_{x}$ and $\mathcal{V}_{p}$
is best performed if we introduce new variables
$\tau=\sqrt{\mathcal{V}_{x}\mathcal{V}_{p}}$ and
$z=\sqrt{\mathcal{V}_{x}/\mathcal{V}_{p}}$, where $\tau\in[1,\nu]$
and $z\in[1,\nu/\tau]$. Then, the task is to minimize $g$ in the
subset $\mathcal{O}$ of the three-dimensional space of the
variables $\varphi,\tau$ and $z$ characterized by the intervals
$\varphi\in[0,\pi]$, $\tau\in[1,\nu]$ and $z\in[1,\nu/\tau]$.
Note, that here and in what follows we admit for the sake of simplicity
also phase $\varphi=\pi$, although it is not necessary because the function
$g$ is $\pi$-periodic. Calculating now the extremal equations $\partial
g/\partial\varphi=0$ and $\partial g/\partial z=0$ and taking into
account inequality $c_{1}\geq0$ and inequality $a^2-c_{1}^2\geq1$
which has to be satisfied for any CM of a physical quantum state
\cite{Giedke_01b}, one finds that the equations are equivalent
to the extremal equations $\partial a/\partial\varphi=0$ and
$\partial a/\partial z=0$. The first extremal equation $\partial
a/\partial\varphi=0$ is satisfied if either $\varphi=0,\pi/2,\pi$
or $z=1$. Since for $\varphi=\pi/2$ the second equation $\partial
a/\partial z=0$ has no solution $z$ in the interval $[1,\nu]$ and
all points with $\varphi=0,\pi$ or $z=1$ lie on the boundary of
the set $\mathcal{O}$ the function $g$ has no stationary points in
the interior of the set $\mathcal{O}$. A detailed analysis of the
behavior of the function $g$ on the boundary of the set
$\mathcal{O}$ reveals that the candidates for extremes will lie on the following
parts of the boundary:

1. The segment $(\tau=\nu,z=1,\varphi\in[0,\pi])$ and the curves
$(\tau\in[1,\nu],z=\nu/\tau,\varphi=0)$ and
$(\tau\in[1,\nu],z=\nu/\tau,\varphi=\pi)$, where
%%%%%%%%%%%%%%%%%%%%%%%%%%%%%%%%%%%%%%%%%%%%%%%%%%%%%%%%%%%%%%%%%%%%%%%
\begin{eqnarray}\label{U1}
U_{1}\equiv\frac{1}{2}\ln\left(\frac{1}{1-g^2}\right)=\ln\left(\frac{e^{r}x_{+}}{\sqrt{x_{-}}}\right)
\end{eqnarray}
%%%%%%%%%%%%%%%%%%%%%%%%%%%%%%%%%%%%%%%%%%%%%%%%%%%%%%%%%%%%%%%%%%%%%%%
in all three cases. The value $U_{1}$ can be obtained in various
ways including homodyne detection of quadrature $p_{E}$ on mode
$E$, i.e.
$\Gamma_{E}=\Gamma_{E}^{p}\equiv\Gamma_{p}^{t\rightarrow+\infty}$,
where $\Gamma_{p}^{t}\equiv\mbox{diag}(e^{2t},e^{-2t})$, or by
tracing out mode $E$.

2. The segment $(\tau=1,z=1,\varphi\in[0,\pi])$ corresponding to
heterodyne detection on mode $E$, i.e. $\Gamma_{E}=\openone_{2}$,
where
%%%%%%%%%%%%%%%%%%%%%%%%%%%%%%%%%%%%%%%%%%%%%%%%%%%%%%%%%%%%%%%%%%%%%%%
\begin{eqnarray}\label{U2}
U_{2}=\ln\left(\frac{e^{r}\sqrt[4]{\frac{x_{-}}{x_{+}}}+e^{-r}\sqrt[4]{\frac{x_{+}}{x_{-}}}}{2}\right).
\end{eqnarray}
%%%%%%%%%%%%%%%%%%%%%%%%%%%%%%%%%%%%%%%%%%%%%%%%%%%%%%%%%%%%%%%%%%%%%%%

3. In the point $\tau=1$, $z=\nu$ and $\varphi=\pi/2$ which
corresponds to homodyne detection of quadrature $x_{E}$ on mode
$E$, i.e.
$\Gamma_{E}=\Gamma_{E}^{x}\equiv\Gamma_{x}^{t\rightarrow+\infty}$,
where $\Gamma_{x}^{t}\equiv\mbox{diag}(e^{-2t},e^{2t})$, and where
%%%%%%%%%%%%%%%%%%%%%%%%%%%%%%%%%%%%%%%%%%%%%%%%%%%%%%%%%%%%%%%%%%%%%%%
\begin{eqnarray}\label{U3}
U_{3}=\ln\left(\frac{x_{-}}{e^{r}\sqrt{x_{+}}}\right).
\end{eqnarray}
%%%%%%%%%%%%%%%%%%%%%%%%%%%%%%%%%%%%%%%%%%%%%%%%%%%%%%%%%%%%%%%%%%%%%%%

It remains to find the smallest of the three quantities
$U_{1},U_{2}$ and $U_{3}$. For this purpose it is convenient to
express them as $U_{j}=\ln[\cosh(p_{j})]$, $j=1,2,3$, where
$p_{1}=\ln(e^r\sqrt{x_{-}})$,
$p_{2}=\ln(e^r\sqrt[4]{x_{-}/x_{+}})$ and
$p_{3}=\ln(e^r/\sqrt{x_{+}})$. As for $r>0$ it holds that $\nu>1$,
we have $p_{1}-p_{3}=\ln\nu>0$ and therefore $p_1>p_{3}$ which
implies $U_{1}>U_{3}$. Similarly, one gets
$p_{2}-p_{3}=\ln\sqrt{\nu}>0$ and therefore $p_{2}>p_{3}$ which
gives finally $U_{2}>U_{3}$. Consequently, the sought upper bound
(\ref{U}) is equal to $U_{3}$, i.e.
%%%%%%%%%%%%%%%%%%%%%%%%%%%%%%%%%%%%%%%%%%%%%%%%%%%%%%%%%%%%%%%%%%%%%%%
\begin{eqnarray}\label{Ufinal}
U\left(\rho_{AB}^{GHZ}\right)=\ln\left(\frac{x_{-}}{e^{r}\sqrt{x_{+}}}\right)
\end{eqnarray}
%%%%%%%%%%%%%%%%%%%%%%%%%%%%%%%%%%%%%%%%%%%%%%%%%%%%%%%%%%%%%%%%%%%%%%%
and is achieved by triple homodyne detection of
$x$-quadratures.

In the final step of evaluation of the GIE we find for some fixed
measurements with CMs $\Gamma_{A}$ and $\Gamma_{B}$ on modes $A$
and $B$ of the purification with CM (\ref{gammaGHZ}) an infimum
over all CMs $\Gamma_{E}$ which saturates the upper bound
(\ref{Ufinal}),
$\mathop{\mbox{inf}}_{\Gamma_{E}}f(\gamma_{\pi},\Gamma_{A},\Gamma_{B},\Gamma_{E})=U\left(\rho_{AB}^{GHZ}\right)$.
This means that this is the largest infimum and hence GIE is equal
to the upper bound (\ref{Ufinal}). Let us denote as
$\Gamma_{j}^{x'}=S^{-1}\Gamma_{j}^{x}(S^{T})^{-1}$, $j=A,B$, where
the CM $\Gamma_{j}^{x}$ describes homodyne detection of quadrature
$x$ on mode $j$ and the single-mode symplectic matrix $S$ brings
the CM (\ref{gammaABcondEblock}) to the standard form
(\ref{gammaABcondE}), i.e. $(S\oplus S)\gamma_{AB|E}(S^{T}\oplus
S^{T})=\gamma_{AB|E}^{\rm st}$. Then
$\mathcal{I}_{c}^{G}\left(\rho_{AB|E}\right)=f(\gamma_{\pi},\Gamma_{A}^{x'},\Gamma_{B}^{x'},\Gamma_{E})$
and as we have shown above
$\mathop{\mbox{inf}}_{\Gamma_{E}}f(\gamma_{\pi},\Gamma_{A}^{x'},\Gamma_{B}^{x'},\Gamma_{E})=
f(\gamma_{\pi},\Gamma_{A}^{x'},\Gamma_{B}^{x'},\Gamma_{E}^{x'})=U\left(\rho_{AB}^{GHZ}\right)$,
where $\Gamma_{E}^{x'}=S_{E}\Gamma_{E}^{x}S_{E}^{T}$. Thus for
measurements with CMs $\Gamma_{A}^{x'}$ and $\Gamma_{B}^{x'}$ on
modes $A$ and $B$ of the purification with CM (\ref{gammaGHZ}) the
measurement on mode E with CM $\Gamma_{E}^{x'}$ gives the minimal
mutual information
$f(\gamma_{\pi},\Gamma_{A}^{x'},\Gamma_{B}^{x'},\Gamma_{E})$ which
is at the same time largest with respect to the CMs $\Gamma_{A}$
and $\Gamma_{B}$ as it saturates the upper bound (\ref{Ufinal}).
Consequently,
%%%%%%%%%%%%%%%%%%%%%%%%%%%%%%%%%%%%%%%%%%%%%%%%%%%%%%%%%%%%%%%%%%%%%%%
\begin{eqnarray}\label{GIEGHZ2}
E_{\downarrow}^{G}\left(\rho_{AB}^{GHZ}\right)=U\left(\rho_{AB}^{GHZ}\right)=\ln\left(\frac{x_{-}}{e^{r}\sqrt{x_{+}}}\right)
\end{eqnarray}
%%%%%%%%%%%%%%%%%%%%%%%%%%%%%%%%%%%%%%%%%%%%%%%%%%%%%%%%%%%%%%%%%%%%%%%
as we wanted to prove.

In the course of the derivation of the formula (\ref{GIEGHZ2}) we
have used the equality (\ref{IcGsym}) which was shown to be valid
for all CMs $\Gamma_{E}$ when the inequality (\ref{rmax}) is
fulfilled. Hence, the analytical expression of GIE in
Eq.~(\ref{GIEGHZ2}) is also valid for all states $\rho_{AB}^{GHZ}$
for which $r\leq0.684$. However, by repeating the previous
minimization of the ratio $g=c_1/a$, Eq.~(\ref{c1a}), in the
subset $\mathcal{O}$ for function $1/a-s$ on the LHS of inequality
(\ref{IcGcondition2}), we find that the inequality
(\ref{IcGcondition2}) and therefore also the formula
(\ref{GIEGHZ2}) holds for all $r\geq0$.

In order to show this, consider first the case when $r=0$. From
the previous results it then follows that $a=1$ and $c_1=0$ which
implies fulfillment of the inequality (\ref{IcGcondition2}). For
$r>0$ we can proceed as follows. Note first, that the minimization
of $1/a$, which is the first part of the function $1/a-s$, has
already been done by maximization of $a$. This gave the minimum
$1/a_{\rm max}=1/\nu=1/\sqrt{x_{+}x_{-}}$ which is attained if Eve
projects her mode onto an infinitely hot thermal state which is
equivalent to dropping of mode $E$. Now, if it happens that the
function $s$ defined below Eq.~(\ref{IcGcondition2}) attains its
maximum ($\equiv s_{\rm max}$) also when Eve drops her mode, then
$1/a_{\rm max}-s_{\rm max}$ represents the sought lower bound for
the function $1/a-s$. If we derive the function $s$ with respect
to $\varphi$ and $z$ and we use the expressions (\ref{a}) and
(\ref{c1a}), we arrive after some algebra at the following
expressions:
%%%%%%%%%%%%%%%%%%%%%%%%%%%%%%%%%%%%%%%%%%%%%%%%%%%%%%%%%%%%%%%%%%%%%%%%%%%%%%
\begin{eqnarray}\label{ders}
\frac{\partial s}{\partial
x}=-2\frac{(\tau^2-1)(a^2-c_{1}^2)}{4a^2c_1-(\tau^2-1)a}\frac{\partial
a}{\partial x},\quad x=\varphi,z.
\end{eqnarray}
%%%%%%%%%%%%%%%%%%%%%%%%%%%%%%%%%%%%%%%%%%%%%%%%%%%%%%%%%%%%%%%%%%%%%%%%%%%%%%%
Consequently, for $\tau>1$ the extremal equations $\frac{\partial
s}{\partial \varphi}=0$ and $\frac{\partial s}{\partial z}=0$ are
equivalent to the equations $\frac{\partial a}{\partial \varphi}=0$
and $\frac{\partial a}{\partial z}=0$. However, as it was shown
before, the latter equations have no solution in the interior of
the set $\mathcal{O}$ and thus the extremes will lie on the
boundary of the set $\mathcal{O}$. On the boundary plane $z=1$,
$\varphi\in[0,\pi]$ and $\tau\in[1,\nu]$ the function $s$ is
independent of $\varphi$ and it monotonously increases with $\tau$
attaining the maximum
%%%%%%%%%%%%%%%%%%%%%%%%%%%%%%%%%%%%%%%%%%%%%%%%%%%%%%%%%%%%%%%%%%%%%%%%%%%%%%
\begin{eqnarray}\label{smax}
s_{\rm max}=\frac{x_{-}}{e^{r}\sqrt{x_{+}}}
\end{eqnarray}
%%%%%%%%%%%%%%%%%%%%%%%%%%%%%%%%%%%%%%%%%%%%%%%%%%%%%%%%%%%%%%%%%%%%%%%%%%%%%%%
at $\tau=\nu$ which corresponds to dropping Eve's mode $E$. The
second boundary plane $\tau=1$, $\varphi\in[0,\pi]$ and
$z\in[1,\nu]$ corresponds to pure-state Gaussian measurements on
mode $E$ which yield pure conditional states $\rho_{AB|E}$ for
which $s=1$. On the boundary planes $\varphi=0$ and $\pi$,
$\tau\in[1,\nu]$ and $z\in[1,\nu/\tau]$ the extremal equation
$\frac{\partial a}{\partial z}=0$ does not have any solution for
$z\in[1,\nu/\tau]$ and therefore the extremes of $s$ will lie on
the boundary of the plane. Likewise, for the last boundary surface
$z=\nu/\tau$, $\varphi\in[0,\pi]$ and $\tau\in[1,\nu]$ the
extremal equations $\frac{\partial s}{\partial \varphi}=0$ and
$\frac{\partial s}{\partial \tau}=0$ have no solution in the
interior of the surface and therefore also in this case the
extremes will be on the boundary. We have already calculated the
extremes of $s$ on the boundary curves of the surface except for
the curves $z=\nu/\tau$, $\varphi=0,\pi$ and $\tau\in[1,\nu]$,
where $s$ attains the maximum (\ref{smax}) for $\tau=\nu$. In
summary, there are two extremes of the function $s$ on the set
$\mathcal{O}$. One is equal to $s=1$ and it is localized on the
boundary plane $\tau=1$, and the other one is equal to $s_{\rm
max}$, Eq.~(\ref{smax}), which lies on the segment $\tau=\nu$,
$z=1$ and $\varphi\in[0,\pi]$ which corresponds to dropping Eve's
mode E. Since one can easily show that $s_{\rm max}\geq1$ we
finally find that the function $s$ attains the maximum value
(\ref{smax}) exactly in the same points where the function also
$a$ is maximized. Thus, the function $1/a-s$ on the LHS of
inequality (\ref{IcGcondition2}) has the lower bound given in
inequality (\ref{lowerbound1}) which is further restricted from
below as in inequality (\ref{lowerbound2}). From that it follows
finally, that the inequality (\ref{IcGcondition2}) and hence also
the formula (\ref{GIEGHZ2}) for GIE of the state $\rho_{AB}^{GHZ}$
is indeed satisfied for all $r\geq0$ as we wanted to prove.

\begin{figure}[t]
\includegraphics[width=0.9\columnwidth]{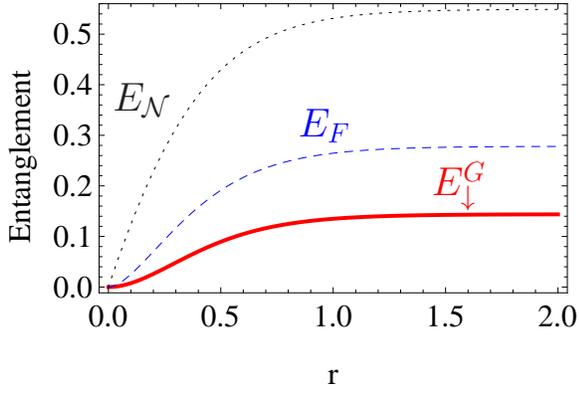}% This imports a EPS figure
\caption{(Color online) GIE $E_{\downarrow}^{G}$ (solid red
curve), entanglement of formation $E_{F}$ (dashed blue curve), and
logarithmic negativity $E_{\cal{N}}$ (dotted black curve) versus
the squeezing parameter $r$ for CM (\ref{gammaABGHZ}).}
\label{fig_CVGHZ}
\end{figure}

It might again be of interest to compare GIE for state $\rho_{AB}^{GHZ}$ with the GR2 entanglement. For a generally mixed
two-mode Gaussian state $\rho_{AB}$ with CM $\gamma_{AB}$ the GR2 entanglement is defined as \cite{Adesso_12}
%%%%%%%%%%%%%%%%%%%%%%%%%%%%%%%%%%%%%%%%%%%%%%%%%%%%%%%%%%%%%%%%%%%%
\begin{equation}\label{GR2}
E_{2}\left(\rho_{AB}\right)=\mathop{\mbox{inf}}_{\substack{\theta_{AB}\leq\gamma_{AB} \\ \mbox{\small det}\theta_{AB}=1}}\frac{1}{2}\ln\left(\mbox{det}\theta_{A}\right),
\end{equation}
%%%%%%%%%%%%%%%%%%%%%%%%%%%%%%%%%%%%%%%%%%%%%%%%%%%%%%%%%%%%%%%%%%%%
where the minimization is carried over all pure two-mode Gaussian states with CM $\theta_{AB}$ smaller than $\gamma_{AB}$.
The considered state $\rho_{AB}^{GHZ}$ is a reduced state of a pure three-mode state and therefore it
belongs to the class of Gaussian states with minimal partial uncertainty \cite{Adesso_06}
for which GR2 entanglement can be expressed analytically \cite{Adesso_12}. Making use of the fact that the state $\rho_{AB}^{GHZ}$ is a
reduction of the fully symmetric state with CM (\ref{gammaGHZ}) with local symplectic eigenvalue $\nu=\sqrt{x_{+}x_{-}}$, Eq.~(\ref{nu}),
GR2 entanglement reads explicitly as
%%%%%%%%%%%%%%%%%%%%%%%%%%%%%%%%%%%%%%%%%%%%%%%%%%%%%%%%%%%%%%%%%%%%
\begin{equation}\label{GR2GHZ}
E_{2}\left(\rho_{AB}^{GHZ}\right)=\frac{1}{2}\ln g'
\end{equation}
%%%%%%%%%%%%%%%%%%%%%%%%%%%%%%%%%%%%%%%%%%%%%%%%%%%%%%%%%%%%%%%%%%%%
with
%%%%%%%%%%%%%%%%%%%%%%%%%%%%%%%%%%%%%%%%%%%%%%%%%%%%%%%%%%%%%%%%%%%%%%%%%%%%%%%%%%%%%%%%%%%%%%%%%%%%%%%%%%%%%%%%%%%%%%%%%%%%%%%%%%%%%%%%%%%%%%
\begin{equation}
\label{gprimed}
g'=\left\{\begin{array}{lll} 1, & \textrm{if} & \nu=1;\\
\frac{\zeta}{8\nu^2}, & \textrm{if} & \nu>1,\\
\end{array}\right.
\end{equation}
%%%%%%%%%%%%%%%%%%%%%%%%%%%%%%%%%%%%%%%%%%%%%%%%%%%%%%%%%%%%%%%%%%%%%%%%%%%%%%%%%%%%%%%%%%%%%%%%%%%%%%%%%%%%%%%%%%%%%%%%%%%%%%%%%%%%%%%%%%%%%
where
%%%%%%%%%%%%%%%%%%%%%%%%%%%%%%%%%%%%%%%%%%%%%%%%%%%%%%%%%%%%%%%%%%%%%%%%%%%%%%%%%%%%%%%%%%%%%%%%%%%%%%%%%%%%%%%%%%%%%%%%%%%%%%%%%%%%%%%
\begin{equation}\label{tildebeta}
\zeta=3\nu^4+6\nu^2-1-\sqrt{(\nu^2-1)^3(9\nu^2-1)}.
\end{equation}
%%%%%%%%%%%%%%%%%%%%%%%%%%%%%%%%%%%%%%%%%%%%%%%%%%%%%%%%%%%%%%%%%%%%%%%%%%%%%%%%%%%%%%%%%%%%%%%%%%%%%%%%%%%%%%%%%%%%%%%%%%%%%%%%%%
Consider first the case $\nu=1$. From Eqs.~(\ref{GR2GHZ}) and
(\ref{gprimed}) it then follows that
$E_{2}\left(\rho_{AB}^{GHZ}\right)=0$. Equation (\ref{nu}) further
reveals that the equality $\nu=1$ is equivalent with the equality
$r=0$ which implies
$E_{\downarrow}^{G}\left(\rho_{AB}^{GHZ}\right)=0$ and thus GIE
coincides with GR2 entanglement. Moving to the case $\nu>1$ we see
that GR2 entanglement is equal to the RHS of Eq.~(\ref{GR2GHZ})
where $g'=\zeta/(8\nu^2)$ whereas from Eq.~(\ref{GIEGHZ2}) it
follows that
$E_{\downarrow}^{G}\left(\rho_{AB}^{GHZ}\right)=(\ln\tilde{g})/2$,
where $\tilde{g}\equiv x_{-}^2/(e^{2r}x_{+})$. Expressing now
$e^{\pm2r}$ using Eq.~(\ref{nu}) one gets
%%%%%%%%%%%%%%%%%%%%%%%%%%%%%%%%%%%%%%%%%%%%%%%%%%%%%%%%%%%%%%%%%%%%%%%%%%%%%%%%%%%%%%%%%%%%%%%%%%%%%%%%%%%%%%%%%%%%%%%%%%%%%%%%%%%%%%%
\begin{equation}\label{exppm2r}
e^{\pm2r}=\frac{\sqrt{9\nu^2-1}\pm3\sqrt{\nu^2-1}}{2\sqrt{2}},
\end{equation}
%%%%%%%%%%%%%%%%%%%%%%%%%%%%%%%%%%%%%%%%%%%%%%%%%%%%%%%%%%%%%%%%%%%%%%%%%%%%%%%%%%%%%%%%%%%%%%%%%%%%%%%%%%%%%%%%%%%%%%%%%%%%%%%%%%
which further gives
%%%%%%%%%%%%%%%%%%%%%%%%%%%%%%%%%%%%%%%%%%%%%%%%%%%%%%%%%%%%%%%%%%%%%%%%%%%%%%%%%%%%%%%%%%%%%%%%%%%%%%%%%%%%%%%%%%%%%%%%%%%%%%%%%%%%%%%
\begin{equation}\label{xpm}
x_{\pm}=\frac{e^{\pm 2r}+2e^{\mp 2r}}{3}=\frac{\sqrt{9\nu^2-1}\mp\sqrt{\nu^2-1}}{2\sqrt{2}}.
\end{equation}
%%%%%%%%%%%%%%%%%%%%%%%%%%%%%%%%%%%%%%%%%%%%%%%%%%%%%%%%%%%%%%%%%%%%%%%%%%%%%%%%%%%%%%%%%%%%%%%%%%%%%%%%%%%%%%%%%%%%%%%%%%%%%%%%%%
If we now rewrite the quantity $\tilde{g}$ as
$\tilde{g}=x_{-}^{2}(2\nu^2-x_{-}^2)/\nu^{2}$ and substitute to
the RHS for $x_{-}$ from Eq.~(\ref{xpm}) we finally find that
$\tilde{g}=\zeta/(8\nu^2)=g'$. In this way we have arrived at a
surprising result: GIE also coincides with the GR2 entanglement
for a one-parametric family of mixed two-mode Gaussian states
$\rho_{AB}^{GHZ}$, i.e.,
$E_{2}\left(\rho_{AB}^{GHZ}\right)=E_{\downarrow}^{G}\left(\rho_{AB}^{GHZ}\right)$.
A comparison of $E^G_\downarrow\left({\rho}_{AB}^{GHZ}\right)$,
Eq.~(\ref{GIEGHZ2}), with other entanglement measures is depicted
in Fig.~\ref{fig_CVGHZ}.

%%%%%%%%%%%%%%%%%%%%%%%%%%%%%%%%%%%%%%%%%%%%%%%%%%%%%%%%%%%%%%%%%%%%%%%%%%%%%%%%%%%%%%%%%%%%%%%%%%
%%%%%%%%%%%%%%%%%%%%%%%%%%%%%%%%%%%%%%%%%%%%%%%%%%%%%%%%%%%%%%%%%%%%%%%%%%%%%%%%%%%%%%%%%%%%%%%%%%%

The results presented in this section lay the foundations for further exploration of GIE which is deferred for further research.
This may include analytical or numerical evaluation of GIE for other two-mode Gaussian states with a three-mode purification or
states with some symmetry such as two-mode squeezed thermal states with standard-form CM (\ref{gammaABadd}), where $a=b$ and
$c_{2}=-c_{1}$. With these new results in hands we can also begin to explore the exciting question of the relation of
two seemingly very different quantities; GIE and GR2 entanglement.

%%%%%%%%%%%%%%%%%%%%%%%%%%%%%%%%%%%%%%%%%%%%%%%%%%%%%%%%%%%%%%%%%%%%%%%%%%%%%%%%%%%%%%%%%%%%%%%%%%%%%
\section{Lower bound on IE for the continuous-variable non-Gaussian Werner state}\label{sec_7}

So far, we have investigated the properties of IE, Eq.~(\ref{Edownarrow}), only in the
Gaussian scenario. Owing to the relative simplicity of Gaussian states
and measurements we were able to calculate IE analytically for some
nontrivial mixed Gaussian states and there in principle do not seem to be any
obstacles preventing its evaluation, at least numerically, for other
two-mode Gaussian states. A natural question that then arises is
whether IE can be calculated also for some non-Gaussian states. It is apparent that
this case will be much more complicated. Indeed, the calculation of IE for non-Gaussian
states involves optimization over all general non-Gaussian measurements and purifications
and therefore one is led to the apprehension that it will be infeasible, both
analytically and numerically. In this section we show that despite this complexity
a nontrivial analytical lower bound on IE can be found even in the case of some mixed
two-mode non-Gaussian states.

The states which we have in mind form the following two-parametric subfamily of the continuous-variable
Werner states \cite{Mista_02},
%%%%%%%%%%%%%%%%%%%%%%%%%%%%%%%%%%%%%%%%%%%%%%%%%%%%%%%%%%%%%%%%%%
\begin{equation}\label{Wernerrho0}
\rho_{0}=p|\psi(\lambda)\rangle_{AB}\langle \psi(\lambda)|+(1-p)|00\rangle_{AB}\langle00|,
\end{equation}
%%%%%%%%%%%%%%%%%%%%%%%%%%%%%%%%%%%%%%%%%%%%%%%%%%%%%%%%%%%%%%%%%%
where $0\leq p\leq 1$, which is just a mixture of a two-mode squeezed vacuum state (\ref{TMSV}) with the vacuum.
Making use of the partial transposition separability criterion \cite{PPT} one can show easily \cite{Mista_02} that for
$p>0$ the state (\ref{Wernerrho0}) is entangled. For calculation of IE we first need to find
a purification of the state (\ref{Wernerrho0}), which can be taken in the form
%%%%%%%%%%%%%%%%%%%%%%%%%%%%%%%%%%%%%%%%%%%%%%%%%%%%%%%%%%%%%%%%%%
\begin{equation}\label{psiWerner}
|\Psi\rangle_{ABE}=\sqrt{p}|\psi(\lambda)\rangle_{AB}|0\rangle_{E}+\sqrt{1-p}|00\rangle_{AB}|1\rangle_{E},
\end{equation}
%%%%%%%%%%%%%%%%%%%%%%%%%%%%%%%%%%%%%%%%%%%%%%%%%%%%%%%%%%%%%%%%%%
where Eve's purifying system is obviously a two-level quantum system (qubit) with basis vectors
$|0\rangle_{E}$ and $|1\rangle_{E}$. As the definition (\ref{Edownarrow}) of IE involves
minimization with respect to all purifications of the state (\ref{Wernerrho0}), we need to know the
form of an arbitrary purification which can be expressed as
%%%%%%%%%%%%%%%%%%%%%%%%%%%%%%%%%%%%%%%%%%%%%%%%%%%%%%%%%%%%%%%%%%
\begin{eqnarray}\label{psitildeWerner}
|\Psi'\rangle_{ABE'}&=&(\openone_{AB}\otimes V)|\Psi\rangle_{ABE}\nonumber\\
&=&\sqrt{p}|\Psi(\lambda)\rangle_{AB}V|0\rangle_{E}+\sqrt{1-p}|00\rangle_{AB}V|1\rangle_{E},\nonumber\\
\end{eqnarray}
%%%%%%%%%%%%%%%%%%%%%%%%%%%%%%%%%%%%%%%%%%%%%%%%%%%%%%%%%%%%%%%%%%
where $V$ is an isometry from a qubit Hilbert space
$\mathcal{H}_{E}$ to a Hilbert space $\mathcal{H}_{E'}$ of another
purifying system $E'$ and $\openone_{AB}$ is the identity operator
on modes $A$ and $B$. Instead of calculating the full IE for the
state (\ref{Wernerrho0}), here we will calculate its lower bound
%%%%%%%%%%%%%%%%%%%%%%%%%%%%%%%%%%%%%%%%%%%%%%%%%%%%%%%%%%%%%%%%%%%%%%%%
\begin{equation}\label{IElowerbound}
\mathcal{L}_{\downarrow}(\rho_{0})=\mathop{\mbox{inf}}_{\left\{\Pi_{E},|\Psi\rangle\right\}}\left[I\left(A;
B\downarrow E\right)\right]
\end{equation}
%%%%%%%%%%%%%%%%%%%%%%%%%%%%%%%%%%%%%%%%%%%%%%%%%%%%%%%%%%%%%%%%%%%%%%%%
for fixed photon counting measurements on modes $A$ and $B$. Assume therefore, that the projective measurements
$\{|m\rangle_{A}\langle m|,m=0,1,\ldots\}$ and $\{|n\rangle_{B}\langle n|,n=0,1,\ldots\}$ are carried out on
modes $A$ and $B$ of the purification (\ref{psitildeWerner}), whereas the subsystem $E'$ is exposed to
some generalized measurement $\{\Pi_{E'}(k)\}$. The outcomes of the measurements are then distributed according to
the probability distribution
%%%%%%%%%%%%%%%%%%%%%%%%%%%%%%%%%%%%%%%%%%%%%%%%%%%%%%%%%%%%%%%%%%%%%%%%%%%%%%%%%%%%%%%%%%%%%%%%%%%%%%%%%%%%%%%%%%%%%%%%%%%%%%%%%%%%%%%%%%%%%%
\begin{equation}
\label{pmnk}
p(m,n,k)=\left\{\begin{array}{lll} p_{E}(k)-\lambda^{2}p\Pi_{00}(k), & \textrm{if} & m=n=0;\\
p(1-\lambda^2)\lambda^{2m}\delta_{mn}\Pi_{00}(k), & & \textrm{otherwise},\\
\end{array}\right.
\end{equation}
%%%%%%%%%%%%%%%%%%%%%%%%%%%%%%%%%%%%%%%%%%%%%%%%%%%%%%%%%%%%%%%%%%%%%%%%%%%%%%%%%%%%%%%%%%%%%%%%%%%%%%%%%%%%%%%%%%%%%%%%%%%%%%%%%%%%%%%%%%%%%
where
%%%%%%%%%%%%%%%%%%%%%%%%%%%%%%%%%%%%%%%%%%%%%%%%%%%%%%%%%%%%%%%%%%
\begin{eqnarray}\label{pE}
p_{E}(k)&=&p\Pi_{00}(k)+\sqrt{p(1-p)(1-\lambda^{2})}[\Pi_{10}(k)+\Pi_{01}(k)]\nonumber\\
&&+(1-p)\Pi_{11}(k)
\end{eqnarray}
%%%%%%%%%%%%%%%%%%%%%%%%%%%%%%%%%%%%%%%%%%%%%%%%%%%%%%%%%%%%%%%%%%
is the probability distribution of measurement outcome $k$, where
%%%%%%%%%%%%%%%%%%%%%%%%%%%%%%%%%%%%%%%%%%%%%%%%%%%%%%%%%%%%%%%%%%
\begin{eqnarray}\label{Piijk}
\Pi_{ij}(k)\equiv\langle i|V^{\dag}\Pi_{E'}(k)V|j\rangle,\quad
i,j=0,1.
\end{eqnarray}
%%%%%%%%%%%%%%%%%%%%%%%%%%%%%%%%%%%%%%%%%%%%%%%%%%%%%%%%%%%%%%%%%%
By calculating the entropies $H(A,B,E),H(A,E)$ and $H(B,E)$ for
the distribution (\ref{pmnk}) and the marginal distributions
$p_{AE}(m,k)\equiv\sum_{n=0}^{\infty}p(m,n,k)$ and
$p_{BE}(n,k)\equiv\sum_{m=0}^{\infty}p(m,n,k)$, we further
observe, that $H(A,B,E)=H(A,E)=H(B,E)$ and the conditional mutual
information (\ref{conditionalI}) simplifies to
%%%%%%%%%%%%%%%%%%%%%%%%%%%%%%%%%%%%%%%%%%%%%%%%%%%%%%%%%%%%%%%%%%%%
\begin{equation}\label{conditionalIW}
I(A;B|E)=H(A)-I(A;E),
\end{equation}
%%%%%%%%%%%%%%%%%%%%%%%%%%%%%%%%%%%%%%%%%%%%%%%%%%%%%%%%%%%%%%%%%%%%
where $I(A;E)=H(A)+H(E)-H(A,E)$ is the mutual information of the
marginal distribution $p_{AE}(m,k)$.

\begin{figure}[ht]
\includegraphics[width=0.9\columnwidth]{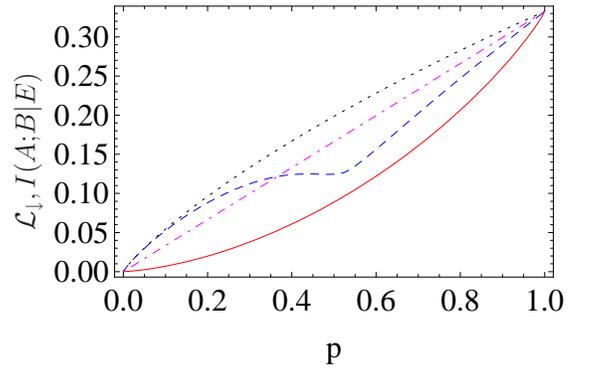}
\caption{(Color online) Lower bound $\mathcal{L}_{\downarrow}$,
Eq.~(\ref{LW}), (solid red curve) and $I(A;B|E)$ versus the
parameter $p$ for measurement in the $\{|0\rangle,|1\rangle\}$
basis (dash-dotted magenta curve) and the $\{|\pm\rangle\}$ basis
(dashed blue curve) and when Eve's qubit $E$ is dropped (dotted
black curve) for $\lambda=0.3$.} \label{fig_IEbound}
\end{figure}

Moving to the minimizations in Eq.~(\ref{IElowerbound}) we see
from Eq.~(\ref{conditionalIW}), that it boils down to the
maximization of the mutual information $I(A;E)$ over all channels
$E\rightarrow\tilde{E}$, isometries $V$, and measurements
$\{\Pi_{E'}(k)\}$ on purifying subsystem $E'$. Since sending a
random variable $E$ over a channel $P(\tilde{E}|E)$ cannot
increase the mutual information, i.e. $I(A;\tilde{E})\leq I(A;E)$,
it is best for Eve to not apply any channel to her
measurement outcomes. Further, as the operators
$V^{\dag}\Pi_{E'}(k)V$ appearing in Eq.~(\ref{Piijk}) are
Hermitian, positive semi-definite, and sum to a qubit identity
operator, they comprise a qubit generalized measurement.
Therefore, in Eq.~(\ref{IElowerbound}) we can omit minimization
with respect to all purifications and we can minimize only over
single-qubit measurements on the fixed purification
(\ref{psiWerner}). The latter minimization can be carried out with
the help of the following upper bound on the classical mutual
information \cite{Wu_09}
%%%%%%%%%%%%%%%%%%%%%%%%%%%%%%%%%%%%%%%%%%%%%%%%%%%%%%%%%%%%%%%%%%%%%%%%%%%%%%%%%%%%%%%%%%%%%%%%%%%
\begin{eqnarray}\label{IAEbound}
I(A;E)\leq\mbox{min}\left\{{\cal S}(\rho_{A}),{\cal S}(\rho_{E}),{\cal I}_{q}\left(\rho_{AE}\right)\right\},
\end{eqnarray}
%%%%%%%%%%%%%%%%%%%%%%%%%%%%%%%%%%%%%%%%%%%%%%%%%%%%%%%%%%%%%%%%%%%%%%%%%%%%%%%%%%%%%%%%%%%%%%%%%%%
where ${\cal S}(\rho_{A})$ and ${\cal S}(\rho_{E})$ are marginal
von Neumann entropies of the reduced states $\rho_{A}$ and
$\rho_{E}$, respectively, of subsystems $A$ and $E$ of the state
(\ref{psiWerner}) and ${\cal I}_{q}\left(\rho_{AE}\right)={\cal
S}(\rho_{A})+{\cal S}(\rho_{E})-{\cal S}(\rho_{AE})$ is the
quantum mutual information of the reduced state $\rho_{AE}$ of the
subsystem $(AE)$. From the purity of the state (\ref{psiWerner})
it further follows that ${\cal S}(\rho_{AE})={\cal S}(\rho_{B})$
whereas the symmetry of the state (\ref{Wernerrho0}) under the
exchange of modes $A$ and $B$ implies ${\cal S}(\rho_{A})={\cal
S}(\rho_{B})$. As a consequence, we get ${\cal
I}_{q}\left(\rho_{AE}\right)={\cal S}(\rho_{E})$ and for finding
of the minimum on the RHS of the inequality (\ref{IAEbound}) we
have to compare the marginal entropies ${\cal S}(\rho_{A})$ and
${\cal S}(\rho_{E})$. Using once again the purity argument we get
${\cal S}(\rho_{E})={\cal S}(\rho_{0})$ and therefore we need to
compare ${\cal S}(\rho_{A})$ with ${\cal S}(\rho_{0})$. In
Ref.~\cite{Mista_12} it was already shown with the help of the
majorization theory \cite{Wehrl_74} that ${\cal
S}(\rho_{A})\geq{\cal S}(\rho_{0})$ and the entropy ${\cal
S}(\rho_{0})$ has been calculated in the form:
%%%%%%%%%%%%%%%%%%%%%%%%%%%%%%%%%%%%%%%%%%%%%%%%%%%%%%%%%%%%%%%%%%%%%%%%%%%%%%%%%%%%%%%%%%%%%%%%%%%%%%
\begin{equation}\label{Srho0}
{\cal S}(\rho_{0})=-\sum_{i=1}^{2}e_{i}\ln e_{i},
\end{equation}
%%%%%%%%%%%%%%%%%%%%%%%%%%%%%%%%%%%%%%%%%%%%%%%%%%%%%%%%%%%%%%%%%%%%%%%%%%%%%%%%%%%%%%%%%%%%%%%%%%%%%%
where
%%%%%%%%%%%%%%%%%%%%%%%%%%%%%%%%%%%%%%%%%%%%%%%%%%%%%%%%%%%%%%%%%%%%%%%%%%%
\begin{eqnarray}\label{nurho0}
e_{1,2}=\frac{1\pm\sqrt{1-4p(1-p)\lambda^2}}{2}
\end{eqnarray}
%%%%%%%%%%%%%%%%%%%%%%%%%%%%%%%%%%%%%%%%%%%%%%%%%%%%%%%%%%%%%%%%%%%%%%%%%%%%%%%%%%
are the eigenvalues of the state (\ref{Wernerrho0}). Therefore, from Eq.~(\ref{IAEbound}) it follows that
the mutual information $I(A;E)$ has an upper bound equal to ${\cal S}(\rho_{E})={\cal S}(\rho_{0})$, Eq.~(\ref{Srho0}),
which is achieved by a measurement of the qubit $E$ in the eigenbasis
of the reduced state
%%%%%%%%%%%%%%%%%%%%%%%%%%%%%%%%%%%%%%%%%%%%%%%%%%%%%%%%%%%%%%%%%%%%%%%%%%%%%%%%%%%%%%%%
\begin{eqnarray}\label{rhoEW}
\rho_{E}&=&p|0\rangle_{E}\langle0|+\sqrt{p(1-p)(1-\lambda^2)}(|0\rangle_{E}\langle1|+|1\rangle_{E}\langle0|)\nonumber\\
&&+(1-p)|1\rangle_{E}\langle1|.
\end{eqnarray}
%%%%%%%%%%%%%%%%%%%%%%%%%%%%%%%%%%%%%%%%%%%%%%%%%%%%%%%%%%%%%%%%%%%%%%%%%%%%%%%%%%%%%%%%
Consequently, we get finally from Eqs.~(\ref{IElowerbound}) and (\ref{conditionalIW}) the analytical form of the lower bound on
IE
%%%%%%%%%%%%%%%%%%%%%%%%%%%%%%%%%%%%%%%%%%%%%%%%%%%%%%%%%%%%%%%%%%%%%%%%
\begin{equation}\label{LW}
\mathcal{L}_{\downarrow}(\rho_{0})=H(A)-{\cal S}(\rho_{E}),
\end{equation}
%%%%%%%%%%%%%%%%%%%%%%%%%%%%%%%%%%%%%%%%%%%%%%%%%%%%%%%%%%%%%%%%%%%%%%%%
where ${\cal S}(\rho_{E})$ is given by the RHS of Eq.~(\ref{Srho0}) and $H(A)$ is the
Shannon entropy of the photon-number distribution in mode $A$ of the state (\ref{Wernerrho0}) \cite{Mista_12}
 %%%%%%%%%%%%%%%%%%%%%%%%%%%%%%%%%%%%%%%%%%%%%%%%%%%%%%%%%%%%%%%%%%%%%%%%%%%%%%%%%%%
\begin{eqnarray}\label{SrhoB0}
H(A)=\mathcal{S}(\rho_{A})&=&-\left\{\ln(1-p\lambda^2)+p\lambda^2\ln\left[\frac{p(1-\lambda^2)}{1-p\lambda^2}\right]\right.\nonumber\\
&&\left.+\frac{2p\lambda^2\ln\lambda}{1-\lambda^2}\right\}.
\end{eqnarray}
%%%%%%%%%%%%%%%%%%%%%%%%%%%%%%%%%%%%%%%%%%%%%%%%%%%%%%%%%%%%%%%%%%%%%%%%%%%%%%
The lower bound (\ref{LW}) is depicted by a solid red curve in
Fig.~\ref{fig_IEbound}. For comparison, we have plotted into the
figure also cases when Eve just drops her qubit $E$ or she
measures it in the $\{|0\rangle,|1\rangle\}$ and
$\{|\pm\rangle=(|0\rangle\pm|1\rangle)/\sqrt{2})\}$ bases.

In the previous text we have performed minimization on the RHS of
Eq.~(\ref{Edownarrow}) for a particular fixed  measurement on
modes $A$ and $B$ of the purification (\ref{psiWerner}), which was
given by photon counting. In order to calculate the true IE, we
would have to carry out the minimization for arbitrary local
projective measurements on modes $A$ and $B$ and then we would
have to perform maximization over the measurements. Our derivation
given above thus yields only a lower bound on IE the actual value
of which can in fact be larger and may not be reached by photon
counting. However, photon counting on modes $A$ and $B$ of the
state (\ref{Wernerrho0}) gives $I(A;B)=\mathcal{S}(\rho_{A})$
\cite{Mista_12} which is the highest classical mutual information
one can get by locally measuring the state. This leads us to the
conjecture that this measurement is in fact optimal and therefore
the lower bound (\ref{LW}) coincides with IE. The proof or
disproof of this conjecture as well as further analysis of IE for
other non-Gaussian states is already beyond the scope of the
present paper and will be given elsewhere.

%%%%%%%%%%%%%%%%%%%%%%%%%%%%%%%%%%%%%%%%%%%%%%%%%%%%%%%%%%%%%%%%%%%%%%%%%%%%%%%%%%%%%%%%%%%%%%%%%%
%%%%%%%%%%%%%%%%%%%%%%%%%%%%%%%%%%%%%%%%%%%%%%%%%%%%%%%%%%%%%%%%%%%%%%%%%%%%%%%%%%%%%%%%%%%%%%%%%%%

%%%%%%%%%%%%%%%%%%%%%%%%%%%%%%%%%%%%%%%%%%%%%%%%%%%%%%%%%%%%%%%%%%%%%%%%%%%%%%%%%%%%%%%%%%%%%%%%%%%%%
\section{Conclusions}\label{sec_8}

In this paper we gave a detailed analysis of the properties of GIE, which is a new quantifier of
bipartite Gaussian entanglement introduced in Ref.~\cite{Mista_14}. The GIE is a Gaussian version
of a more general quantity IE which is a lower bound to the ``classical measure of entanglement''
\cite{Gisin_00} obtained by commuting the order of optimization in the definition of IE.

Initially, we have shown that the assumption of Gaussianity of all channels, states and measurements
greatly simplifies IE. First, we have proved that the classical channel on Eve's measurement outcomes can be integrated into
her measurement. In the next step, we have demonstrated that in the definition of IE we can use an arbitrary fixed
purification of a considered state and that we can omit the minimization over all purifications. As a result of these
simplifications, the GIE boils down to the optimized mutual information of a distribution of outcomes of Gaussian
measurements on subsystems $A$ and $B$ of a conditional state obtained by a Gaussian measurement on subsystem
$E$ of a Gaussian purification of the considered state.

Next, the simple form of GIE enabled us to show that it satisfies
some properties of a Gaussian entanglement measure. For this
purpose we have constructed for any Gaussian separable state a
Gaussian purification and a Gaussian measurement on the purifying
part $E$, which projects the state onto a product of states of
subsystems $A$ and $B$. This allowed us to prove two important
properties of GIE. First, making use of the result we have shown
that if a Gaussian state is separable then GIE vanishes. Second,
combining the result with the realization of LOCC operations by
teleportation with a separable shared state we have arrived to an
important observation that GIE does not increase under the
GLTPOCC. In particular, the monotonicity property implies that GIE
is invariant with respect to all local Gaussian unitary
operations.

Finally, we have calculated analytically GIE for two simple
classes of two-mode Gaussian states. For pure Gaussian states GIE
is equal to the GR2 entanglement \cite{Adesso_12} whereas equality
to the entropy of entanglement is established provided that Alice
and Bob are allowed to perform non-Gaussian measurements. An
analytical formula for GIE has been also derived for
one-parametric family of two-mode reductions of the three-mode CV
GHZ state, which was also found to be equal to the GR2
entanglement. Last but not least, we have also extended our
analysis of the proposed entanglement quantifier to a non-Gaussian
case by calculating a lower bound on IE for a particular subset of
a set of two-mode continuous-variable Werner states.

The results obtained in the present paper rise several questions
which remain open for further research. First, it is imperative to
know, whether GIE is monotonic under all (including
trace-decreasing) GLOCC operations. If answered in affirmative, we
could call GIE a Gaussian entanglement measure. Another important
question concerns computability of GIE on other Gaussian states.
Knowing GIE for other Gaussian states, one can then further
investigate a rather surprising finding that GIE and GR2
entanglement are equal on some Gaussian states. A proof showing
the equality of the two quantities on all bipartite Gaussian
states would link GR2 entanglement with the secret-key agreement
protocol \cite{Maurer_93} and what is more, this would also mean,
that GIE possesses all the properties of GR2 entanglement
including, e.g., monogamy. Finally, GIE is a faithful quantity
\cite{Mista_14} which is nonzero on all entangled states and
therefore it opens a possibility to quantify the amount of
entanglement in Gaussian bound entangled states \cite{Werner_01}.

We hope that the results presented here will further stimulate
research in the field of the computable and physically meaningful
entanglement measures.
%%%%%%%%%%%%%%%%%%%%%%%%%%%%%%%%%%%%%%%%%%%%%%%%%%%%%%%%%%%%%%%%%%%%%%%%%%%%%%%%%%%%%%%%%%%%%%%%%%%%%
\acknowledgments

We would like to thank J. Fiur\'a\v{s}ek and G. Adesso for
fruitful discussions. L. M. acknowledges the Project No.
P205/12/0694 of GA\v{C}R.
%%%%%%%%%%%%%%%%%%%%%%%%%%%%%%%%%%%%%%%%%%%%%%%%%%%%%%%%%%%%%%%%%%%%%%%%%%%%%%%%%%%%%%%

%%%%%%%%%%%%%%%%%%%%%%%%%%%%%%%%%%%%%%%%%%%%%%%%%%%%%%%%%%%%%%%%%%%%%%%%%%%%
%%%%%%%%%%%%%%%%%%%%%%%%%%%%%%%%%%%%%%%%%%%%%%%%%%%%%%%%%%%%%%%%%%%%%%%%%%%%

\begin{thebibliography}{99}
%%%%%%%%%%%%%%%%%%%%%%%%%%%%%%%%%%%%%%%%%%%%%%%%%%%%%%%%%%%%%%%%%%
\bibitem{Bennett_96a} C. H. Bennett, D. P. DiVincenzo, J. A. Smolin, and W. K.
Wootters, Phys. Rev. A {\bf 54}, 3824 (1996).

\bibitem{Shor_95} P. Shor, Phys. Rev. A {\bf 52}, 2493 (1995).

\bibitem{Nielsen_00} M. A. Nielsen and I. L. Chuang, {\it Quantum Computation and Quantum Information} (Cambridge University Press, Cambridge, UK, 2000).

\bibitem{MHorodecki_98} M. Horodecki, P. Horodecki, and R. Horodecki, Phys. Rev. Lett. {\bf 80}, 5239 (1998).

\bibitem{Cubitt_03} T. S. Cubitt, F. Verstraete, W. D\"{u}r, and J. I. Cirac, Phys. Rev. Lett. {\bf 91}, 037902 (2003).

\bibitem{Shor_03} P. W. Shor, J. A. Smolin, and A. V. Thapliyal, Phys. Rev. Lett. {\bf 90}, 107901 (2003).

\bibitem{Gisin_00} N. Gisin and S. Wolf, in {\it Proceedings of CRYPTO 2000}, Lecture Notes in
Computer Science Vol. 1880 (Springer-Verlag, Berlin, 2000), p. 482.

\bibitem{Acin_04} A. Ac\'{\i}n, J. I. Cirac, and Ll. Masanes, Phys. Rev. Lett. {\bf 92}, 107903 (2004).

\bibitem{Bae_09} J. Bae, T. Cubitt, and A. Ac\'{\i}n, Phys. Rev. A {\bf 79}, 032304 (2009).

\bibitem{Pretticio_11} G. Prettico and J. Bae, Phys. Rev. A {\bf
83}, 042336 (2011).

\bibitem{Maurer_93} U. Maurer, IEEE Trans. Inf. Theory {\bf 39}, 733 (1993).

\bibitem{Renner_03} R. Renner and S. Wolf, {\it Advances in Cryptology}, {\it EUROCRYPT 2003},
Lecture Notes in Computer Science Vol. 2656 (Springer-Verlag, Berlin, 2003), p. 562.

\bibitem{PC} By public communication we mean communication of honest parties via
an authentic but otherwise insecure classical communication
channel such that Eve hears the whole communication between the
parties but cannot tamper with it \cite{Maurer_99}.

\bibitem{Collins_02} D. Collins and S. Popescu, Phys. Rev. A {\bf 65}, 032321 (2002).

\bibitem{Acin_05} A. Ac\'{\i}n and N. Gisin, Phys. Rev. Lett. {\bf 94}, 020501 (2005).

\bibitem{Maurer_99} U. M. Maurer and S. Wolf, IEEE Trans. Inf. Theory {\bf 45}, 499 (1999).

\bibitem{Shannon_48} C. E. Shannon, Bell Syst. Tech. J. {\bf 27}, 379 (1948).

\bibitem{secret bit} A secret bit is a probability distribution satisfying
$P(A,B,E)=P(A,B)P(E)$ and $P(A=B=0)=P(A=B=1)=1/2$.

\bibitem{Vidal_00} G. Vidal, J. Mod. Opt. {\bf 47}, 355 (2000).

\bibitem{Christandl_04} M. Christandl and A. Winter, J. Math.
Phys. {\bf 45}, 829 (2004).

\bibitem{Mista_14} L. Mi\v{s}ta Jr. and R. Tatham, submitted.

\bibitem{Boyd_04} S. Boyd and L. Vandenberghe, {\it Convex
Optimization} (Cambridge University Press, Cambridge, 2004).

\bibitem{Loock_00} P. van Loock and S. L. Braunstein, Phys. Rev. Lett. {\bf 84},
3482 (2000).

\bibitem{Adesso_12} G. Adesso, D. Girolami, and A. Serafini, Phys. Rev. Lett. {\bf 109}, 190502 (2012).

\bibitem{Mista_02} L. Mi\v{s}ta Jr., R. Filip, and J. Fiur\'a\v{s}ek, Phys. Rev. A {\bf 65}, 062315 (2002).

\bibitem{Williamson_36} J. Williamson, Am. J. Math. {\bf 58}, 141 (1936).

\bibitem{Fiurasek_07} J. Fiur\'a\v{s}ek and L. Mi\v{s}ta Jr., Phys. Rev. A {\bf 75}, 060302(R) (2007).

\bibitem{CCM} A CCM is a real, symmetric and postive-semidefinite matrix.

\bibitem{Horn_85} R. A. Horn and C. R. Johnson, {\it Matrix Analysis} (Cambridge University Press, Cambridge, England, 1985).

\bibitem{Caruso_11} F. Caruso, J. Eisert, V. Giovannetti, and A.
S. Holevo, Phys. Rev. A {\bf 84}, 022306 (2011).

\bibitem{Giedke_03} G. Giedke, J. Eisert, J. I. Cirac, and M. B.
Plenio, Quantum. Inf. Comput. {\bf 3}, 211 (2003).

\bibitem{Magnin_10} L. Magnin, F. Magniez, A. Leverrier, and N. J.
Cerf, Phys. Rev. A {\bf 81}, 010302(R) (2010).

\bibitem{Gelfand_57} I. M. Gelfand and A. M. Yaglom, Usp. Mat. Nauk {\bf 12}, 3 (1957).

\bibitem{Horodecki_09} R. Horodecki, P. Horodecki, M. Horodecki,
and K. Horodecki, Rev. Mod. Phys. {\bf 81}, 865 (2009).

\bibitem{Werner_01} R. F. Werner and M. M. Wolf, Phys. Rev. Lett. {\bf 86}, 3658 (2001).

\bibitem{Giedke_02} G. Giedke and J. I. Cirac, Phys. Rev. A {\bf 66}, 032316 (2002).

\bibitem{Wolf_04} M. M. Wolf, G. Giedke, O. Kr\"{u}ger, R. F.
Werner, and J. I. Cirac, Phys. Rev. A {\bf 69}, 052320 (2004).

\bibitem{Jamiolkowski_72} A. Jamio{\l}kowski, Rep. Math. Phys. {\bf 3}, 275 (1972); M.-D. Choi, Lin. Alg. Appl. {\bf 10}, 285 (1975).

\bibitem{Fiurasek_02} J. Fiur\'a\v{s}ek, Phys. Rev. Lett. {\bf 89}, 137904 (2002).

\bibitem{Braunstein_98} S. L. Braunstein and H. J. Kimble, Phys. Rev. Lett. {\bf 80}, 869 (1998).

\bibitem{Hofmann_00} H. F. Hofmann, T. Ide, T. Kobayashi, and A. Furusawa, Phys. Rev. A
{\bf 62}, 062304 (2000).

\bibitem{Giedke_01} G. Giedke, B. Kraus, M. Lewenstein, and J. I. Cirac, Phys. Rev. A {\bf
64}, 052303 (2001).

\bibitem{Holevo_11} A. S. Holevo, arXiv:1004.0196.

\bibitem{Eisert_03} J. Eisert and M. B. Plenio, Int. J. Quant. Inf. {\bf 1}, 479 (2003).
\bibitem{Weedbrook_12} Ch. Weedbrook, S. Pirandola, R.
Garc\'{\i}a-Patr\'{o}n, N. J. Cerf, T. C. Ralph, J. H. Shapiro,
and S. Lloyd, Rev. Mod. Phys. {\bf 84}, 621 (2012).

\bibitem{Cover_06} T. M. Cover and J. A. Thomas, {\it Elements of Information
Theory} (Wiley, New Jersey, 2006).

\bibitem{Simon_00} R. Simon, Phys. Rev. Lett. {\bf 84}, 2726 (2000).

\bibitem{Terhal_02} B. M. Terhal, M. Horodecki, D. W. Leung, and D. P.DiVincenzo,  J. Math. Phys. {\bf 43}, 4286 (2002);
D. P. DiVincenzo, M. Horodecki, D. Leung, J. Smolin, and B. M.
Terhal,  Phys. Rev. Lett. {\bf 92}, 067902 (2004).

\bibitem{Mista_11} L. Mi\v{s}ta Jr., R. Tatham, D. Girolami, N. Korolkova, and G. Adesso, Phys. Rev. A {\bf 83}, 042325 (2011).

\bibitem{Bennett_96} C. H. Bennett, G. Brassard, S. Popescu, B. Schumacher, J. A. Smolin, and W. K. Wootters, Phys. Rev. Lett. {\bf 76}, 722
(1996).

\bibitem{Parker_00} S. Parker, S. Bose, and M. B. Plenio, Phys.
Rev. A {\bf 61}, 032305 (2000).

\bibitem{Wu_09} S. Wu, U. V. Poulsen, and K. M{\o}lmer, Phys. Rev. A {\bf 80}, 032319 (2009).

\bibitem{Vidal_02} G. Vidal and R. F. Werner, Phys. Rev. A {\bf
65}, 032314 (2002).

\bibitem{Eisert_PhD} J. Eisert, Ph.D. thesis, University of
Potsdam, 2001.

\bibitem{Ohliger_10} M. Ohliger, K. Kieling, and J. Eisert, Phys.
Rev. A {\bf 82}, 042336 (2010).

\bibitem{Giedke_01b} G. Giedke, L.-M. Duan, J. I. Cirac, and P. Zoller, Quantum. Inf. Comput. {\bf 3}, 79 (2001).

\bibitem{Adesso_06} G. Adesso, A. Serafini, and F. Illuminati, Phys. Rev. A {\bf 73}, 032345 (2006).

\bibitem{PPT} A. Peres, Phys. Rev. Lett. {\bf 77}, 1413 (1996); M. Horodecki, P. Horodecki, and R. Horodecki, Phys. Lett. A \textbf{223}, 1 (1996).

\bibitem{Wehrl_74} A. Wehrl, Rep. Math. Phys. {\bf 6}, 15 (1974).

\bibitem{Mista_12} R. Tatham, L. Mi\v{s}ta Jr., G. Adesso, and N. Korolkova, Phys. Rev. A {\bf 85}, 022326 (2012).

\end{thebibliography}
\end{document}